\documentclass[iop]{emulateapj}
\usepackage{graphicx}
\usepackage{natbib}
\newcommand{\rff}[1]{Fig.\ \ref{#1}}
\newcommand{\Rff}[1]{Figure \ref{#1}}
\renewcommand{\vec}[1]{\mbox{\protect\boldmath$#1$}}

\begin{document}
\title{Transition from Regular to Chaotic Circulation in Magnetized Coronae near Compact Objects}
\shorttitle{Transition from Regular to Chaotic Circulation\ldots}
\shortauthors{Kop\'{a}\v{c}ek et al.}

\author{O. Kop\'{a}\v{c}ek, V. Karas}\affil{Astronomical Institute, Academy of Sciences, Bo\v{c}n\'{i} II 1401, CZ-141\,31~Prague, Czech~Republic.}
\author{J. Kov\'{a}\v{r}, Z. Stuchl\'{\i}k}\affil{Institute of Physics, Faculty of Philosophy and Science, Silesian University in Opava, Bezru\v{c}ovo n\'{a}m.~13, CZ-746\,01~Opava, Czech~Republic.}


\begin{abstract}
Accretion onto black holes and compact stars brings material in a zone
of strong gravitational and electromagnetic fields. We study dynamical
properties of motion of electrically charged particles forming a 
highly diluted medium (a corona) in the regime of strong gravity and
large-scale (ordered) magnetic field.

We start our work from a system that allows regular motion, then we
focus on the onset of chaos. To this end,
we investigate the case of a rotating black hole immersed in a
weak, asymptotically uniform magnetic field. We also consider
a magnetic star, approximated by the Schwarzschild metric and a test
magnetic field of a rotating dipole. These are two model examples of
systems permitting energetically bound, off-equatorial motion of matter
confined to the halo lobes that encircle the central body. Our approach
allows us to address the question of whether the spin parameter of the
black hole plays any major role in determining the degree of the
chaoticness.

To characterize the motion, we construct the Recurrence Plots (RP) and
we compare them with Poincar\'e surfaces
of section. We describe the Recurrence Plots in terms of
the Recurrence Quantification Analysis (RQA), which allows us to
identify the transition between different dynamical regimes. We demonstrate
that this new technique is able to detect the chaos onset very efficiently,
and to provide its quantitative measure. The chaos
typically occurs when the conserved energy is raised to a sufficiently
high level that allows the particles to traverse the equatorial plane.
We find that the role of the black-hole spin in setting the chaos is
more complicated than initially thought.
\end{abstract}

\keywords{acceleration of particles, black hole physics, magnetic fields, methods: numerical}

\section{Introduction}\label{Sec:Intro}
The role of magnetic fields near strongly gravitating objects has been
subject of many investigations \citep[e.g.][]{punsly08}. They are relevant for 
accretion disks that may be embedded in large-scale magnetic fields, for
example when the accretion flow penetrates close to a neutron star
\citep{lipunov92,halo2_11}. Outside the main body of the accretion
disk, i.e.\ above and below the equatorial plane, the accreted material
forms a highly diluted environment, a `corona', where the density of
matter is low and the mean free path of particles is large in comparison
with the characteristic length-scale, i.e.\ the gravitational radius of
the central body, $R_{\rm g}\equiv GM/c^2\approx 1.5(M/M_\odot)\;$km,
where $M$ is the central mass. The origin of the coronal flows and the
relevant processes governing their structure are still unclear. In this
context we discuss motion of electrically charged particles outside the
equatorial plane.

Off-equatorial, energetically bound motion of charged particles in
strong gravitational and electromagnetic fields is pertinent to the
description of accretion disk coronae around black holes and compact
(neutron) stars. In our previous papers \citep{halo1, halo2}  we
discussed the existence of energetically-bound stable orbits of charged
particles occurring outside the equatorial plane, extending thus a large
variety of complementary studies
\citep[e.g.,][and further references cited therein]{halo2_4,halo2_5,halo2_3,halo2_6,aliev02}. Particles on
off-equatorial stable trajectories form a coronal flow that is possible
at certain radii and for certain combinations of the model parameters,
namely, the specific charge of the particles, the conserved energy and
the angular momentum of the particle motion, the strength and
orientation of the magnetic field, and the spin of the central body.

We assume that the magnetic field permeating the corona has a
large-scale (ordered) component \citep{bisnovatyi07}. In this case,
charged particles can be trapped in toroidal regions, extending symmetrically
above and below the equatorial plane and forming two halo lobes. However,
this trapping happens only for certain combinations of model parameters
\citep{halo2}.

We consider two types of the model setup: a rotating (Kerr) black hole
in an asymptotically uniform magnetic field parallel to the symmetry
axis \citep{wald,tomimatsu01,koide04,koide06}, and a non-rotating star (described
by the Schwarzschild metric) endowed with a rotating magnetic dipole
field \citep{petterson,sengupta}. Both cases can be regarded as integrable 
systems with the electromagnetic field acting as a perturbation.

\begin{figure*}[htb]
\centering
\includegraphics[scale=0.6, clip]{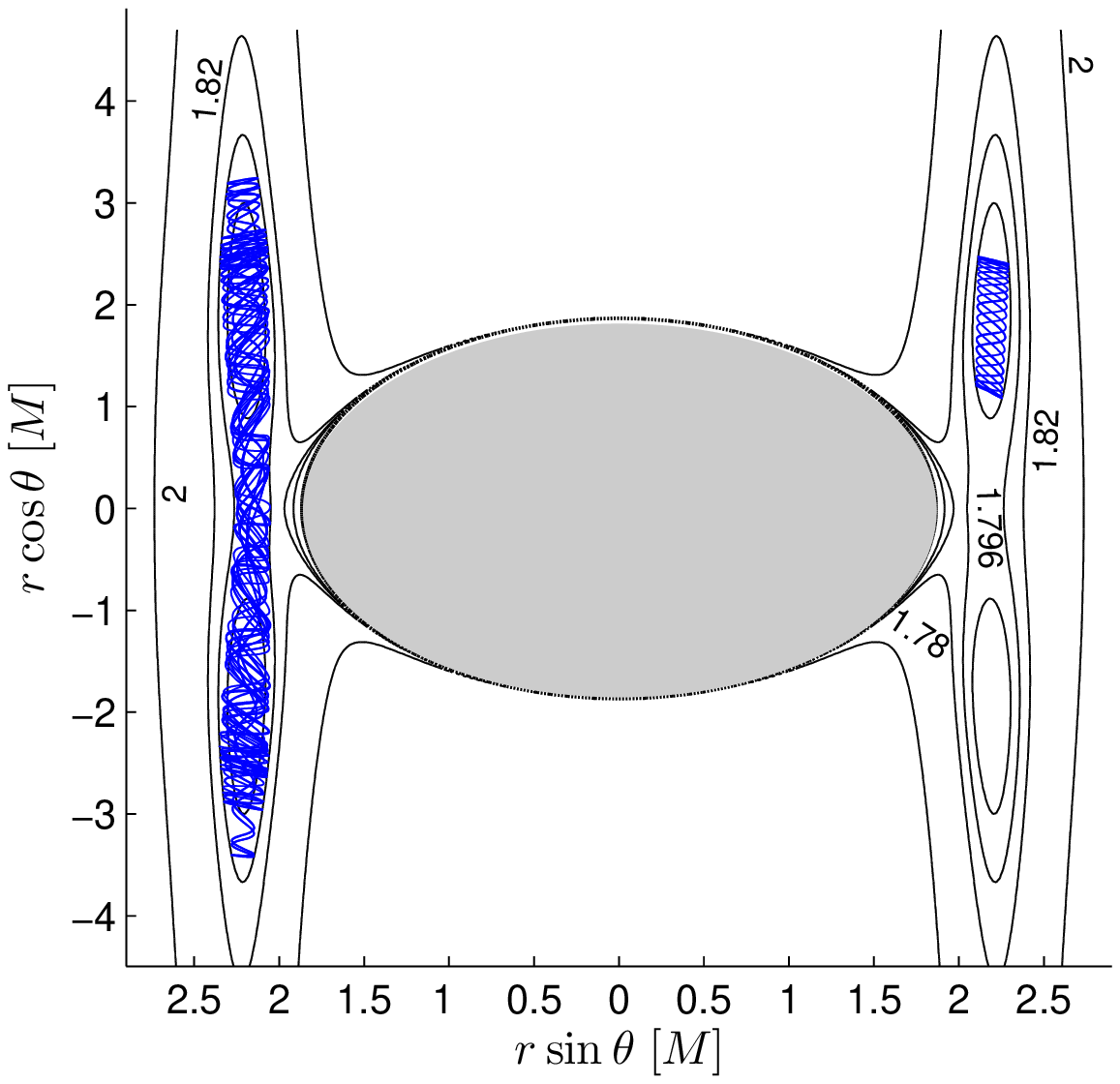}\includegraphics[scale=0.61, clip]{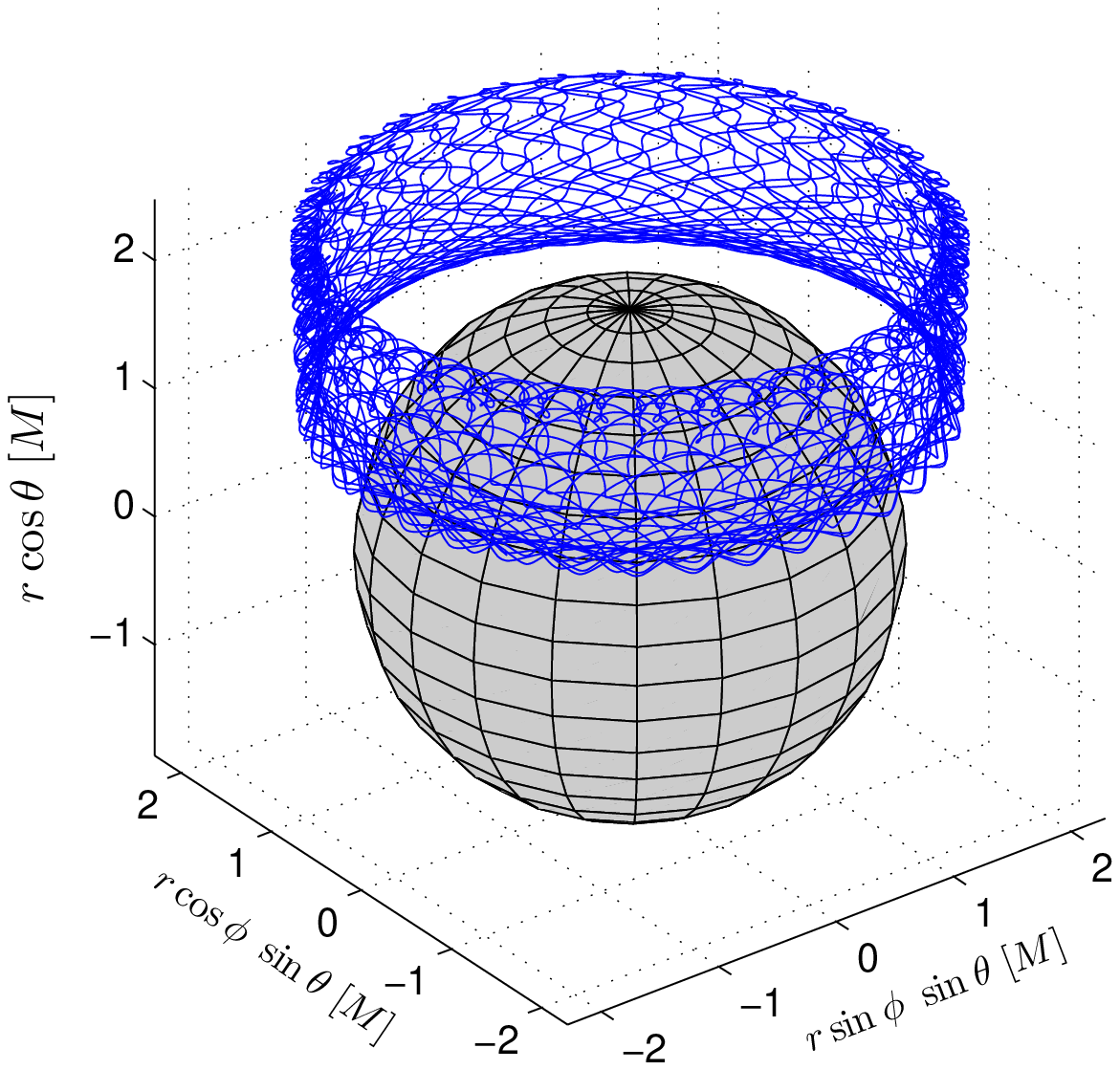}
\caption{In the left panel we present a poloidal section of the
selected isocontours of the effective potential
$V_{\rm{}eff}(r,\theta)$, eq.\ (\ref{effpot}),  for a charged
particle ($\tilde{q}\tilde{Q}=2$, $\tilde{L}=5\;M$) on the Kerr background
($a=0.5\;M$). We assume  the presence of Wald uniform magnetic field
($\tilde{q}B_{0}=2M^{-1}$). The off-equatorial potential lobes
are present, allowing stable motion. Two exemplary trajectories of
test particles are shown -- in the left lobe a chaotic orbit of energy
$\tilde{E}=1.796$, while in the right lobe the regular, purely
off-equatorial trajectory of $\tilde{E}=1.78$. Both particles were
launched at $r(0)=3.11$, $\theta(0)=\pi/4$ with $u^r(0)=0$ and their
trajectories interweave with each other.  We plot the poloidal
$(r,\theta)$ projection of the trajectory; what appears as a lobe in the
poloidal plane is an axially symmetric 3-dimensional rotational
structure. The latter is illustrated in the right panel where the case
of the off-equatorial regular trajectory is shown.}
\label{fig1}
\end{figure*}

The above-mentioned lobes are defined by the figures of the effective
potential in the poloidal plane. These were previously studied in the
context of charge separation that is expected to occur in pulsar
magnetospheres \citep[e.g.][]{neukirch93}. Here, we address whether the
trajectories within these lobes are regular (i.e., whether the system is
integrable), or if they instead exhibit a chaotic behavior. A related
problem was studied recently by \citet{japonci} in an attempt to find a
connection between chaoticness of the motion and the spin of a rotating
black hole residing in the center. These authors suggest that chaotic
behavior occurs for certain values of the black hole spin, while for
others the system is indeed regular.

The idea of investigating the connection between the spin of a black
hole and chaoticness of motion of matter near its horizon is very
interesting for the following reason. Because of high degree of symmetry of the background
spacetime, the unperturbed motion is regular \citep{carter}; no chaos is
present. The electromagnetic perturbation may trigger the chaos,
however, its effect can be expected to diminish very near the horizon,
where strong gravity of the black hole should prevail. This is also the
region where the spin effects are most prominent. Further out various
other influences become important due to distant matter and the
turbulence in accreted material. Therefore the connection between the spin
and the motion chaoticness is best applicable in the immediate vicinity
of the black hole, i.e.\ within the inner parts of corona.

The recurrence analysis \citep{marwan} provides us with a powerful tool
for the investigation of complex dynamical systems. The method examines
the recurrences of the system to the vicinity of previously reached
phase space points. It has been typically adopted to study the
experimental data, where often only some (if not just one) of the phase
space variables are known from the measurements. Takens' embedding
theorems \citep{takens} are then used to reconstruct the phase space
portrait of such a system. In our study we are equipped with the full
phase space trajectory from the numerical integration of the equations
of motion, so that we can use the recurrence analysis directly.

It appears that the method of Recurrence Plots has not been employed in
the context of relativistic astrophysical systems yet. To this end, one
needs a consistent definition of the neighborhood of a point in the
phase space in a curved spacetime. Below, we discuss the phase space distance
and suggest a form of the distance norm suitable in such circumstances.

The paper is organized as follows. In sec.\ \ref{pohrce} we review the
equations of particle motion, which we then integrate to obtain
trajectories. In sec.\ \ref{ra} we introduce the basic properties of
Recurrence Plots. Sec.\ \ref{sectionwald} analyses the motion around a
Kerr black hole endowed with a uniform magnetic test field. We employ
Poincar\'e surfaces of section and Recurrence Plots. The two approaches
allow us to show the onset of chaos in different, complementary ways. We
examine  the motion in off-equatorial lobes, pay special attention to
the spin dependence of the stability of motion, and we notice the
emergence of `potential valleys' that allow the particles to escape from
the equatorial plane along a narrow collimated corridor. Analysis of the
off-equatorial motion around a magnetic star is presented in sec.\
\ref{sectionmagnetized}. We consider a dipole-type magnetic  field,
which sets different limits on the off-equatorial range of allowed
motion of charged particles. It also defines different regimes of
chaoticness in comparison with the uniform magnetic field. Finally,
results of the analysis are summarized in sec.\ \ref{concl}.

\begin{figure*}[htb]
\centering
\includegraphics[scale=0.55, clip]{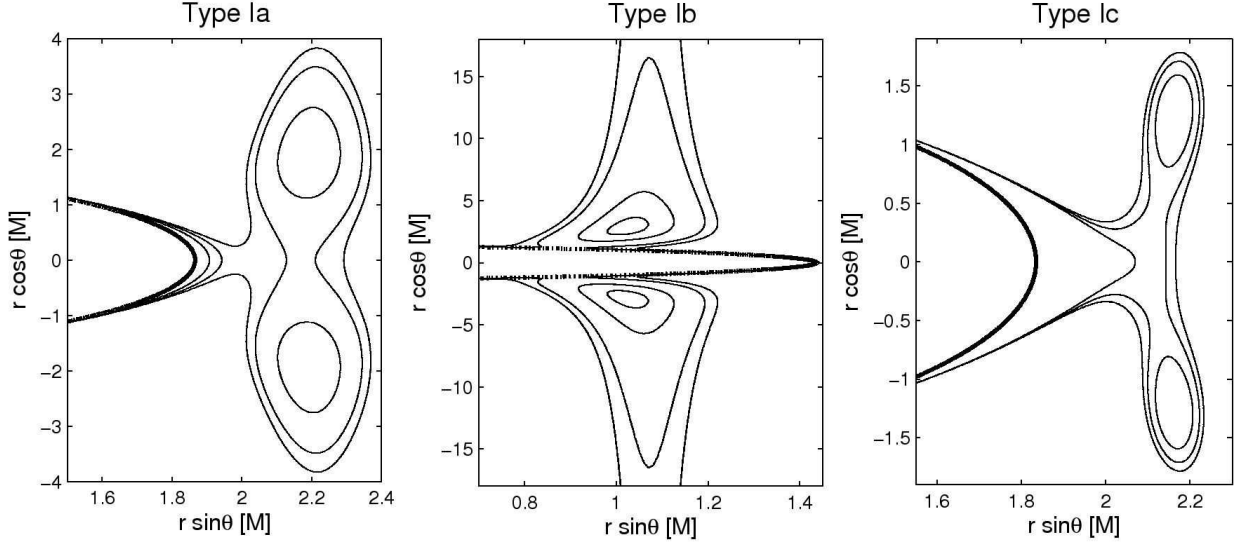}
\caption{The overview of possible topologies of the off-equatorial
potential structure above the event horizon (thick line in plots) of
Kerr black hole endowed with the Wald test field.}
\label{wald_abc}
\end{figure*}

\section{Equations of motion and the effective potential}
\label{pohrce}
The phase space trajectories of integrable systems are regular, meaning
that they are bound to the surface of an $n$-dimensional torus, where
$n$ is the number of degrees of freedom. The torus is determined
uniquely by $n$ constants of motion that are present in such a system.
Its behavior can be explored by Poincar\'e surfaces of section, which
are defined by  intersections of the phase space trajectory with a
$2$-dimensional plane \citep{lieberman}. On the other hand,
non-integrable chaotic systems generally have fewer integrals
than the number of degrees of freedom. In general, both the regular and
the chaotic orbits may coexist in the phase space of a single system.

Chaotic orbits are ergodic on the given hypersurface. Its dimension is
now larger than $n$, and the section points thus fill areas in the plot
of the Poincar\'e surface. However, depending on the initial conditions,
regular orbits can also appear in non-integrable systems. Such orbits
maintain the value of some additional constant of motion, although it is
not generally possible to write this constant in an explicit form. In
the context of motion around black holes perturbed by (weak) external
sources, various aspects of chaos were studied e.g.\ by
\citet{karas92,nakamura93,podolsky98}, and very recently by
\citet{semerak10}.

A standard approach to an integrable system with a non-integrable
perturbation assumes complete control over the strength of the
perturbation (i.e., the perturbation can be set to be arbitrarily weak). If this were
the case, we could first switch the perturbation completely off, analyze the
orbits, and then observe the impact of gradually increasing the
perturbation strength upon these orbits. However, the class of off-equatorial
bound orbits only exists when the electromagnetic term is strong enough
to balance the gravitational attraction of the central body. Then the
(sufficiently strong) perturbation is by itself the cause of the new
kind of the regular motion that happens outside the equatorial plane. 

Having this delicacy on mind, we shall use the usual Hamiltonian formalism 
to express equations of motion
governing the trajectories. We first
construct the super-Hamiltonian $\mathcal{H}$ \citep{mtw},
\begin{equation}
\label{SuperHamiltonian}
\mathcal{H}=\textstyle{\frac{1}{2}}g^{\mu\nu}(\pi_{\mu}-qA_{\mu})(\pi_{\nu}-qA_{\nu}),
\end{equation}
where $m$ and $q$ are the rest mass and charge of the test particle,
$\pi_{\mu}$ is the generalized (canonical) momentum, $g^{\mu\nu}$ is the
metric tensor, and $A_{\mu}$ denotes the vector potential of the
electromagnetic field. The latter is related to the electromagnetic
tensor $F_{\mu\nu}$ by $F_{\mu\nu}=A_{\nu,\mu}-A_{\mu,\nu}$. Unless
otherwise stated, we will use geometrical units, $G=c=1$.

The Hamiltonian equations are given as
\begin{equation}
\label{HamiltonsEquations}
\frac{{\rm d}x^{\mu}}{{\rm d}\lambda}\equiv p^{\mu}=
\frac{\partial \mathcal{H}}{\partial \pi_{\mu}},
\quad 
\frac{d\pi_{\mu}}{d\lambda}=-\frac{\partial\mathcal{H}}{\partial x^{\mu}},
\label{eq:ham}
\end{equation}
where $\lambda=\tau/m$ is the affine parameter, $\tau$ denotes the
proper time, and $p^{\mu}$ is the standard kinematical four-momentum for
which the first equation reads $p^{\mu}=\pi^{\mu}-qA^{\mu}$.

In the case of stationary and axially-symmetric systems, we
identify two constants of motion, namely, the energy $E$ and angular
momentum $L$. From the second Hamiltonian equation (\ref{eq:ham})
we obtain
\begin{eqnarray}
\label{Momenta}
\pi_t&=&p_t+qA_t\equiv-E\\
\label{27}
\pi_{\phi}&=&p_{\phi}+qA_{\phi}\equiv L.
\end{eqnarray}

The trajectory is specified by the integrals of motion $E$, $L$, and 
the initial values $r(0)$, $\theta(0)$ and $u^r(0)$. The initial
$u^{\theta}(0)$ can be calculated from the normalization condition,
$g^{\mu\nu}u_{\mu} u_{\nu}=-1$ (we always choose the non-negative
root).

\begin{figure*}[htb]
\centering
\includegraphics[scale=0.54, clip]{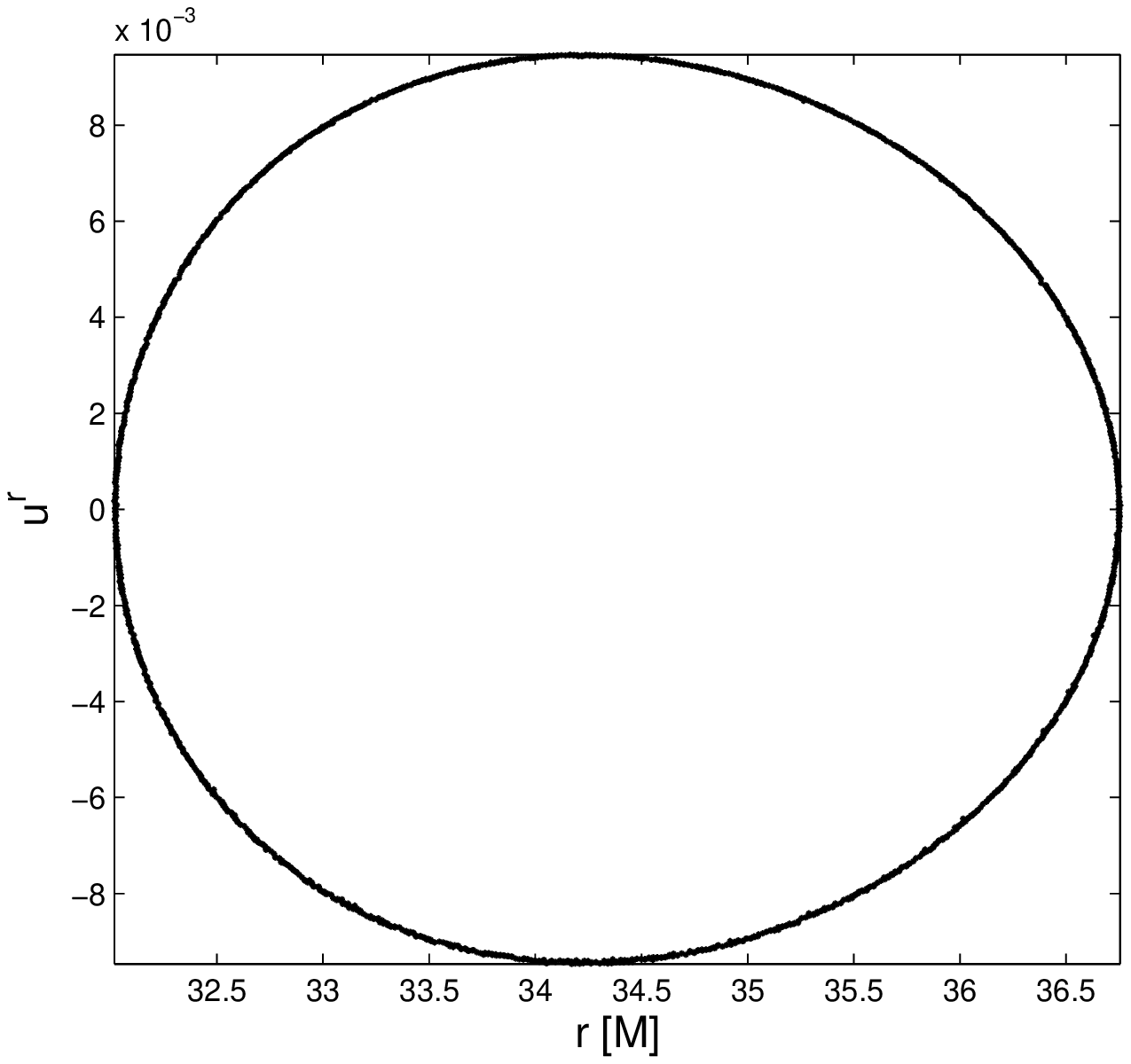}~~
\includegraphics[scale=0.52,  clip]{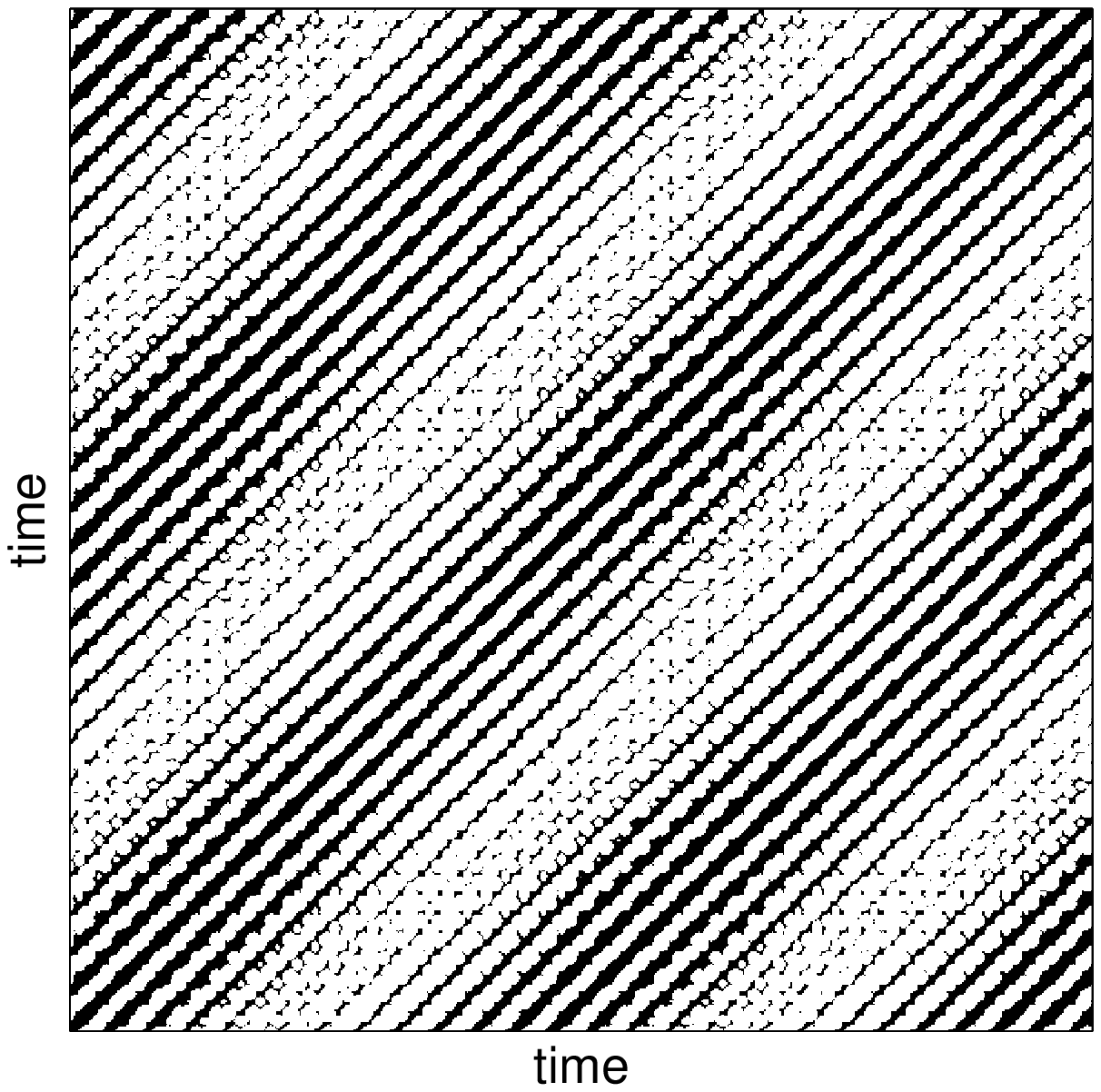}
\caption{Regular 
motion in the equatorial potential lobe in the fully integrable system
of charged test particle ($\tilde{E}=0.99$, $\tilde{L} =5M$,
$\tilde{q}=10^4$, $r(0)=32.02\:M$, $\theta(0)=1.54$) in the pure
Kerr-Newman spacetime ($\tilde{Q}=3\times10^{-5}$, $a=0.5\:M$) endowed
with the fourth Carter constant of motion $\mathcal{L}$. 
Long diagonals parallel to the LOI  are general
hallmark of regularity in the RPs. }
\label{rppoinc1}
\end{figure*}
\begin{figure*}[htb]
\centering
\includegraphics[scale=0.235, clip]{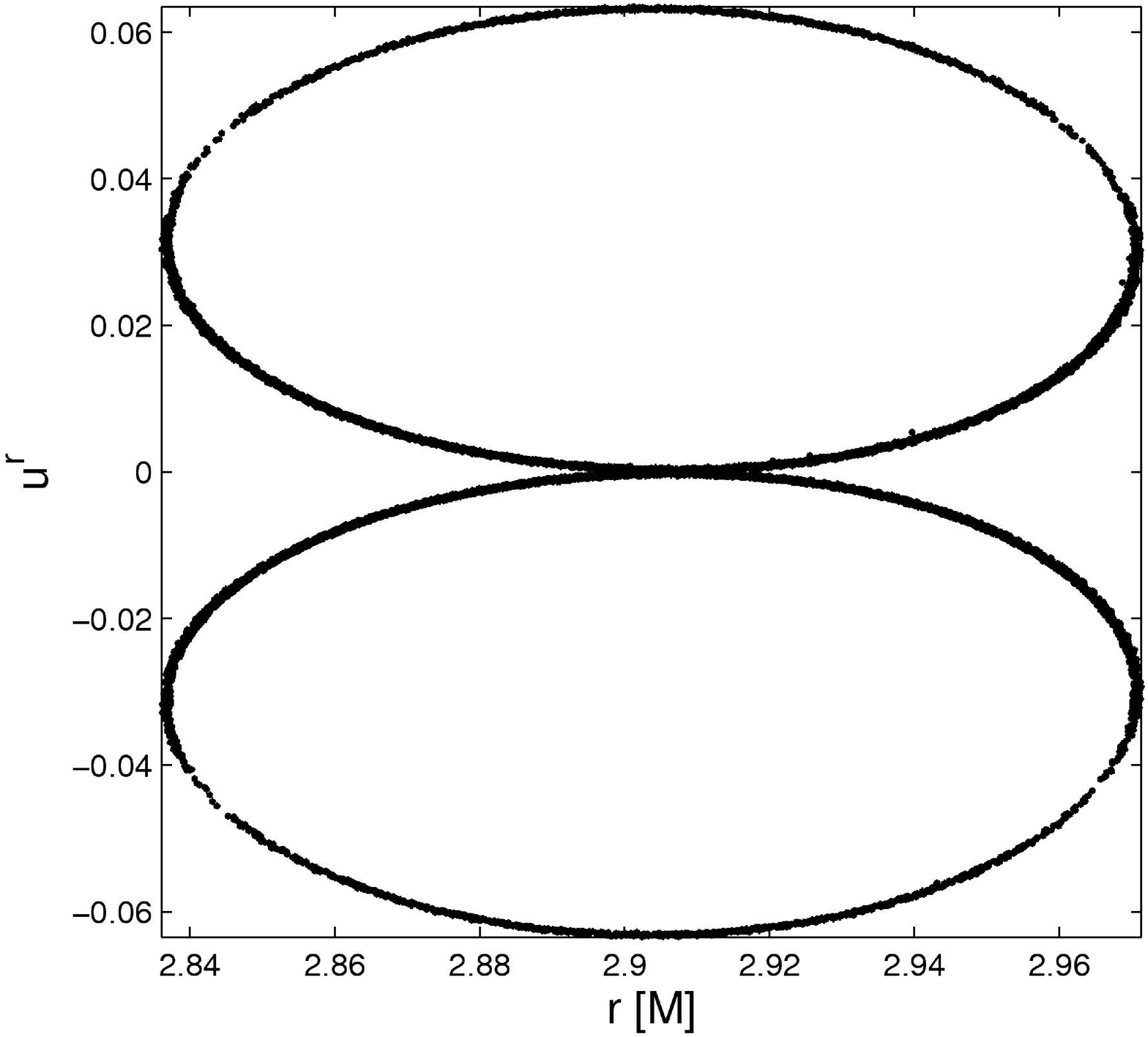}~~
\includegraphics[scale=0.56, clip]{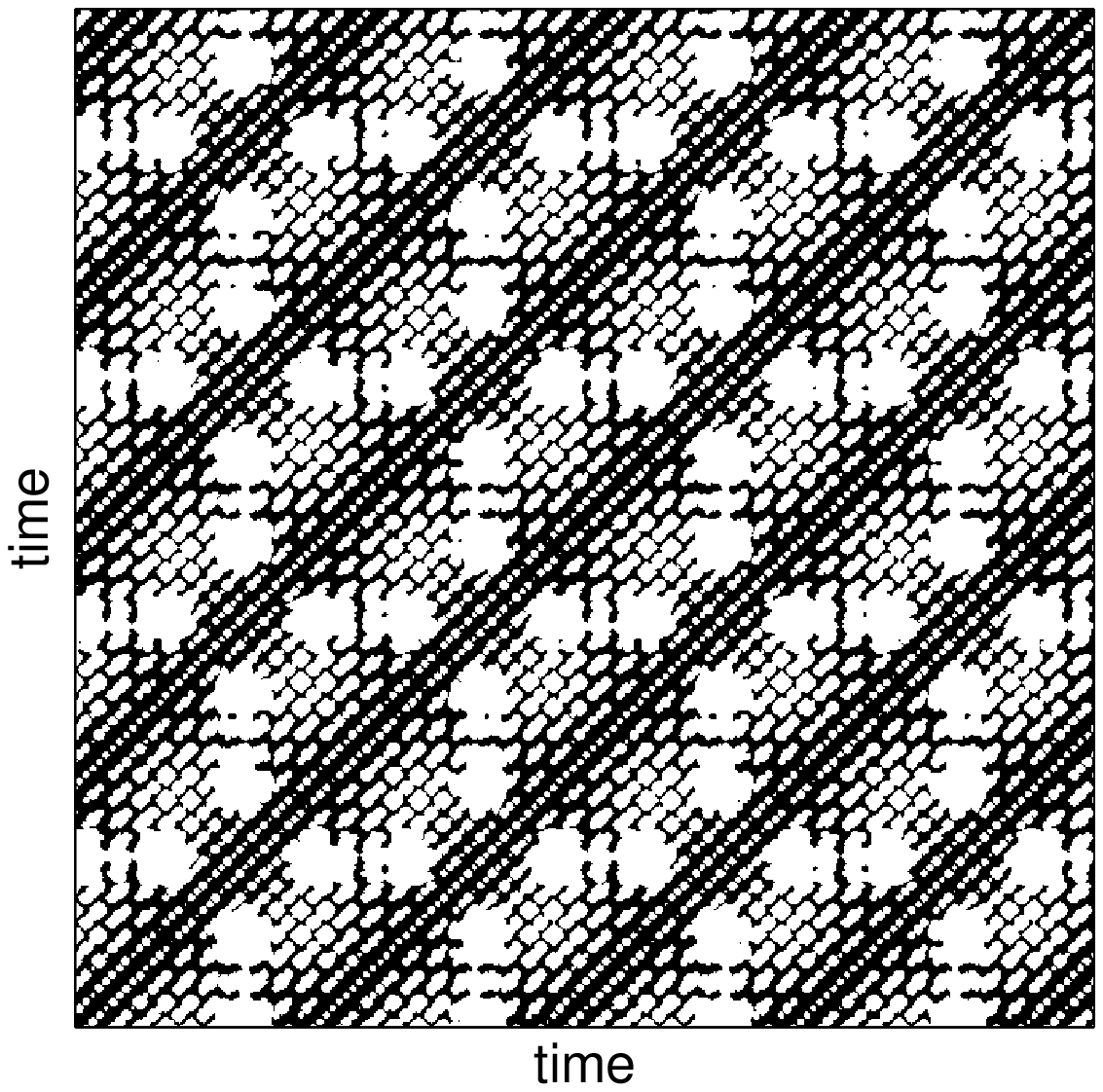}
\caption{Regular off-equatorial motion of a charged test
particle ($\tilde{E}=1.77$, $\tilde{L} =5M$, $r(0)=2.9\:M$, $\theta(0)=0.856$
and $u^r(0)=0$) on the Kerr background ($a=0.5\:M$) enriched with the
Wald test field ($\tilde{q}B_{0}=2M^{-1}$, $\tilde{q}\tilde{Q}=2$). The
diagonal structures typical for trajectories in integrable systems are
preserved, though the pattern is more complicated.}
\label{rppoinc2}
\end{figure*}

The effective potential can be derived in the following form:
\begin{eqnarray}
\label{effpot}
V_{\rm eff}=\frac{-\beta+\sqrt{\beta^2-4\alpha\gamma}}{2\alpha},
\end{eqnarray}
where
\begin{eqnarray}
\label{EfectivePotential_parts}
\alpha&=&-g^{tt},\\
\beta&=&2[g^{t\phi}(\tilde{L}-\tilde{q}A_{\phi})-g^{tt}\tilde{q}A_{t}],\\
\gamma&=&-g^{\phi\phi}(\tilde{L}-\tilde{q}A_{\phi})^2-g^{tt}\tilde{q}^2A_t^2\\
& &+2g^{t\phi}\tilde{q}A_t(\tilde{L}-\tilde{q}A_{\phi})-1\nonumber,
\end{eqnarray}
and where we introduce specific quantities
$\tilde{L}\equiv{}\frac{L}{m}$, $\tilde{E}\equiv{}\frac{E}{m}$ and the
specific charge $\tilde{q}\equiv{}\frac{q}{m}$. Local minima of 
$V_{\rm eff}(r,\theta)$ reflect the
location of stable orbits of test particles. Off-equatorial potential
minima were identified and various types of potential lobes were
discussed elsewhere \citep[see][]{halo2}.
We can express the effective potential (\ref{effpot}) as a function of $r$
and $u^r$, and use it to determine the boundaries of allowed regions in
Poincar\'e surfaces of section for a given value of 
$\theta$.\footnote{We have employed the method of effective potential to
study the stability of the motion, however, we note that the 
force formalism \citep{halo2_26,halo2_28} can
serve as a very efficient alternative tool. In particular, 
the off-equatorial motion of charged particles can be examined
via the procedure described in \citet{halo2}. This allows us to
localize minima of the effective potential around which the stable
orbits occur.}

We employ the Kerr metric in standard Boyer-Lindquist coordinates $t$, $r$, $\theta$,
$\phi$ \citep{mtw}:
\begin{eqnarray}
\label{KerrMetric}
{\rm d}s^2&=&-\frac{\Delta}{\Sigma}\Big({\rm d}t-a\sin{\theta}\,{\rm d}\phi\Big)^2\\& &\nonumber+\frac{\sin^2{\theta}}{\Sigma}\Big[\big(r^2+a^2)\,{\rm d}\phi-a\,{\rm d}t\Big]^2+\frac{\Sigma}{\Delta}\,{\rm d}r^2+\Sigma\, {\rm d}\theta^2,
\end{eqnarray}
where $a$ stands for the spin parameter (the specific angular momentum,
$0\leq a\leq M$), $\Delta\equiv{}r^2-2Mr+a^2$, and
$\Sigma\equiv{}r^2+a^2\sin^2{\theta}$. It is sufficient to consider
positive values of $a$ without loss of generality (the cases of
prograde and retrograde motion are distinguished by the sign of the
particle charge and the orientation of the magnetic field). By setting
$a=0$ the metric reduces to the static one of the Schwarzschild spacetime.

\begin{figure*}[htb]
\centering
\includegraphics[scale=0.51, clip]{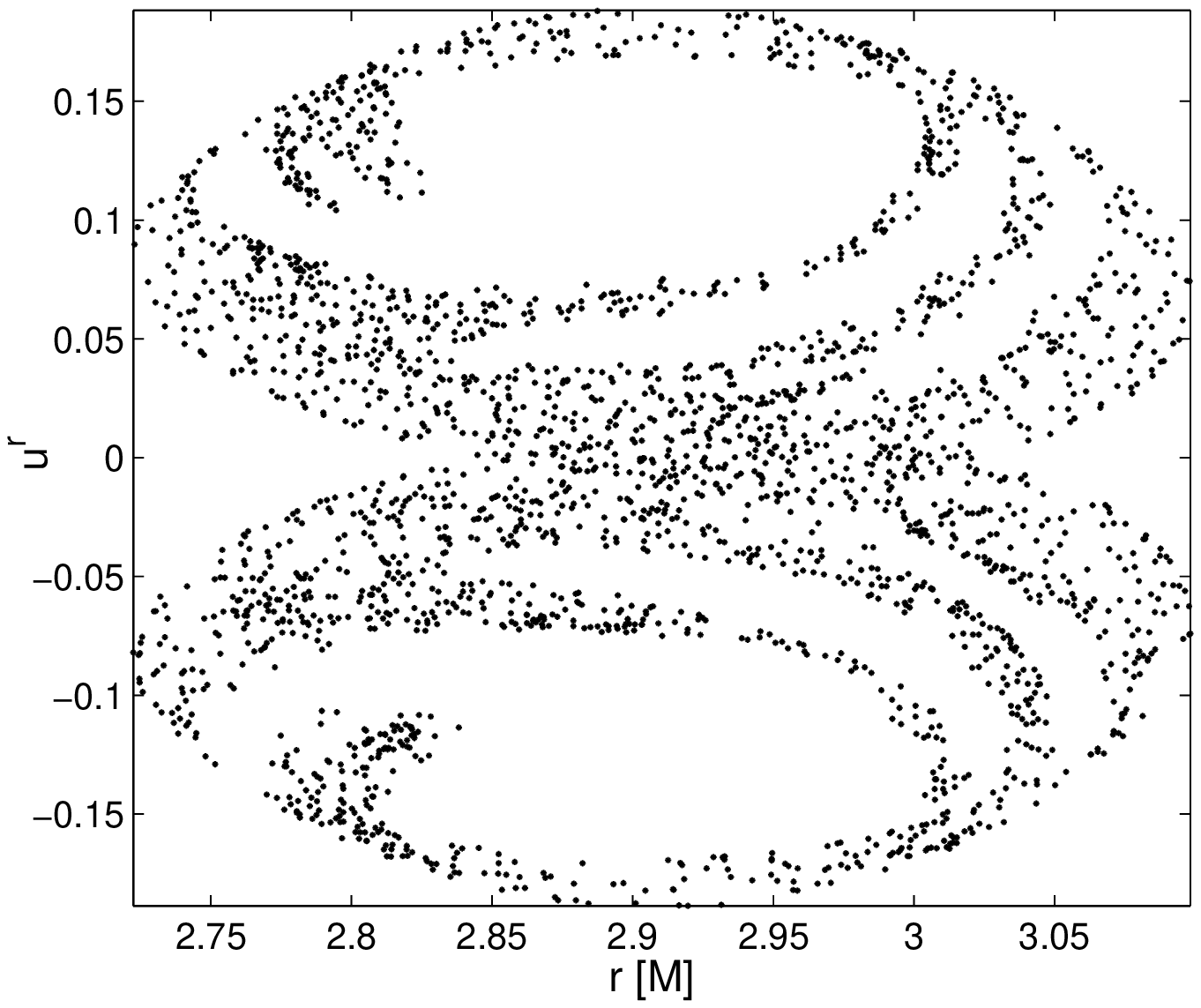}~~\includegraphics[scale=0.51, clip]{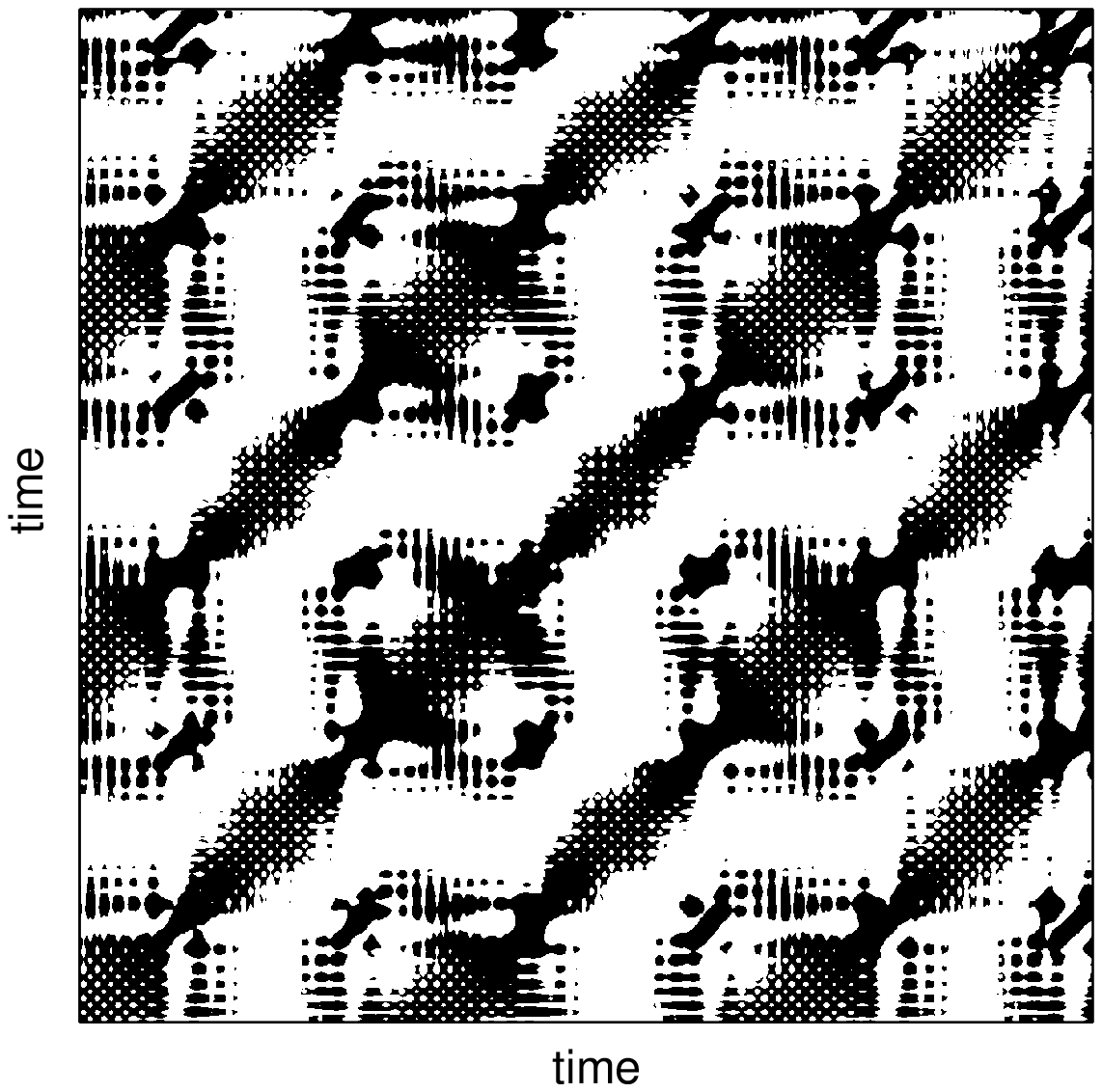}
\caption{A transitional state between the regular and chaotic regimes of
motion of a highly charged test particle which only differs from the
previous case by increasing the energy to $\tilde{E}=1.796$. The diagonal
lines in the RP are partially disrupted, indicating the onset of
chaos.}
\label{rppoinc4}
\end{figure*}
\begin{figure*}[htb]
\centering
\includegraphics[scale=0.53, clip]{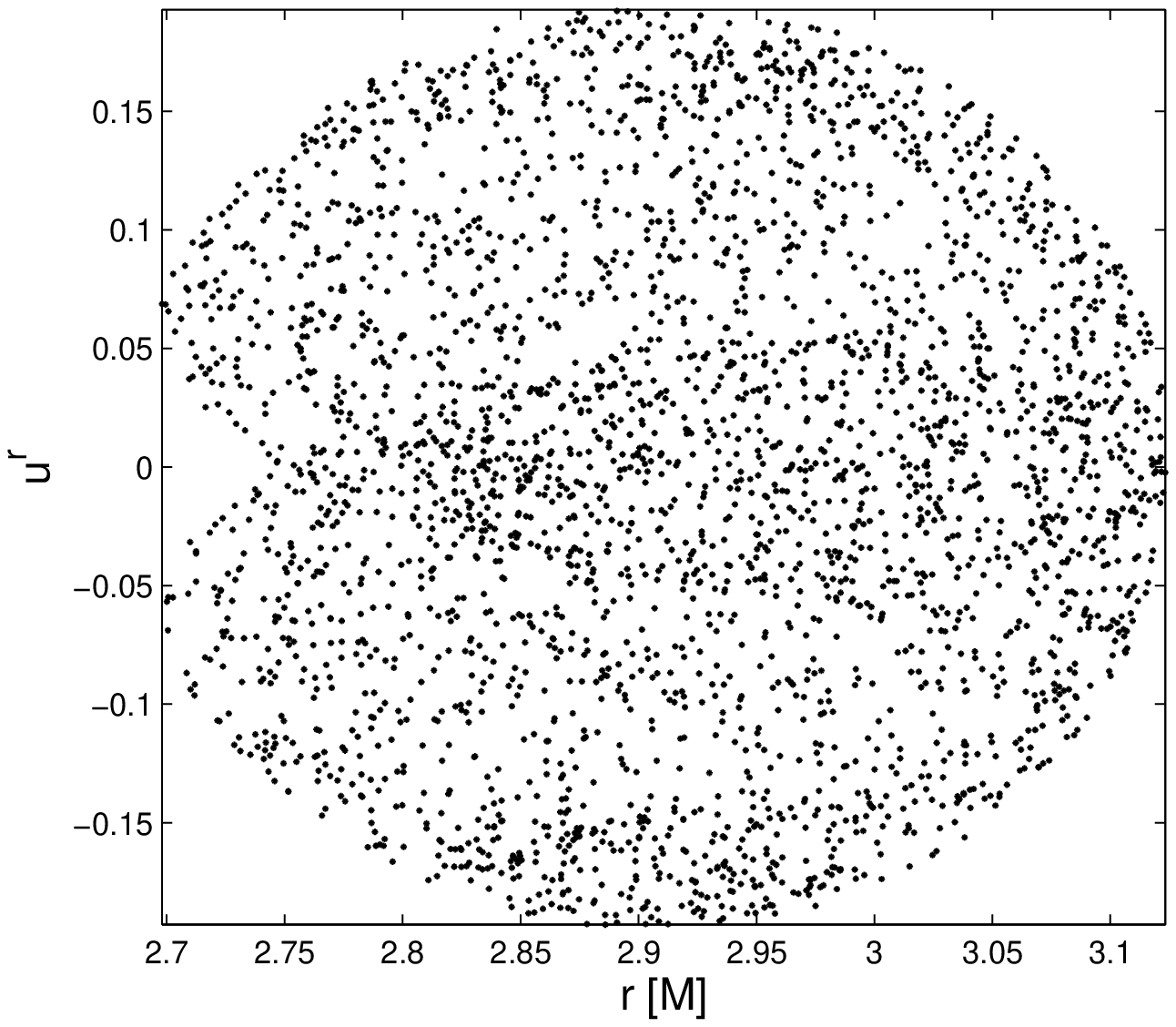}~~\includegraphics[scale=0.52, clip]{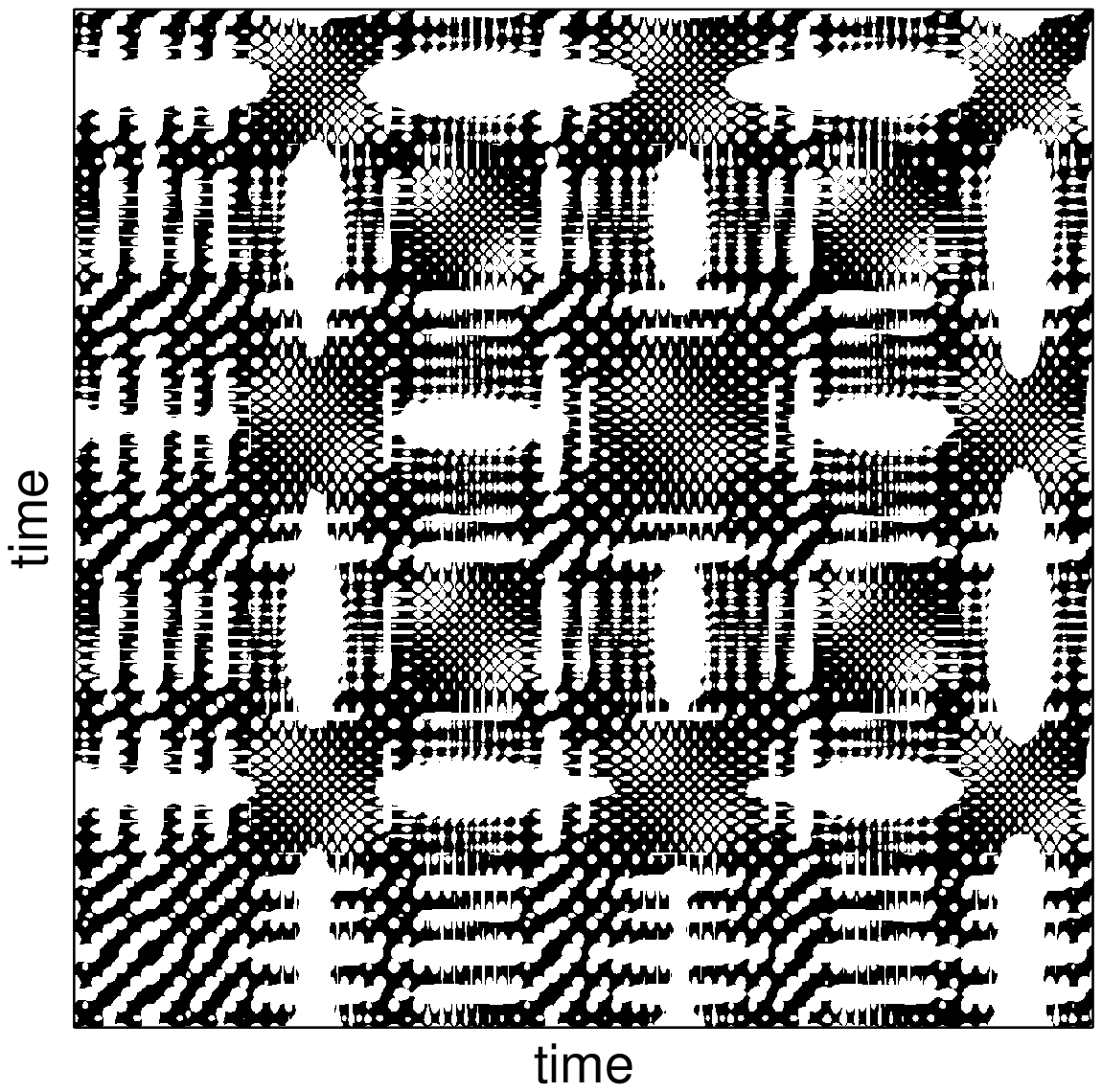}
\caption{The chaotic motion of a highly charged test particle which only
differs from the previous case by increasing the energy to
$\tilde{E}=1.7975$. The diagonal lines in the RP are now disrupted
and complex large-scale structures appear which are a characteristic
indication of deterministic chaos.}
\label{rppoinc3}
\end{figure*}

At this point, a note is worth on the adopted computational scheme which
we have employed to study the trajectories and to detect the chaotical
behavior. In order to reach reliable results, we coded several approaches
and we checked their stability and precision. We employ the multi-step
Adams-Bashforth-Moulton solver to determine the phase-space trajectory
by numerical integration of eqs.\ (\ref{HamiltonsEquations}). In some
cases, when a higher precision is demanded, we use the $7$-$8$th order
Dormand-Prince method that belongs to the family of explicit Runge-Kutta
solvers with adaptive stepsize. This method improves the accuracy
significantly, as can be verified by checking the conservation of the
integrals of motion along the trajectory. However, the improved accuracy
comes at the expense of computational time, as the adopted
Dormand-Prince scheme is more computationally demanding than the
Adams-Bashforth-Moulton solver.

Furthermore, it is well-know that, when dealing with Hamiltonian systems, the 
most appropriate solvers are those which respect the symplectic
nature of Hamiltonian dynamics
\citep[e.g.][]{yoshida93}. We therefore employed also 
the implicit Gauss-Legendre Runge-Kutta (GLRK) method, which is a
symplectic scheme. Indeed, we confirm that this code provides the most
reliable results, especially in the case of long-term integration. The
difference in the accuracy between GLRK and non-symplectic solvers reaches
several orders of magnitude and it is
generally more apparent in the case of chaotic trajectories, as
expected. However, the cost in terms of the computational time is also
non-negligible, and so we only use the GLRK method to achieve very accurate 
long-time determination of the trajectory in several exemplary runs.

\section{Recurrence analysis}
\label{ra}
The Kerr metric is well-known and the analysis of test particle
motion in this spacetime was carried out in many papers \citep{mtw}.
Among important features of the Kerr metric is the fact that the
particle trajectories are integrable, and so the chaos can set in
only when perturbations of the background gravitational field are
introduced or additional electromagnetic interaction with fields of
external sources are allowed. This is also where recurrence analysis
can be helpful.

\begin{figure*}
\centering
\includegraphics[scale=0.62, clip]{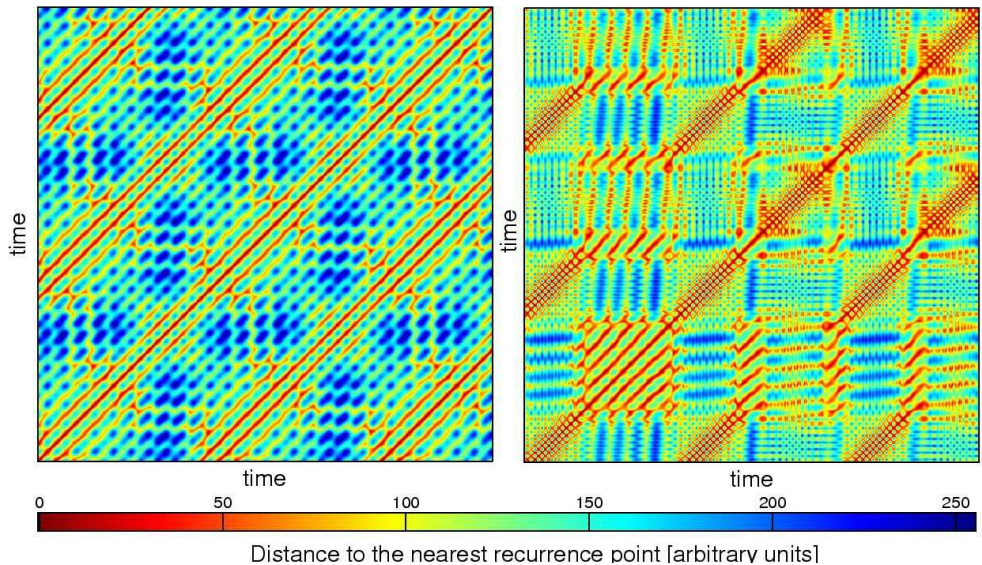}
\caption{Color map of the distance in the phase space to the nearest recurrence point. This example concerns a charged particle  trajectory near the Kerr black hole in the asymptotically uniform magnetic field. Left: the case of regular motion with the energy of $\tilde{E}=1.77$. Right: the case of chaotic motion with $\tilde{E}=1.7975$. In the latter case more complex 
structure appear in the recurrence plot. The common parameters of both panels are $\tilde{L} =5M$, $a=0.5\:M$, $\tilde{q}B_{0}=2M^{-1}$ and $\tilde{q}\tilde{Q}=2$ with the initial condition $r(0)=2.9\:M$, $\theta(0)=0.856$
and $u^r(0)=0$.}
\label{unthresholded}
\end{figure*}

Methods of phase space recurrences have been successfully applied to
a wide range of various empirical data, not only in physics but also
related to physiology, geology, finances and other fields.
Recurrence plots are especially suitable for the investigation of
rather short and nonstationary data. On the other hand, the method
of recurrence analysis has not yet been widely applied to study the
dynamical properties of motion in relativistic systems. We thus
briefly summarize this approach for our context.

Besides more traditional methods of the numerical analysis of
dynamical systems, such as a visual survey of Poincar\'{e} surfaces of
section or the evaluation of the Lyapunov spectra \citep{lce}, the
recurrence analysis is a rather novel technique, based on the
analysis of recurrences of the system into the vicinity of its
previous states.

Recurrence Plots (RP)
are introduced as a tool of visualizing the recurrences of a trajectory
in the phase space \citep{eckmann}. The method is based on examination 
of the binary values that are constructed from the trajectory $\vec{x}(t)$. 
Results of the orbit analysis can be quantified statistically in
terms of the Recurrence Quantification Analysis (RQA).\footnote{We use
the CRP ToolBox \citep[p. 321]{marwan} in Matlab (R2009b) to construct RPs
and to evaluate RQA measures.}

The RP construction is straightforward regardless of the dimension of
the phase space. We only need to evaluate the binary values of the
recurrence matrix $\mathbf{R}_{ij}$, which can be formally expressed as
follows:
\begin{equation}
\label{rpdef}
\mathbf{R}_{ij}(\varepsilon)=\Theta(\varepsilon-||\vec{x}(i)-\vec{x}(j)||),\;\;\;
i,j=1,...,N,
\end{equation}
where $\varepsilon$ is a pre-defined threshold parameter, $\Theta$ the
Heaviside step function, and $N$ specifies the sampling frequency. The
sampling frequency is applied to the time segment of the trajectory
$\vec{x}(t)$ under examination. There is, however, 
no unique prescription for the appropriate definition of the
phase space norm $||\;.\;||$ in eq.\ (\ref{rpdef}). We can consider a
purely abstract vector space and apply one of the elementary
norms $L^1$, $L^2$ (Euclidean norm) or $L^{\infty}$ (maximum norm).
Some aspects of the appropriate choice of the norm are deferred to
Appendix~\ref{appa}.

\begin{figure*}
\centering
\includegraphics[scale=0.62,clip]{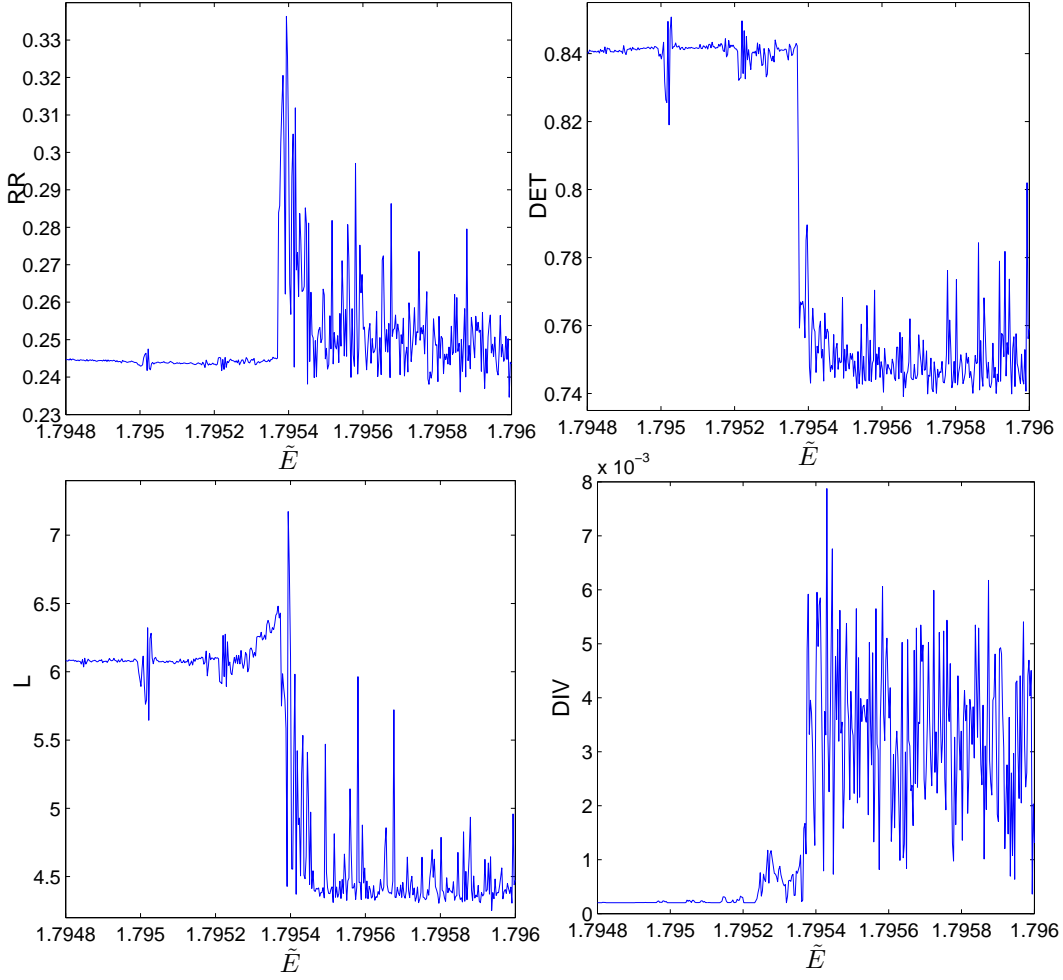}
\caption{Graphs of different RQA
measures based on the diagonal lines in the RP as a function of specific
energy $\tilde{E}$. In each panel, 400 trajectories were analyzed in a
given energetic range. All measures exhibit the evident change of their
behavior at $\tilde{E}\approx{}1.7954$ which we interpret as an onset
of chaos. Other parameters remain fixed at following values: $\tilde{L}
=5M$, $\tilde{q}B_{0}=2$, $r(0)=2.9\:M$, $\theta(0)=0.856$, $u^r(0)=0$,
$a=0.5\:M$ and $\tilde{q}\tilde{Q}=2$.}
\label{rqa1}
\end{figure*}

Finally, we need to specify the value of the threshold parameter
$\varepsilon$. To this end we follow the suggestion of 
\citet[sec. 3.2.2]{marwan} and relate $\varepsilon$ to the standard mean deviation,
$\sigma$, of the given data set. Setting $\varepsilon=k\sigma$ is
advantageous because the proportionality constant $k$,
once adjusted to obtain a properly filled Recurrence Plot, remains valid
(with only minor adjustments) for all data sets of other trajectories.
We therefore normalize the time series of each coordinate separately to
zero mean and $\sigma=1$.

The binary valued matrix $\mathbf{R}_{ij}$ represents the RP which we
get by assigning a black dot where $\mathbf{R}_{ij}=1$ and leaving a
white dot where $\mathbf{R}_{ij}=0$. Both axes represent a time segments
over which the data set (the phase space vector) is being examined. RP
is thus symmetric; the main diagonal is always occupied by the line
of identity (LOI).

Recurrence Plots contain wealth of information about the dynamics of the
system \citep{thiel}. Different pieces of knowledge are encoded in
large-scale and short-scale patterns \citep[sec. 3.2.3]{marwan}. To decide if a
particular trajectory is a regular or a chaotic one, the determining
factor is the presence of diagonal structures in the RP. Diagonal lines
in the RP reflect the time segments of phase space trajectory during
which the system evolution proceeds in a regular way. It captures the
epoch when the trajectory proceeds almost parallel to its previous
segment, i.e.\ within the $\varepsilon$-tube around that segment. Hence,
integrable systems exhibit themselves by diagonally oriented structures
in their RP. On the other hand, if the motion is chaotic the diagonal
lines disappear and the diagonal features become shorter, as the
trajectories tend to diverge quickly. As a result, more complicated
structures appear in the RP. 

\begin{figure*}
\centering
\includegraphics[scale=0.62,clip]{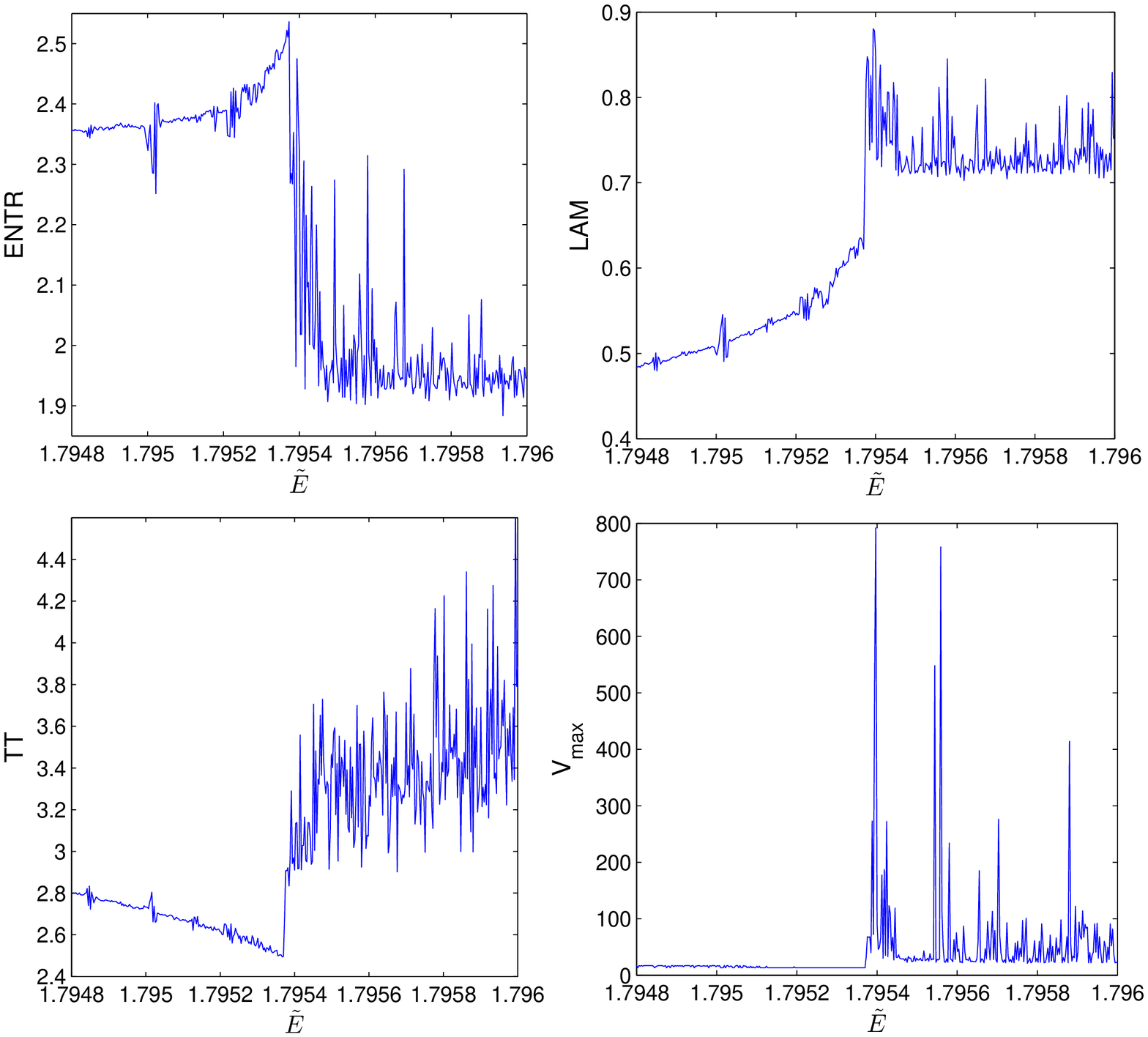}
\caption{Shannon entropy of probability distribution of diagonal
lines lengths $\mathrm{ENTR}$ and three RQA measures based on the vertical
lines of RPs as a function of specific energy $\tilde{E}$ (details in the 
text). To some
surprise, the vertical measures also react dramatically to the onset of
chaos at $\tilde{E}=1.7954.$} \label{rqa2}
\end{figure*}

We stress that the interpretation of the RP is primarily intuitive. We refer to the review paper \citet[sec. 3.2.3]{marwan} where the patterns appearing in the RP and their relation to the current dynamic regime are analyzed in detail. Although basic conclusions may be inferred in general (e.g. distinction between regular versus chaotic regime) the fine structure of the RP depends heavily on the properties of a given dynamic system. In order to gain more insight into the way in which various dynamical regimes  manifest themselves in our system we typically present RPs accompanied by corresponding Poincar\'{e} surface of section throughout this paper.

Visual behavior of RP and it's complexity is
quantitatively reflected in RQA. The RQA evaluates statistical characteristics of the recurrence matrix
$\mathbf{R}_{ij}$. First of all, we define the recurrence rate
($\mathrm{RR}$) as a density of points in RP,
\begin{equation}
 \mathrm{RR}(\varepsilon)\equiv\frac{1}{N^2}\sum_{i,j=1}^N\mathbf{R}_{i,j}(\varepsilon).
\end{equation}
Now we can turn our attention to diagonal segments in RP. Their length
draws distinction between regularity and chaos. The histogram
$P(\varepsilon,l)$ records the number of diagonal lines of length $l$.
It is formally given as follows:
\begin{eqnarray}
\label{diaghist}
P(\varepsilon,l)&=&\sum^N_{i,j=1}(1-\mathbf{R}_{i-1,j-1}(\varepsilon))(1-\mathbf{R}_{i+l,j+l}(\varepsilon))\\
& &\times\prod_{k=0}^{l-1}\mathbf{R}_{i+k,j+k}(\varepsilon)\nonumber.
\end{eqnarray}
This histogram defines the determinism factor ($\mathrm{DET}$), defined
as a fraction of recurrence points, which form the diagonal lines of
length at least $l_{\rm{min}}$ to all recurrence points,
\begin{equation}
\label{det}
\mathrm{DET}\equiv\frac{\sum^{L_{\rm{max}}}_{l=l_{\rm{min}}}lP(\varepsilon,l)}{\sum^{L_{\rm{max}}}_{l=1}lP(\varepsilon,l)}.
\end{equation}
The average length of diagonal lines $L$ (where only lines of length at
least $l_{\rm{min}}$ count) is
\begin{equation}
\label{Lav}
L\equiv\frac{\sum^{L_{\rm{max}}}_{l=l_{\rm{min}}}lP(\varepsilon,l)}{\sum^{L_{\rm{max}}}_{l=l_{\rm{min}}}P(\varepsilon,l)},
\end{equation}
and the corresponding divergence ($\mathrm{DIV}$) is defined as inverse
of the length of the longest diagonal line $L_{\rm{max}}$,
\begin{equation}
\label{div}
\mathrm{DIV}\equiv\frac{1}{L_{\rm{max}}}.
\end{equation}
$\mathrm{DIV}$ is in its very nature closely related to the divergent
features of the phase space trajectory, and so it was originally
\citep{eckmann} claimed to be directly related to the largest
positive Lyapunov characteristic exponent $\lambda_{\rm{max}}$. On the
other hand, theoretical considerations justify the use of $\mathrm{DIV}$
as an estimator only for the lower limit of the sum of the positive
Lyapunov exponents \citep[sec. 3.6]{marwan}. Nevertheless, a strong correlation
between $\mathrm{DIV}$ and $\lambda_{\rm{max}}$ arises in numerical
experiments \citep{trulla}.

\begin{figure*}
\centering
\includegraphics[scale=0.78, clip]{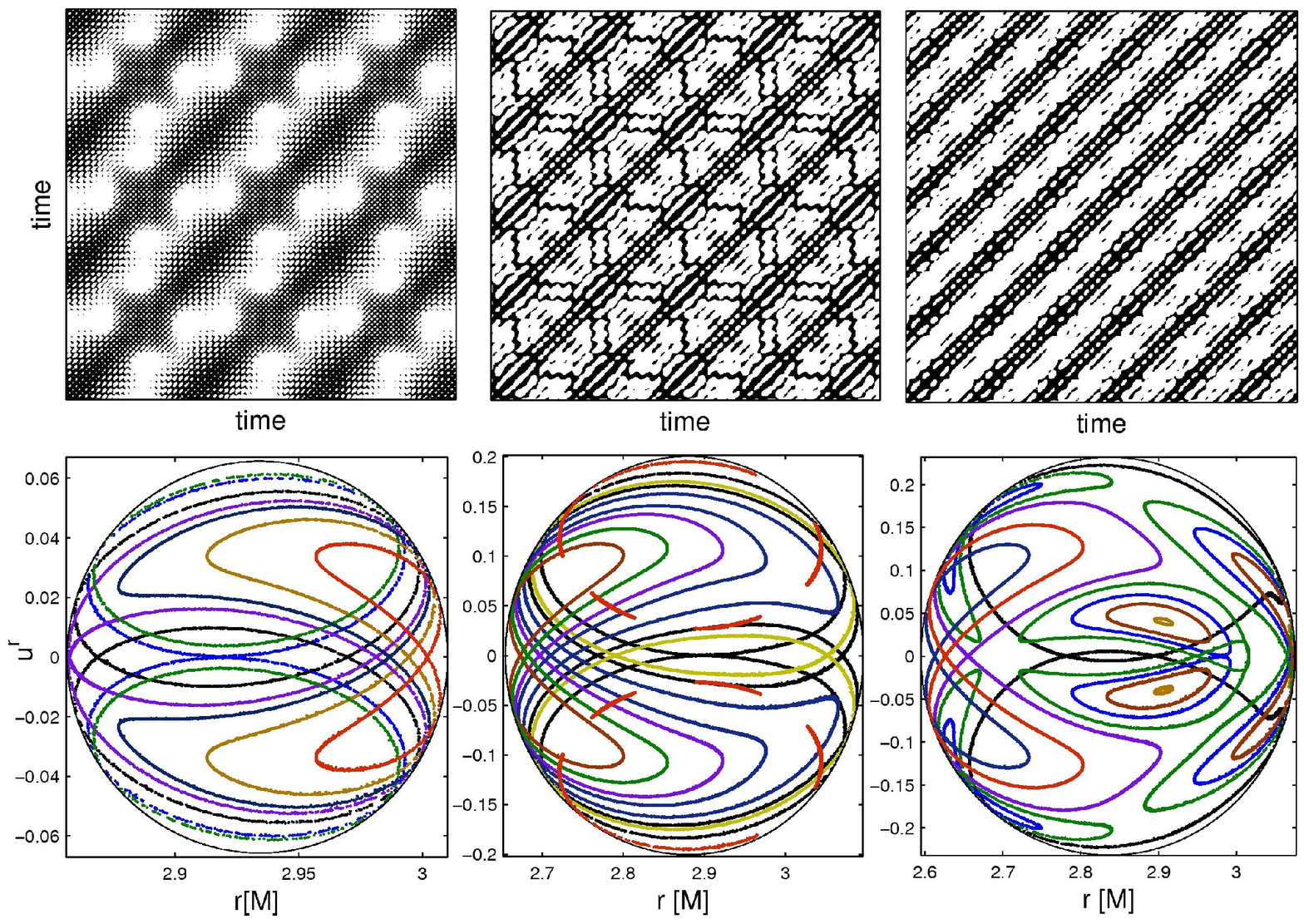}
\caption{Comparison
of purely off-equatorial trajectories in spacetimes differing by the
spin parameter $a$ (left panels: $a=0.3M$; middle: $a=0.6M$; right: $a=M$) 
which is linearly linked to the energy $\tilde{E}$ (left
panels: $\tilde{E}=1.56$; middle: $\tilde{E}=1.9$, and $\tilde{E}=2.35$ in
the right panels). Other parameters remain fixed: $\tilde{L} =5M$, $M^{-1}$,
$\theta(0)=\theta_{\rm{section}}=0.856$, $\tilde{q}B_{0}=2M^{-1}$ and
$\tilde{q}\tilde{Q}=2$. RPs are taken for trajectories with
$r(0)=2.9\:M$ and $u^{r}(0)=0$. Although the structures in the
Recurrence Plots clearly differ from each other,  all of them represent
diagonally oriented patterns that are characteristic of regular motion.
The regularity of the motion is confirmed by  surfaces of section in the
bottom panels, where several trajectories (for each value of spin
 and energy) are presented. Different colours (grey scaled in the 
printed version of the paper) are used to distinguish the orbits originating from
different initial conditions.}
\label{spin_diskuze}
\end{figure*}

The quantification measure $\mathrm{ENTR}$ is defined as the Shannon
entropy of the probability $p(\varepsilon,l)=P(\varepsilon,l)/N_{l}$ of
finding a diagonal line of length $l$ in the Recurrence Plot,
\begin{equation}
\label{entr}
\mathrm{ENTR}\equiv-\sum_{l=l_{\rm{min}}}^{L_{\rm{max}}}p(\varepsilon,l)\ln{p(\varepsilon,l)},
\end{equation}
where $N_{l}$ is a total number of diagonal lines:
$N_{l}(\varepsilon)=\sum_{l\geq{}l_{\rm{min}}}P(\varepsilon,l)$.

Analogous statistics may be performed for vertical as well as the
horizontal segments (RP is symmetric with respect to the main diagonal).
These segments are generally connected with periods in which the system
evolves during its laminar state. To this end, the histogram
$P(\varepsilon,v)$ records the number of vertical lines of length $v$
and it can be constructed as follows:
\begin{equation}
\label{verthist}
P(\varepsilon,v)=\sum^N_{i,j=1}(1-\mathbf{R}_{i,j}(\varepsilon))(1-\mathbf{R}_{i,j+v}(\varepsilon))\prod_{k=0}^{v-1}\mathbf{R}_{i,j+k}(\varepsilon).
\end{equation}

In analogy with the diagonal statistics histogram, $P(\varepsilon,v)$ is
used to define the vertical RQA measures. Laminarity ($\mathrm{LAM}$) is
defined as a fraction of recurrence points that form vertical lines of
length at least $v_{\rm{min}}$ to all recurrence points,
\begin{equation}
\label{lam}
\mathrm{LAM}\equiv\frac{\sum^{V_{\rm{max}}}_{v=v_{\rm{min}}}vP(\varepsilon,v)}{\sum^{V_{\rm{max}}}_{v=1}vP(\varepsilon,v)}.
\end{equation}
The trapping time ($\mathrm{TT}$) is an average length of vertical lines,
\begin{equation}
\label{tt}
\mathrm{TT}\equiv\frac{\sum^{V_{\rm{max}}}_{v=v_{\rm{min}}}vP(\varepsilon,v)}{\sum^{V_{\rm{max}}}_{v=v_{\rm{min}}}P(\varepsilon,v)}.
\end{equation}
Finally, the length of the longest vertical line ($V_{\rm{max}}$) can
also be of some interest.

RQA measures the crucial dependence of RP on the value of the
threshold parameter, $\varepsilon$, which must be adjusted appropriately
to a given data set. This lack of invariance is a drawback of both RPs
and RQA. Nevertheless, it was shown \citep{thiel2} that stable estimates
of various dynamical invariants, such as the second order R\'{e}nyi
entropy and the correlation dimension, can be inferred if $\varepsilon$
is kept within a reasonable range. Since we shall use the standard RQA
measures to compare the dynamics between test particles with different
initial conditions, we have to eliminate the numerical effect of
variances in the range of coordinate values spanned by these
trajectories. We achieve this by fixing the value of $\varepsilon$.

After the brief set of preliminaries we are now prepared to proceed to the
intended application of the recurrence analysis in the next section.

\section{Kerr black hole in uniform magnetic field}
\label{sectionwald}
Large-scale magnetic fields are known to be present in cosmic
conditions. They can exist around black holes, which do not support
their own magnetic field but may be embedded in fields of distant
sources. In the case of neutron stars, dipole-type magnetic fields of
very high strength often arise. We concentrate on black holes in this
section and defer the case of a magnetic star to sec.\
\ref{sectionmagnetized}.

By employing the uniform test field solution \citep{wald} we incorporate
a weak large-scale magnetic field near a rotating black hole. The
vector potential can be expressed in terms of Kerr metric coefficients
(\ref{KerrMetric}) as follows,
\begin{eqnarray}
\label{waldpot1}
A_t&=&\textstyle{\frac{1}{2}}B_{0}\left(g_{t\phi}+2a\,g_{tt}\right)-\textstyle{\frac{1}{2}}\tilde{Q}g_{tt}-\textstyle{\frac{1}{2}}\tilde{Q},\\
\label{waldpot2}
A_{\phi}&=&\textstyle{\frac{1}{2}}B_{0}\left(g_{\phi\phi}+{2a}g_{t\phi}\right)-\textstyle{\frac{1}{2}}\tilde{Q}g_{t\phi},
\end{eqnarray}
where $B_0$ is magnetic intensity and  $\tilde{Q}$ stands for the test
charge on the background of Kerr metric. The terms containing
$\tilde{Q}$ can be identified with the components of the vector
potential of Kerr-Newman solution (although the test charge does not
enter the metric itself). An example of an integrated trajectory is
shown in \rff{fig1}. \citet{wald} has shown that the black hole
selectively accretes charges from its vicinity, until it becomes
itself charged to the equilibrium value
\begin{equation}
\label{waldcharge} \tilde{Q}_{\rm{W}}=2B_{0}a.
\end{equation}

We remark that the particle charge $\tilde{q}$ appears always as a
product with $\tilde{Q}$ or $B_{0}$ in the formula (\ref{effpot}) for
the effective potential, as well as in equations of motion
(\ref{HamiltonsEquations}). Therefore, the simultaneous alteration of
$\tilde{q}$, $\tilde{Q}$ and $B_{0}$ values, preserving the products
$\tilde{q}\tilde{Q}$ and $\tilde{q}B_{0}$, does not affect the particle
dynamics. If we further assume that
$\tilde{Q}=\tilde{Q}_{\rm{W}}=2B_{0}a$ is maintained, we only need to
specify the value of $\tilde{q}B_{0}$ to uniquely determine a particular
trajectory. However, since we do not restrict ourselves to the case
$\tilde{Q}=\tilde{Q}_{\rm{W}}$ we decide to always explicitely specify
the values of $\tilde{q}\tilde{Q}$ and $\tilde{q}B_{0}$.

\begin{figure*}
\centering
\includegraphics[scale=0.53,clip]{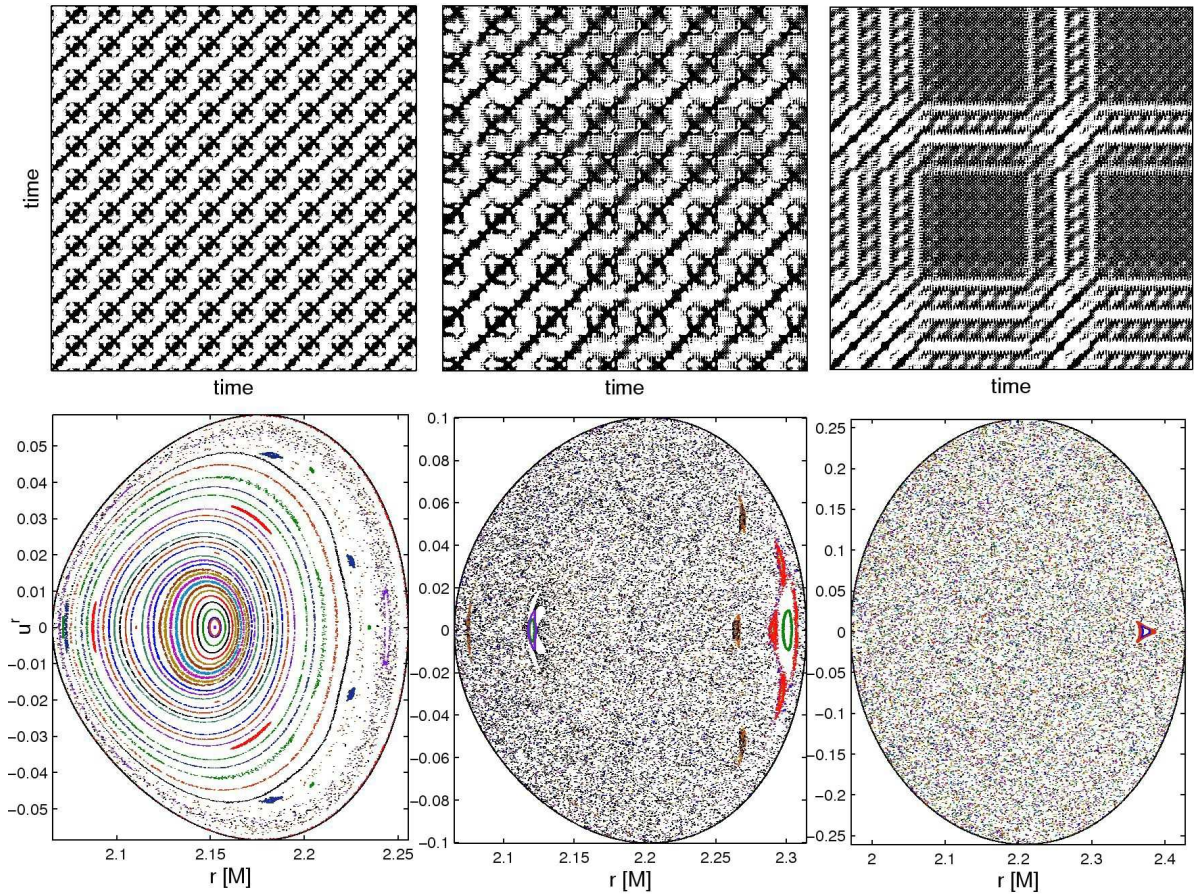}
\caption{Comparison
of trajectories of particles launched from the equatorial plane with
different spin values. Also in this case we have to link linearly the
value of spin $a$ with $\tilde{E}$ in order to maintain the existence of
the potential lobe. In left panels we set $a=0.5M$, $\tilde{E}=1.795$,
in middle panels $a=0.6M$, $\tilde{E}=1.92$ and in right panels $a=M$,
$\tilde{E}=2.42$. For all three cases we show surfaces of section of
several trajectories differing in initial values $r(0)$ and $u^r(0)$.
Recurrence Plots are taken for trajectories with $r(0)=2.15M$,
$u^r(0)=0$. Other parameters remain fixed: $\tilde{L} =5M$,
$\tilde{q}B_{0}=2M^{-1}$, $\theta(0)=\theta_{\rm{section}}=\frac{\pi}{2}$, 
$\tilde{q}\tilde{Q}=2$.}
\label{spin_diskuze_eq}
\end{figure*}

\subsection{\label{wald_lobes}Motion within the potential lobes}
Our previous analysis \citep{halo2} concludes that the off-equatorial
bound orbits are allowed only for test particles obeying simultaneously
the two conditions, ${\rm sgn}(aL)=1$ and ${\rm sgn}(\tilde{q})={\rm
sgn}(aB_{0})$. Three distinct types were found (see \rff{wald_abc}) and we
examined the dynamics of test particles in all of these types. The results 
for different types are comparable, and so we present here only the analysis of one of them,
namely the type Ia. While raising the energy level from the
local minima of the symmetric halo orbits, we observe that the
off-equatorial lobes grow and eventually merge with each other once 
the energy of the saddle point in the equatorial plane is reached. 

The size of the lobes is controlled by the specific energy $\tilde{E}$.
Employing the Poincar\'e surfaces of section and the Recurrence Plots we
investigate how the regime of the particle motion changes with
$\tilde{E}$. This parameter appears as a suitable control parameter
producing a sequence of bound trajectories while all other parameters
(and the initial position) remain fixed. For the sake of comparison we
first present the case of a fully integrable system of a charged test
particle on the pure Kerr-Newman spacetime (\rff{rppoinc1}). The motion
occurs in the potential lobe around the local minimum in the equatorial
plane; there are no halo orbits above the horizon in this case
\citep{halo2, Fel:1979}.

\Rff{rppoinc2} shows regular motion occurring in the
off-equatorial lobe on the Kerr background with Wald test field.
Increasing the energy level (while keeping all other parameters fixed)
above the value in the equatorial saddle point results in a transitional
regime depicted in \rff{rppoinc4}. The onset of chaotic features does
not occur as a direct consequence of the lobe merging. We rather observe
that the orbit bound in merged (cross-equatorial) lobe remains regular
until the particle {\it notices} the possibility of crossing the
equatorial plane which happens when its energy is increased
sufficiently. Once the motion becomes cross-equatorial, chaotic features
appear. By increasing the energy even more we approach the critical
value when the lobe opens and allows the particles to fall onto the
horizon. For energies slightly below this limit we detect a fully
chaotic regime of motion (\rff{rppoinc3}).

\begin{figure*}
\centering
\includegraphics[scale=0.45, clip]{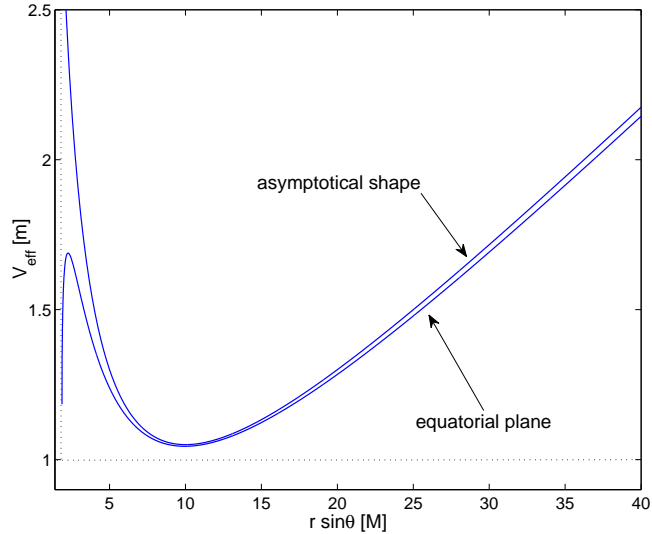}
\caption{Profiles of the effective potential $V_{\rm{eff}}$ for $\tilde{L} =5M$,
$a=0.5\:M$, $\tilde{q}\tilde{Q}=1.03$, $\tilde{q}B_{0}=0.1M^{-1}$, taken
in the equatorial plane and in the asymptotic region
$z=r\cos{\theta}\rightarrow\infty$ (as indicated with the corresponding curves). 
The valley crosses the equatorial
plane, almost unaffected in its bottom parts, although the behavior of
the potential near the symmetry axis is quite different in the
equatorial plane where it approaches the horizon of the black hole
(vertical dotted line at $r=r_+(a)$). The horizontal dotted line at unity measures 
the rest energy of the particle $m$.}
\label{valley}
\end{figure*}

Figure \ref{unthresholded} shows an alternative representation of the recurrence plots, where the 
distance to the nearest neighboring point along the phase space trajectory is encoded by 
different colors. Again, by comparing the two panels one can clearly recognize how the 
diagonal structures disintegrate into scattered points as the chaos sets in.

From the survey of Poincar\'e surfaces of section and the
Recurrence Plots we may only conclude that the transition from the
regular to chaotic regime occurs somewhere close to the
value $\tilde{E}=1.796$. In order to localize this transition more
precisely we evaluate RQA measures for 400 trajectories 
with the energy spread equidistantly over the interval
$\tilde{E}\in(1.7948,1.7968)$. In figs.\ (\ref{rqa1}) and
(\ref{rqa2}), we observe a sudden change of the behavior of
the statistical measure at $\tilde{E}\approx{}1.7954$,
reflecting a dramatic change of the particle dynamics.

Moreover, we know that the divergence $\mathrm{DIV}$ is related to the
Lyapunov exponents, and in \rff{rqa1} we observe that it suddenly rises
at $\tilde{E}\approx{}1.7954$, meaning that the trajectories become more
divergent when this energy is reached. All of these indications combined
lead to the conclusion that this energy level represents a critical value
at which a transition from regular to chaotic regime occurs.

We conclude that the energy of the particle $\tilde{E}$ acts as a
governing factor determining the dynamic regime of motion. Our survey
across various initial conditions has shown that motion in potential
wells of the type Ia in a Wald test field is generally regular. Chaos
appears well after the merging point of the lobes and close to the
critical breaking energy. We have verified that all queried RQA
measures, i.e.\ not only those based on diagonal lines in RP, react to
the transition from the regular to chaotic regime of motion, allowing us
to localize precisely the transition. This was fully confirmed also in
the other two types (classes Ib and Ic of our typology in
\rff{wald_abc}).

\begin{figure*}
\centering
\includegraphics[scale=0.41,clip]{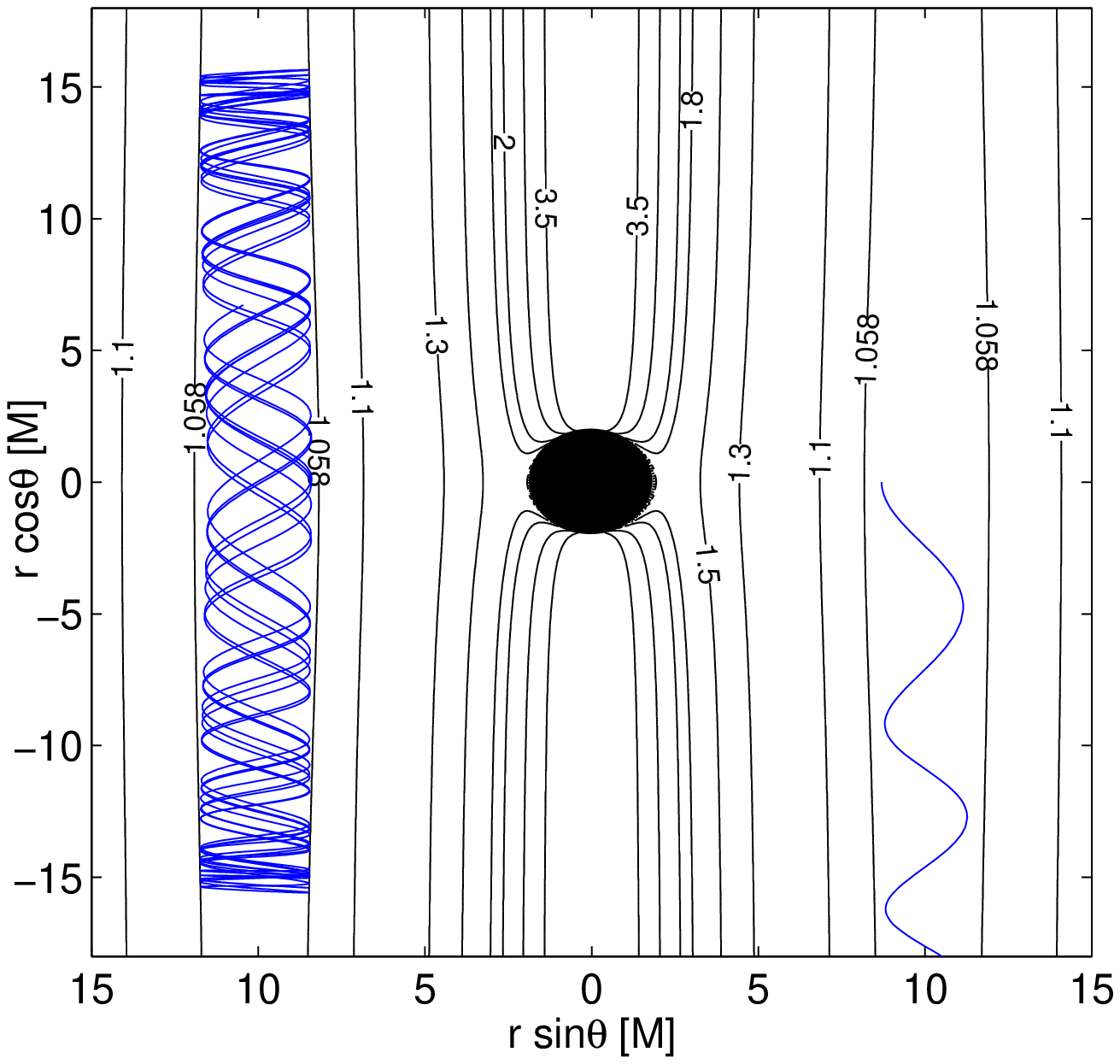}\includegraphics[scale=0.178,clip]{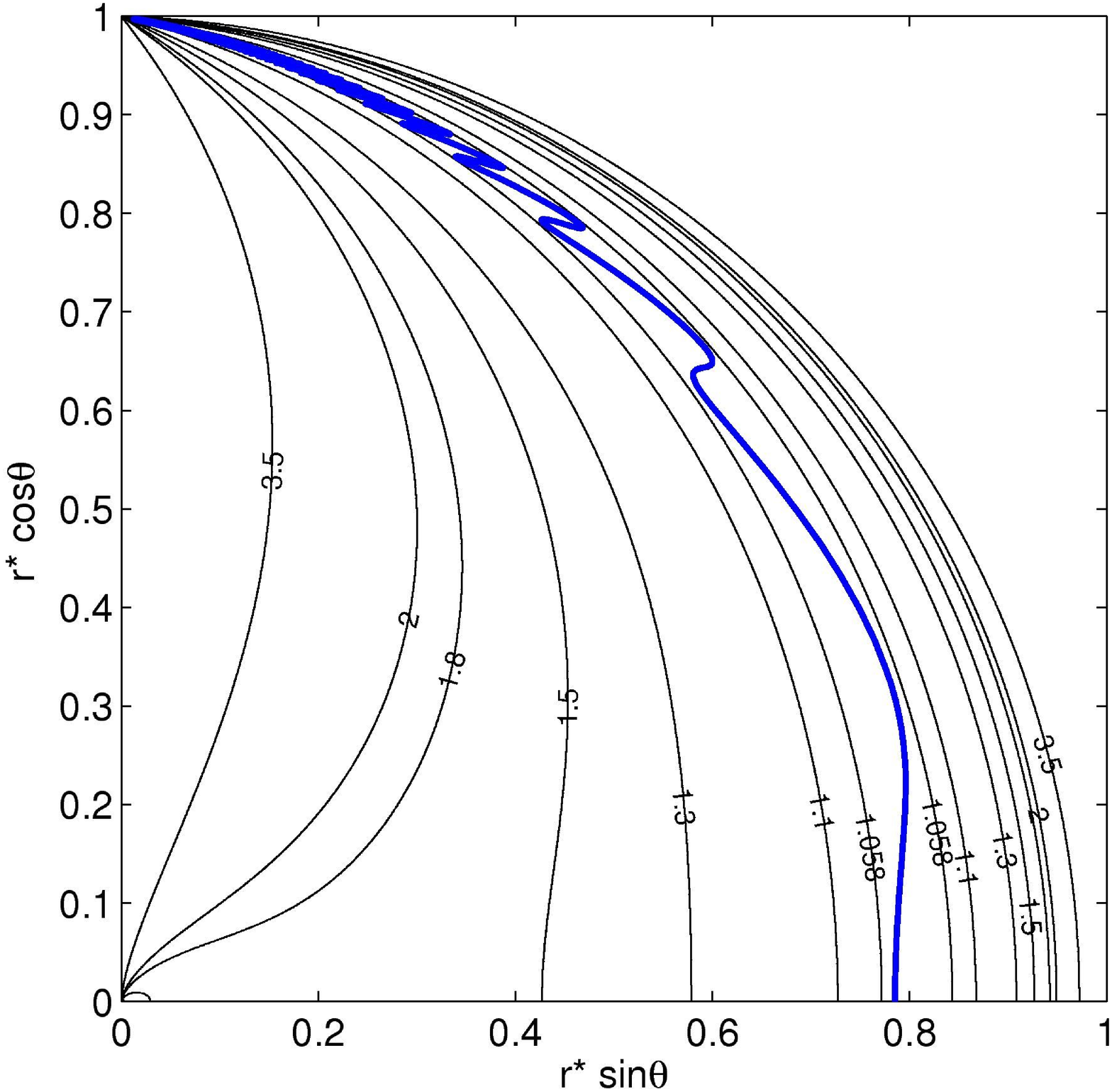}~\includegraphics[scale=0.176,clip]{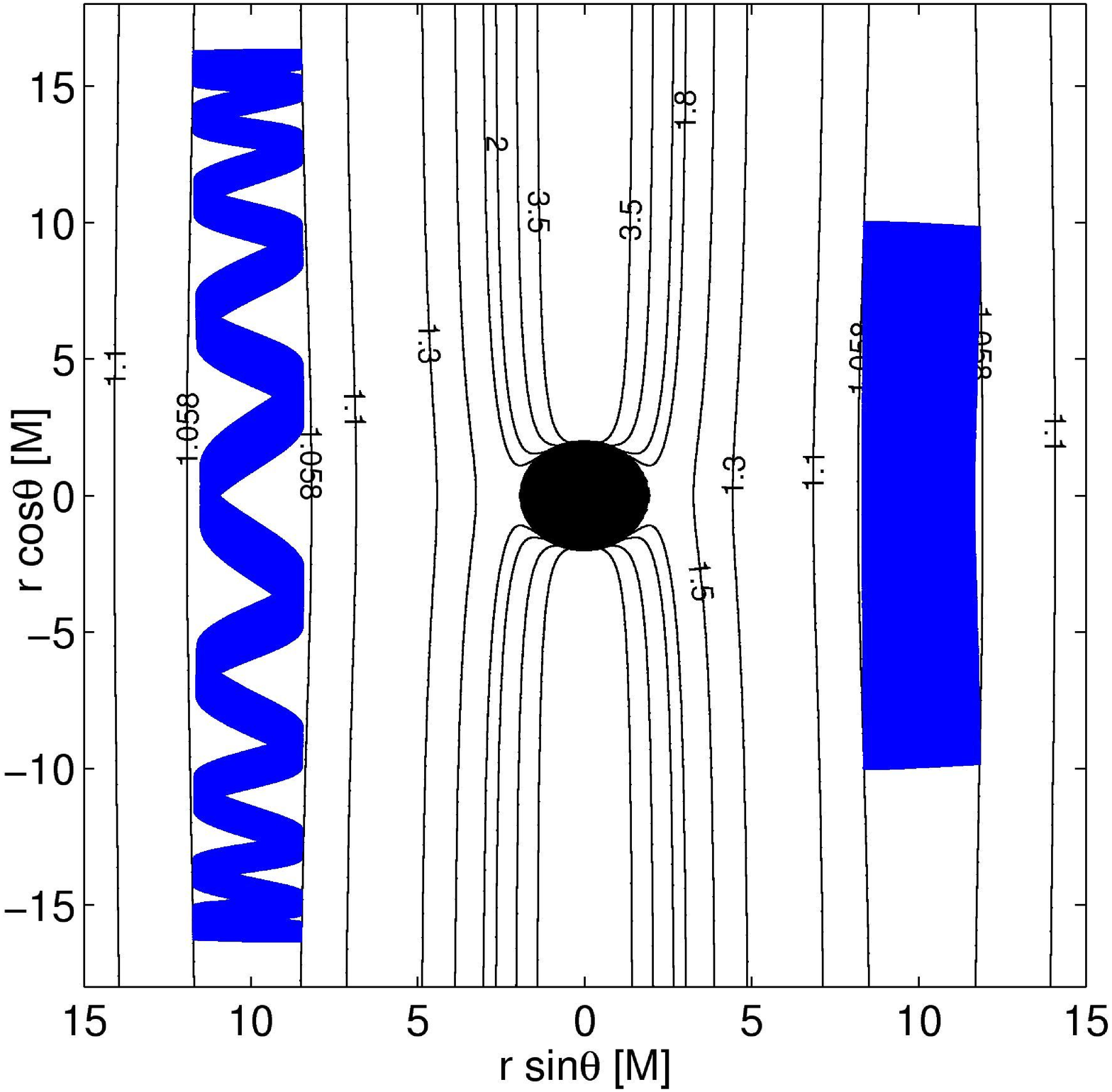}\\[10pt]
\includegraphics[scale=0.47,clip]{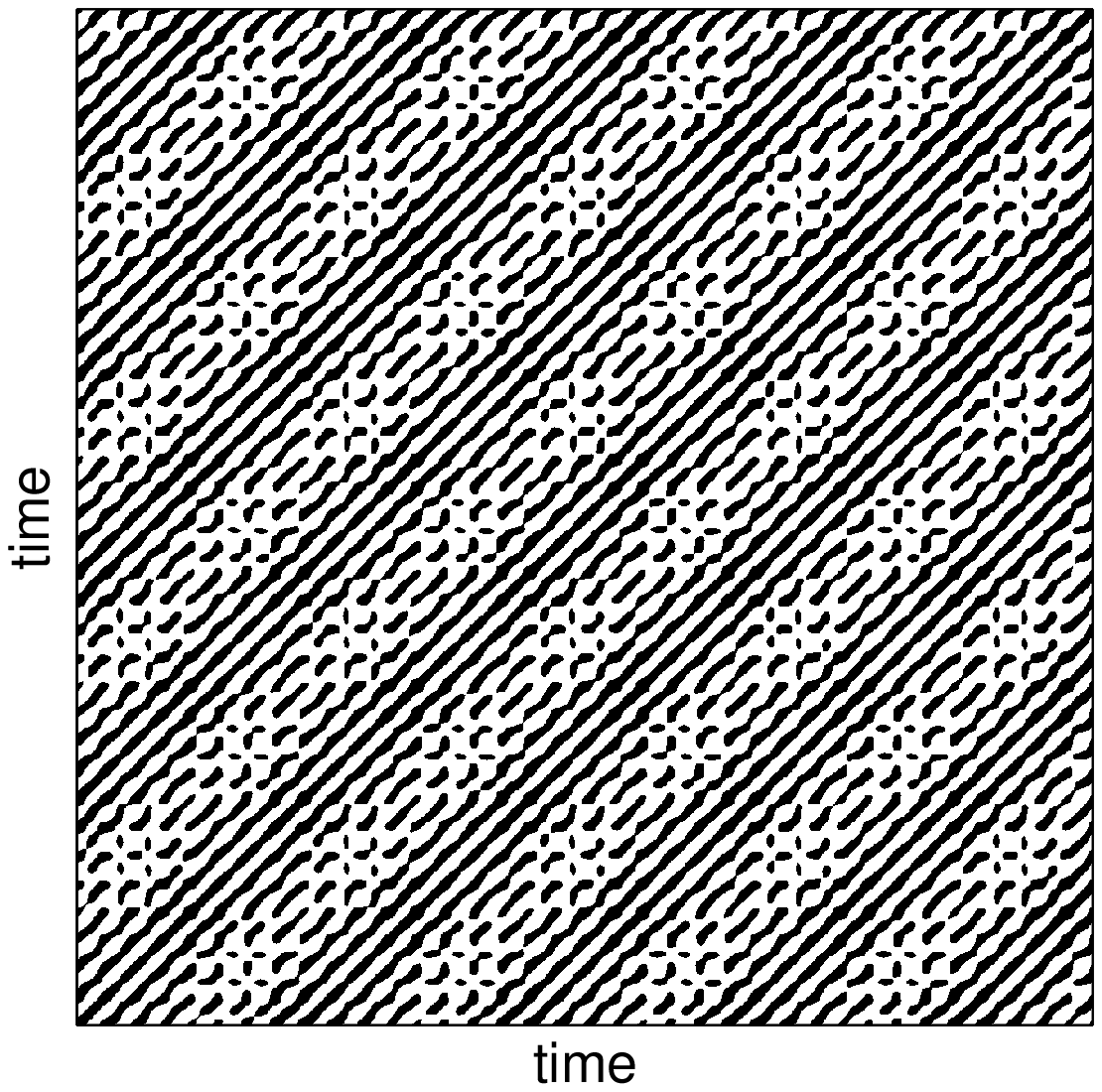}~\includegraphics[scale=0.178, clip]{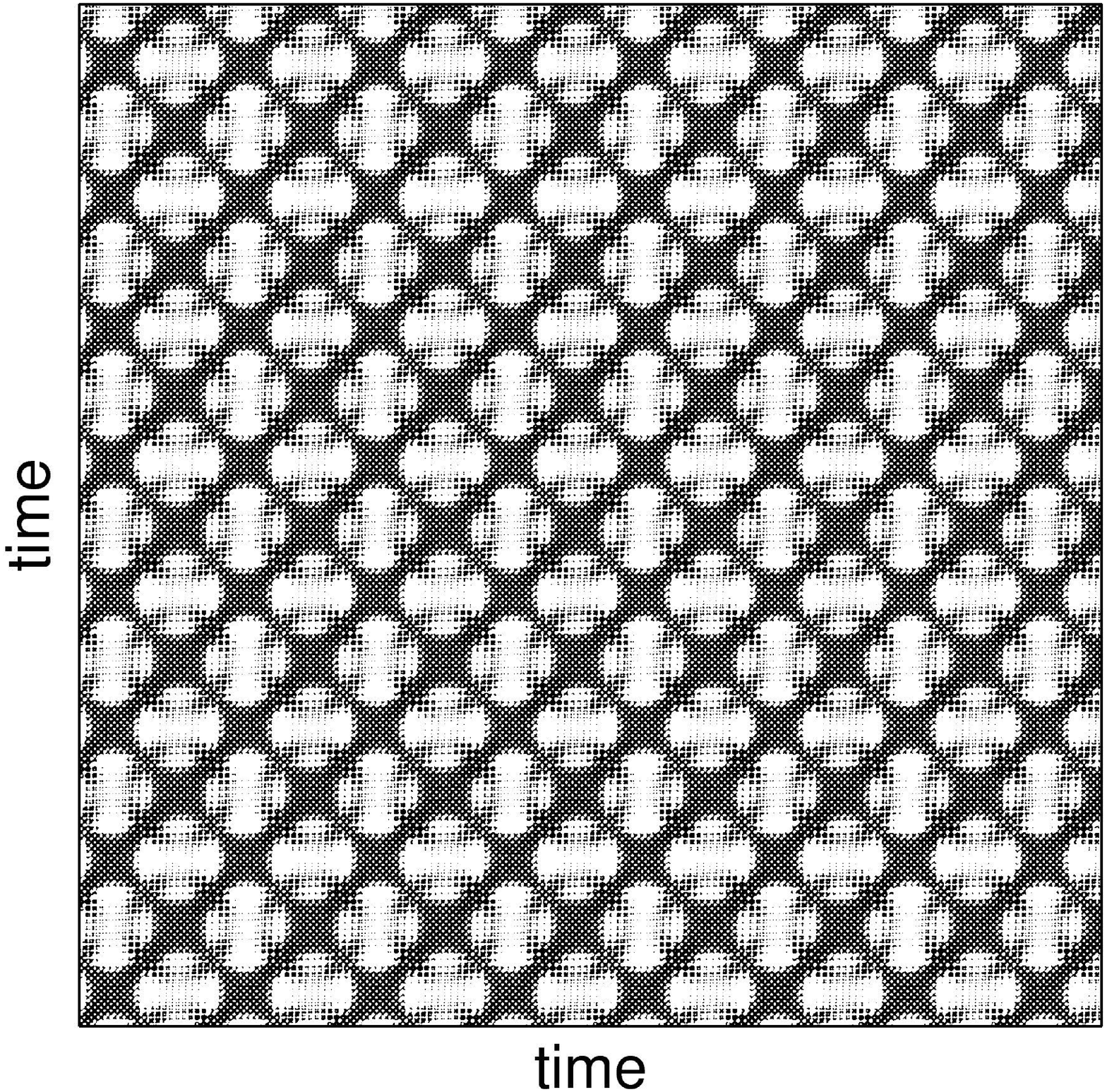}\includegraphics[scale=0.48,clip]{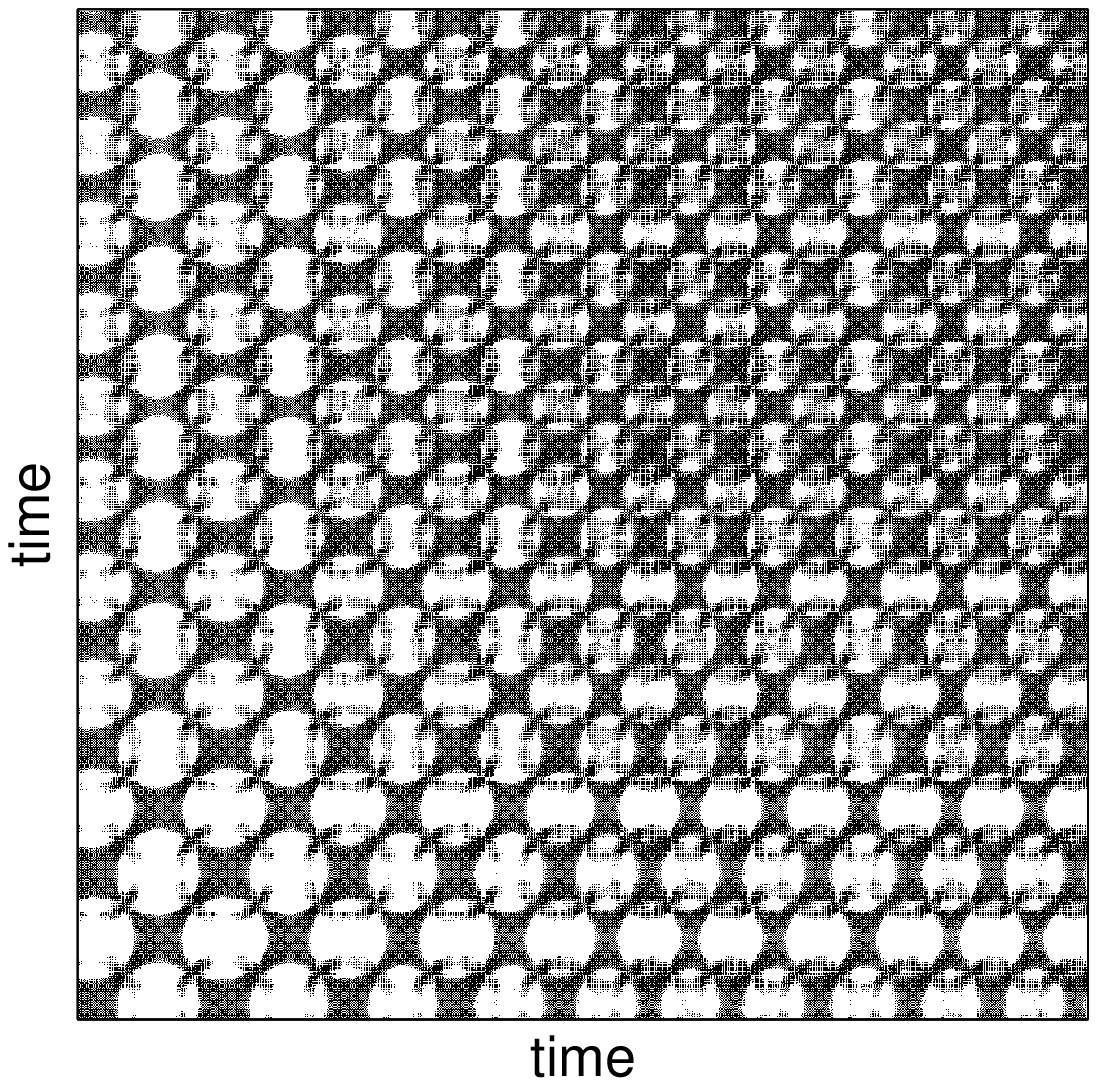}
\caption{An exemplary trajectory ($\tilde{E}=1.058$, $\tilde{L} =5M$)
is launched from the equatorial plane
$\theta(0)=\frac{\pi}{2}$ with $u^r(0)=0$. Parameters of the
background are $a=0.5M$, $\tilde{q}B_{0}=0.1M^{-1}$, $\tilde{q}\tilde{Q}=1.03$. In the upper left panel we observe
that setting $r(0)=8.4M$ results in oscillations around the
equatorial plane while launching it at $r(0)=8.7M$ makes it escape.
In the upper middle panel we examine the trajectory of the escaping
particle in terms of the rescaled radial coordinate
$r^*\equiv\frac{r-r_{+}}{r}$. In the case of oscillating trajectories
two distinct modes of motion are observed (upper right panel). The
first particle ($r(0)=11.5M$) shows a complex ``ribbon-like'' trajectory;
the other one ($r(0)=8.4M$) fills uniformly the given portion of
the potential valley. The Recurrence Plots are also shown (bottom panels). 
We observe a highly ordered regular pattern for the particle with $r(0)=8.4M$
(left panel), a more complicated diagonal pattern of the ribbon--like
trajectory (launched at $r(0)=11.5M$, middle panel), and a
disrupted diagonal pattern of the transitional trajectory ($r(0)=11.4M$,
right panel).}
\label{osc_unik1}
\end{figure*}

\subsection{\label{spindep}The effect of spin on the chaoticness of motion}
Much attention has been recently focused towards the problem of
determining the black hole spin from the properties of motion of
surrounding matter \citep{narayan,reynolds03}. The astrophysical
motivation to address these issues arises from the fact that cosmic
black holes are fully described by three parameters -- mass, electric
charge and spin. While the methods of mass determination have been
widely discussed \citep[e.g.][]{casares07,vestergaard10,czerny10}, the
electric charge is considered to be negligible because of rapid
neutralization of black holes via selective accretion. However,
determining the spin is a much more challenging task: the spin is
important, but its influence is apparent only {\em very} near the black hole
horizon \citep{murphy09}.

One can raise a question of whether the value of spin parameter $a$ of
the Kerr black hole affects the dynamical regime of motion in the
immediate neighborhood of the black hole. In other words, we ask if the
spin parameter $a$ triggers or diminishes the chaoticness of the system.
Answering this question is not straightforward because by altering the
spin across an interval of values ($a^2\leq{}M^2$) we inevitably have
to change some other variables of the system; otherwise the different
cases could not be directly compared. Moreover, the location and the
very existence of the potential lobes is not automatically ensured  over
the whole range of spin because of strong $V_{\rm{eff}}(a)$ dependence.

We found that in order to keep the off-equatorial lobe at the initial
position and the original size we need to increase the energy
$\tilde{E}$, roughly proportionally to the increment of $a$. The effect
of increasing $a$ exhibits itself by lifting the hyperplane of effective
potential. To compensate for this effect we have to elevate the
$\tilde{E}$-plane at which we cut the potential, so that we obtain
(roughly) the original closed contour (the potential lobe) inside of
which the motion is confined. It turns out that this can be achieved by
linking both quantities linearly.

\begin{figure*}[htb]
\centering
\includegraphics[scale=0.55, clip]{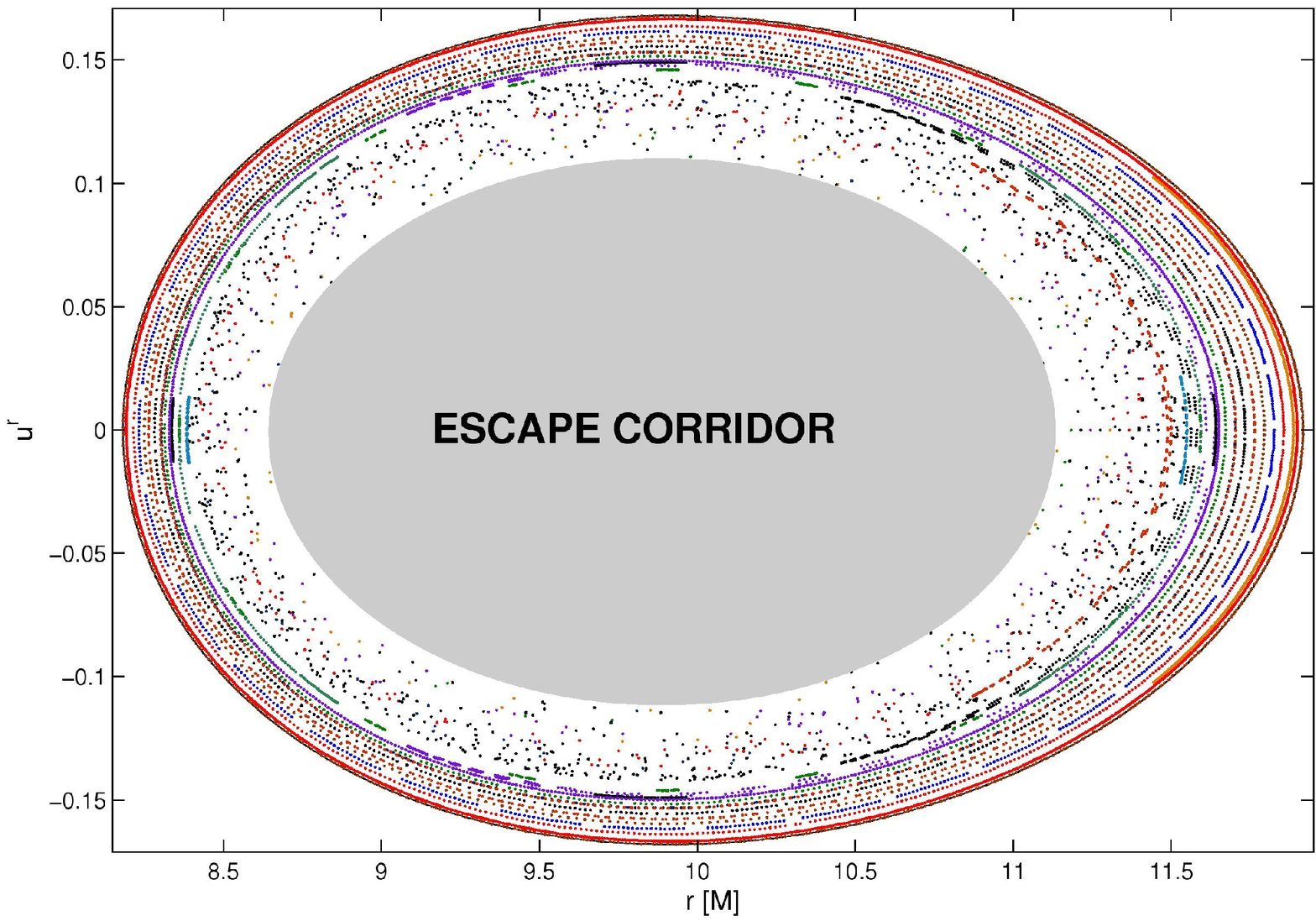}
\caption{The Poincar\'e
surface of section of several trajectories
($\tilde{E}=1.058$, $\tilde{L} =5M$, $\tilde{q}B_{0}=0.1M^{-1}$, 
$\tilde{q}\tilde{Q}=1.03$, $a=0.5M$, $\theta=\frac{\pi}{2}$)
launched from the equatorial plane with
various values of $r(0)$ and  $u^r(0)=0$. Grey colour indicates the
escape corridor which lets the particles escape from the equatorial
plane (see details in the text).} \label{corridor}
\end{figure*}

First we compare the dynamics in the off-equatorial lobe in the range
$\frac{a}{M}\in{\langle}0.3,1{\rangle}$ (for $a\lesssim{}0.3M$ the
topology of the effective potential changes) to which we linearly relate
the energy range $\tilde{E}\in{\langle}1.56, 2.35{\rangle}$ (whilst
other parameters are kept fixed as follows: $\tilde{L} =5M$,
$\tilde{q}B_{0}=2M^{-1}$, $r(0)=2.9\:M$, $\theta(0)=0.856$,
$u^{r}(0)=0$, $\tilde{q}\tilde{Q}=2$). By inspecting the Poincar\'e
surfaces of section and performing the recurrence analysis for a large
number of trajectories across the given range of $a$ and $\tilde{E}$
(exemplary cases presented in
\rff{spin_diskuze}) we come to the conclusion that there is no overall
trend that could suggest that $a$ is the unique driving agent affecting
the regime of motion. All trajectories in our survey exhibit a regular
behavior, which is also in agreement with the previous conclusion that
the motion in off-equatorial potential lobes associated with the Wald
test field is generally regular.

We also examined the dynamics of test particles launched from the
equatorial plane whose trajectories occupy the potential lobe extending
symmetrically above and below the equatorial plane. A given lobe
maintains its size for spin values 
$\frac{a}{M}\in{\langle}0.5,1{\rangle}$ and the related interval of
energy $\tilde{E}\in{\langle}1.795, 2.42{\rangle}$. A survey across the
given range of spin (energy) values reveals for this class of
trajectories both chaotic and regular regimes. In
\rff{spin_diskuze_eq} we observe that for the lowest inspected spin,
$a=0.5M$ ($\tilde{E}=1.795$), the regular motion dominates, although
islands of chaotic behavior are also present. Increasing the spin
(energy) we observe that regular trajectories gradually diminish. For
$a=0.6M$ ($\tilde{E}=1.92$) some regular orbits still appear, but they
are already dominated by chaotic trajectories. For higher spins
the traces of regular motion further diminish. We present the extreme case $a=M$
($\tilde{E}=2.42$) in \rff{spin_diskuze_eq} to illustrate this apparent
chaotic takeover. 

\begin{figure*}
\centering
\includegraphics[scale=0.65, clip]{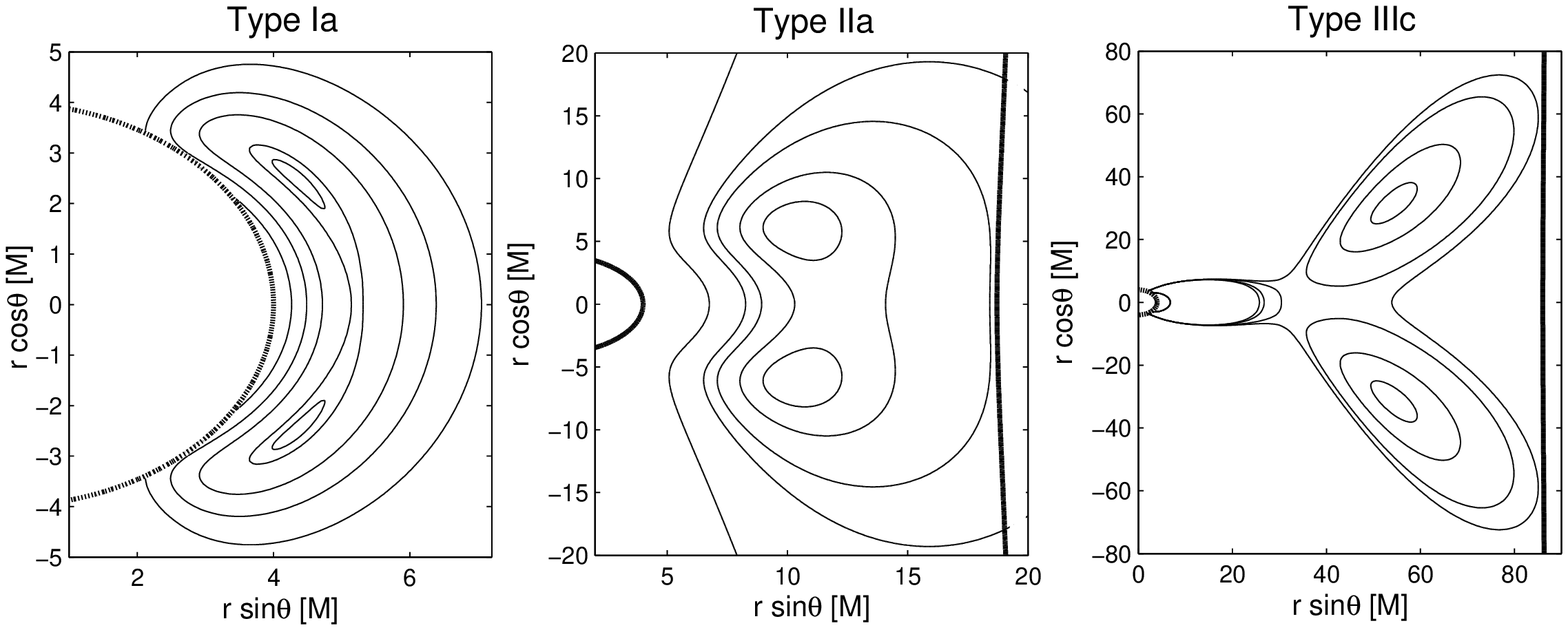}
\caption{Selected 
types of the effective potential behavior in the vicinity of
off-equatorial halo orbits above the surface of magnetic star with
rotating dipole magnetic field. The inner bold line signifies the
surface of the star at $r=4M$. The outer line is the light
surface.}
\label{rotdip_abc}
\end{figure*}

\begin{figure*}
\centering
\begin{tabular}{rl}
\includegraphics[scale=0.233, clip=true]{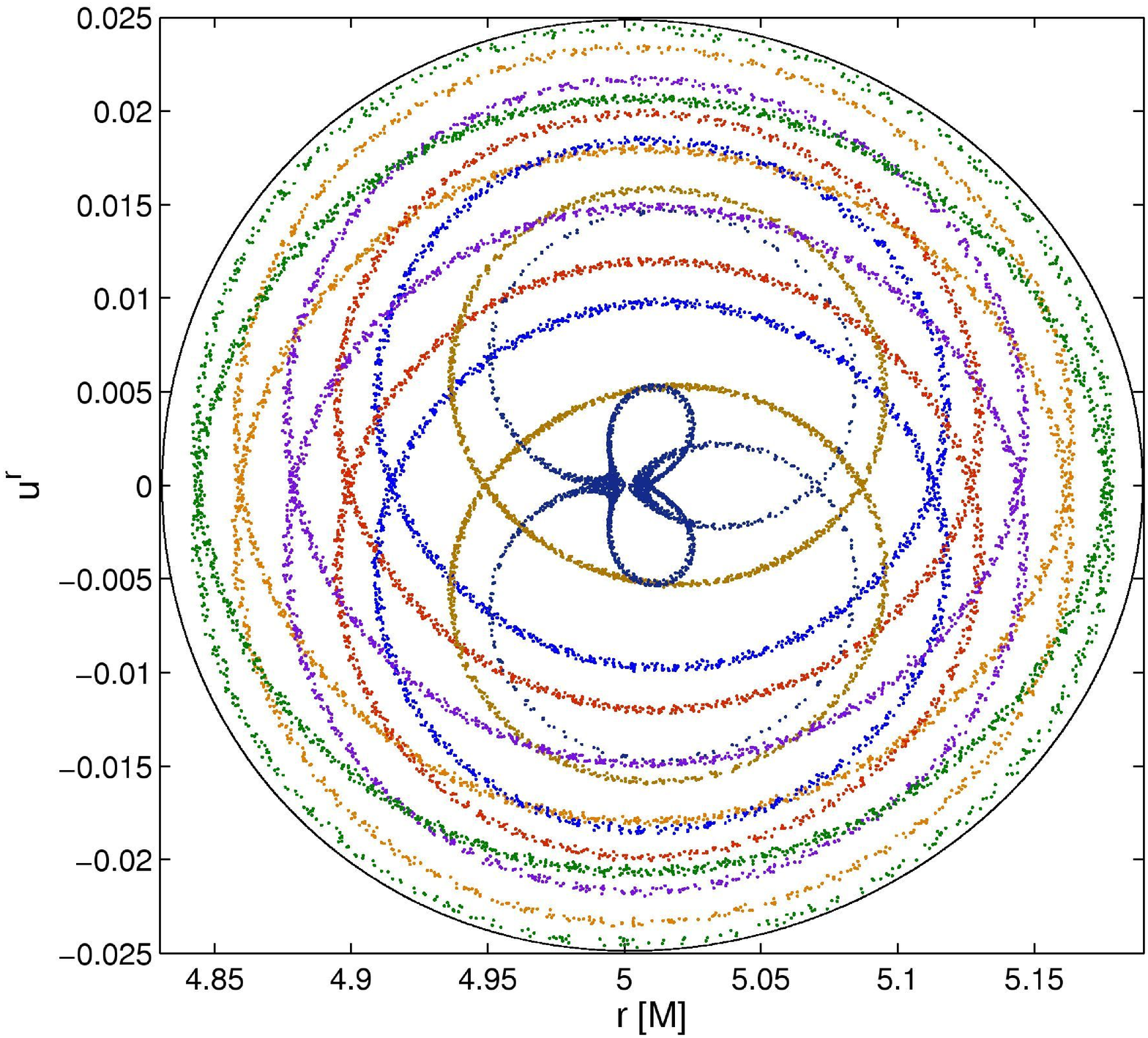} &\includegraphics[scale=0.36,clip=true]{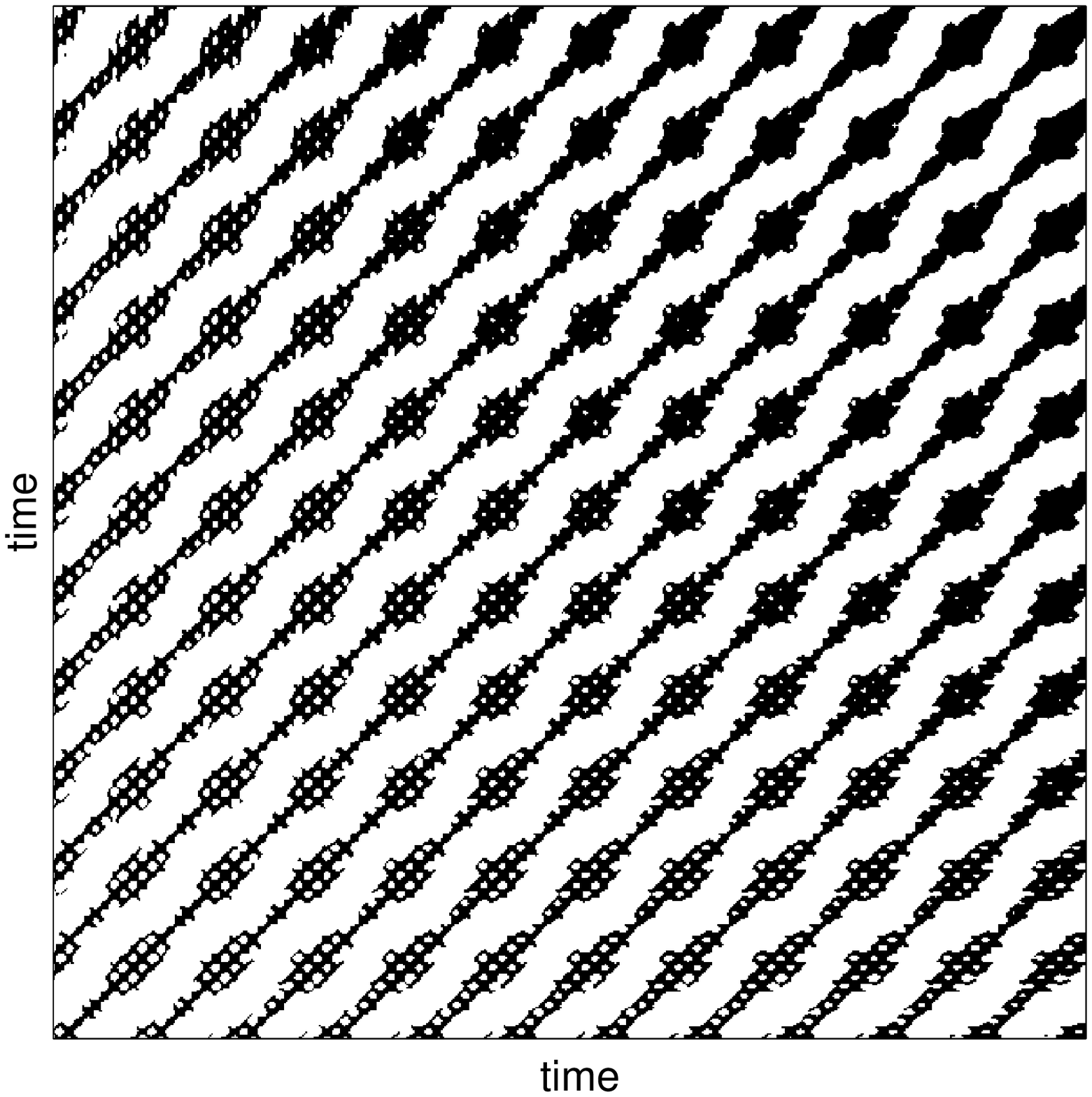}\\
\end{tabular}
\caption{Regular motion in the off-equatorial potential lobe at the
energy level $\tilde{E}=0.8482$. Parameters of the system are
$\tilde{q}\mathcal{M}=-5.71576\; M^2$, $\tilde{L}=0.87643\; M$,
$\Omega=0.011485\; M^{-1}$. The left panel shows sections of several
trajectories launched from
$\theta(0)=\theta_{\rm{section}}=1.0492$. One of them ($r(0)=5.02\;
M$ and $u^r{}(0)=0$) is visualized in the Recurrence Plot in the
right panel. The motion is regular (RP remains diagonal), however,
the density of recurrence points clearly grows
during the analyzed period.} \label{rotdip1_1p}
\end{figure*}

We conclude that for the class of orbits originating
in the equatorial plane, spin $a$ could possibly act as a destabilization factor
which triggers the chaotic motion if enhanced sufficiently. However
since the energy $\tilde{E}$ is increased simultaneously it is not
possible to attribute the observed dependence to the spin {\em itself}.
On the other hand we have already seen in the above-given discussion (sec.
\ref{wald_lobes}) that energy $\tilde{E}$ may {\em itself} act as a key factor
determining the dynamic regime of motion. Thus we suggest to attribute the observed triggering of the chaos to the increase of energy $\tilde{E}$ rather then spin $a$.

We remark that a similar problem concerning the spin dependence of the
motion chaoticness was addressed recently by \citet{japonci}. Authors of
the quoted paper employ a dipole magnetic test field upon a Kerr
background and perform a study of test particle trajectories, concluding
that increasing the value of the spin parameter stabilizes the motion in
a given setup. Unlike our case, the topology of the effective potential
in their scenario allows the chosen potential lobe to be maintained at a
given location and roughly the same size (but not the same depth) even
if $a$ varies while $\tilde{E}$ is kept constant. However increasing the
spin allows the authors to set gradually lower and lower energies, for
which more regular trajectories are found. This is not at all
surprising in perspective of our results, where the energy $\tilde{E}$
proved to play a key role in determining the stability of motion.
Although a direct comparison of presented surfaces of section differing
only in $a$ value may suggest the spin dependence, there is no
clear and unambiguous correlation. We thus suggest attributing the
observed dependence primarily to the level of energy, which acted as a
motion destabilizer also in our setup. In fact, even if a real trend
with the spin is present, it is hard to disentangle it from a
simultaneous change of $\tilde{E}$.

The question of the dependence of the dynamics upon the other parameters of the system ($\tilde{L}$, $\tilde{q}B_{\rm{0}}$, $\tilde{q}\tilde{Q}$) was also addressed during the analysis. It appeared that above mentioned difficulties accompanying the analysis of the spin dependence became even more serious in this case. Namely, neither it was possible to maintain the given potential lobe for a reasonable range of values of selected parameter nor we were able to fix this problem by binding this parameter in some simple manner to some other parameter (e.g. energy $\tilde{E}$). In other words, none of these parameters itself may be regarded as a trigger for chaos.

\subsection{\label{sec_valley}Motion in potential valleys}
Besides off-equatorial lobes, the potential may form another remarkable
structure -- an endless potential valley of almost constant depth which
runs parallely to the symmetry axis (\rff{valley}). The poloidal
orientation and asymptotical form of the valley are due to the fact that
the Wald test field does not vanish at spatial infinity and it
approaches the uniform magnetic field parallel to the symmetry axis.

The existence of such a potential corridor suggests that test
particles with a particular range of parameter values could escape from
the equatorial plane to large distance. We observe that a test particle
in the potential valley can keep oscillating around the equatorial
plane or it may escape from this plane completely, depending on its
initial position in the phase space (see the upper left panel of
\rff{osc_unik1}). We can examine the motion in the
asymptotic region by rescaling the radial coordinate (upper middle panel
of \rff{osc_unik1}). To achieve this, we use
$r^*\equiv\frac{r-r_{+}}{r}$, where $r_{+}=M+\sqrt{M^2-a^2}$ is the
position of the outer horizon of the black hole.

The normalization condition allows one to express $u^{\theta}$ as a
function of other phase space variables and parameters of the system.
Intuition suggests that, considering particles launched from the
equatorial plane ($\theta(0)=\frac{\pi}{2}$), the initial value
$u^{\theta}(0)$ is a governing parameter which decides whether the
particle remains oscillating around $\theta=\frac{\pi}{2}$ or leaves it
once forever. We can draw the isolines of selected $u^{\theta}$ values
in the ($r$, $u^r$)-plane which we use as a surface of section
($\theta_{\rm{section}}=\frac{\pi}{2}$) for the inspection of the test
particle dynamics. Comparing acquired isolines for various values of
$u^{\theta}$ with the empirically stated escape corridor of
\rff{corridor} leads to the conclusion that they never coincide
perfectly, although the correlation is quite high. In other words there
is no definite threshold value of $u^{\theta}(0)$ which would determine
whether a selected combination of $r(0)$, $u^r(0)$ (while other
parameters are fixed) lets the particle launched at equatorial
plane leave or oscillate around the plane.

In \rff{corridor} we observe several qualitatively different types of
possible particle dynamics. Next to the effective potential contour we
find closed and well-defined curves which represent the regular motion.
Going further inside we notice that fragmented curves are present. Below
them we find progressively more and more blurred patterns, which again
signifies the onset of chaos. The inner parts of the potential lobe are
occupied by the ``escape corridor'' where the particles can stream
freely from the equatorial plane, as seen in \rff{osc_unik1}.

The trajectories represented by the closed curves in the surface of
section differ profoundly from the disconnected curve orbits. The
difference can be seen also in the direct projection onto the poloidal
plane (upper right panel of \rff{osc_unik1}). While the trajectories of
the first type gradually fill each particular compact region of given
section, the latter forms bundles which curl through the projection plane
resembling ribbons that bound regions which are never reached by the
particle.

\begin{figure*}
\centering
\includegraphics[scale=0.6, clip=true]{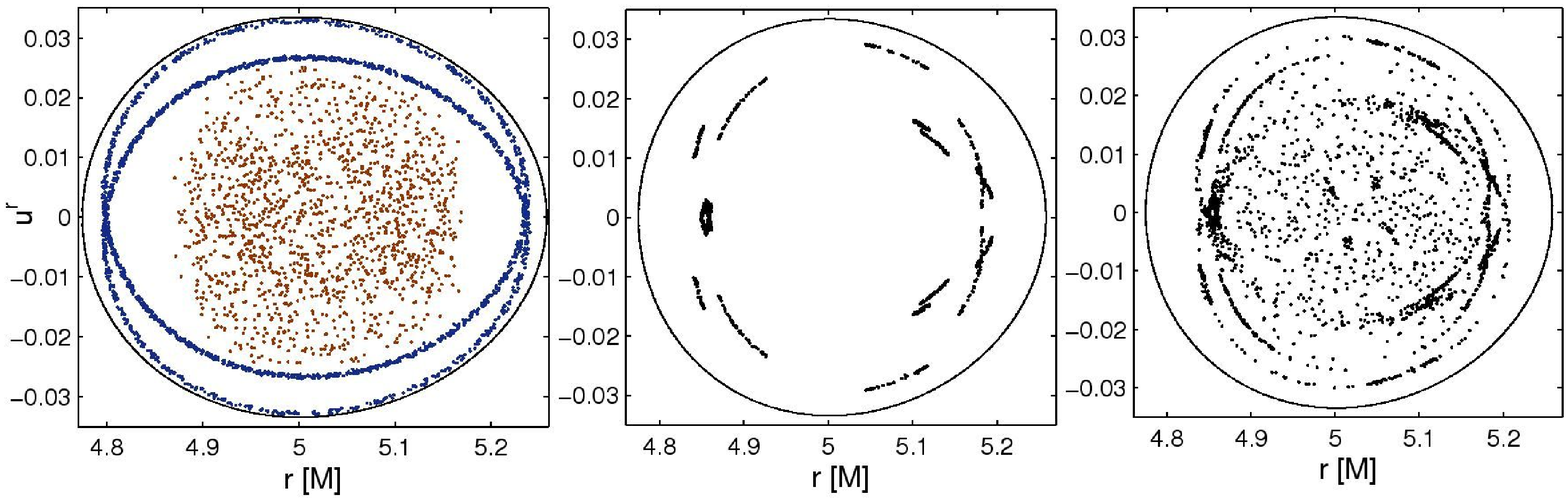}\\[10pt]
\includegraphics[scale=0.325, clip=true]{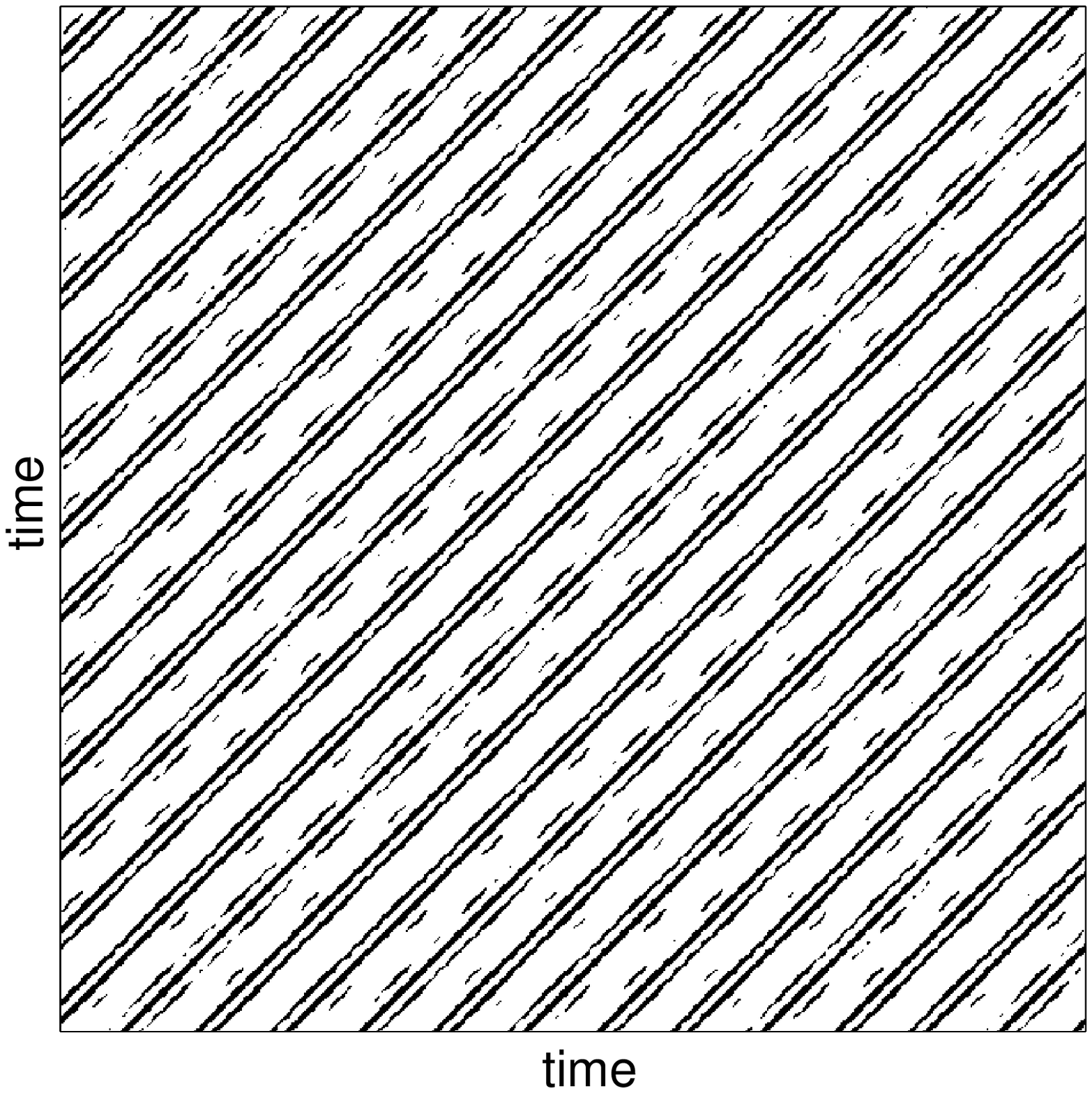}~~
\includegraphics[scale=0.3, clip=true]{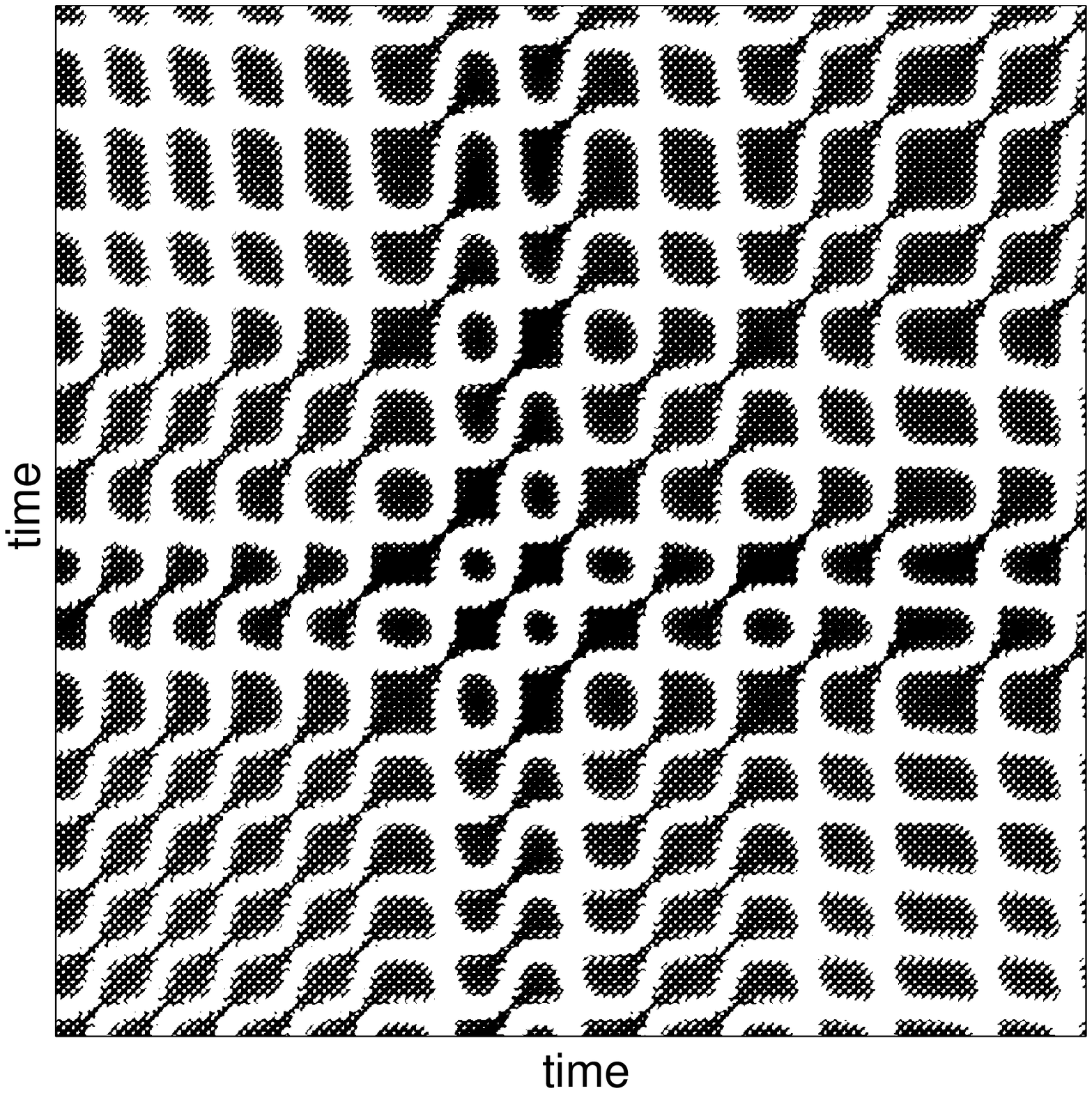} 
\caption{For an energy value of
$\tilde{E}=0.8485$, both off-equatorial lobes merge via the equatorial
plane. The upper-left panel shows Poincar\'e sections of two
trajectories ($\theta(0)=\theta_{\rm{section}}=1.0492$, $u^r(0)=0$). A
particle launched at $r(0)=4.8 \;M$ never crosses the equatorial plane
and moves regularly. Setting $r(0)=5\; M$ we observe a chaotic motion 
crossing the equatorial plane repeatedly. All
particles launched with $r(0)$, $u^r(0)$, corresponding to the inner
parts of the potential curve, move in the same chaotic manner. The
outskirts are occupied by regular trajectories. The upper-middle panel
shows the transient trajectory ($r(0)=4.85\; M$), regular during
the integration period of $\lambda=10^5$. In the upper-right panel, the
integration time is prolonged to
$\lambda=3\times{}10^5$. Here, the onset of chaos is connected with the first
passage through the equatorial plane. The RP of the regular trajectory
with $r(0)=4.8 \;M$ is presented in the bottom-left panel; the RP
on the right belongs to the chaotic trajectory with $r(0)=5 \;M$.}
\label{rotdip1_2}
\end{figure*}

By employing the Recurrence Plots (bottom panels of \rff{osc_unik1}) we
confirm that the dynamics differs significantly in these distinct modes
of motion. Not surprisingly we obtain a typical regular pattern in the
case of trajectory which forms a closed sharp curve in the surface of
section (bottom left panel of \rff{osc_unik1}). A ``ribbon--like''
trajectory (fragmented curve in \rff{corridor}) results in an ordered
checkerboard pattern (bottom middle panel), which is known to be typical
for periodic and quasi--periodic systems \citep{marwan}. Finally, in the
bottom right panel of \rff{osc_unik1} we observe that a blurred curve
trajectory exhibits slight chaotic behavior in its RP. The diagonal
structures are partially disrupted and we notice that diagonal lines
become bent as they approach the line of identity.

Authors of the recent paper \citep{vlny10} observed patterns similar to our fragmented curves in Poincar\'{e} sections  of phase space occupied by geodesic trajectories around slightly perturbed Kerr spacetime described by Manko-Novikov metric. They assume that such fragments can be identified with the Birkhoff islands anticipated by the Poincar\'{e}--Birkhoff theorem. A~given chain of Birkhoff islands (i.e. the fragmented curve in the section) originates from the resonant torus of an unperturbed system whose orbits have characteristic frequencies in a given rational ratio. The most prominent chains of islands should belong to the resonances characterized by simple integer ratios $\frac{1}{2}$, $\frac{2}{3}$ etc. Authors of the cited paper conclude that the presence of the Birkhoff islands might be in principle detected (at least in the context of gravitational radiation emitted by extreme mass ratio inspiraling sources) which would prove the system to be perturbed. However, since the assumption of the weakness of the perturbation is generally not fulfilled in our case, we do not explicitely identify fragmented tori in the surfaces of section with the Birkhoff chains.

The class of escaping trajectories in the Kerr background was recently
discussed by \citet{preti}. The author suggests that a Wald
electromagnetic field employed in this setup could serve as a charge
separation mechanism for astrophysical black holes since the sign of the
particle charge may determine whether a given particle escapes from the
equatorial plane or becomes trapped in cross-equatorial confinement (or
falls into the horizon).

\section{A magnetic star}
\label{sectionmagnetized}
We describe the gravitational field outside a compact star
by the Schwarzschild metric,
\begin{eqnarray}
\label{MetricSchw}
{\rm d}s^2&=&-\left(1-\frac{2M}{r}\right){\rm d}t^2
+\left(1-\frac{2M}{r}\right)^{-1}{\rm d}r^2\\& &+r^2({\rm d}\theta^2+\sin^2{\theta}{\rm d}\phi^2)\nonumber.
\end{eqnarray}
The associated magnetic field is modeled as a dipole rotating
at angular velocity $\Omega$ \citep{sengupta}:
\begin{eqnarray}
\label{rotdippot}
A_{t}&=&-\Omega{}A_{\phi}=\frac{3\mathcal{M}\Omega \mathcal{R}\sin^2{\theta}}{8M^3},\\ A_{\phi}&=&-\frac{3\mathcal{M}\mathcal{R}\sin^2{\theta}}{8M^3},
\end{eqnarray}
where
\begin{eqnarray}
\mathcal{R}=2M^2+2Mr+r^2\log{\left(1-\frac{2M}{r}\right)}.
\end{eqnarray}
The related dipole moment $\mathcal{M}$ is given by \citep{halo2_31}
\begin{equation}
\label{Magnetic}
\mathcal{M}=\frac{4M^3r_{\star}^{3/2}\left(r_{\star}-2M\right)^{1/2}\;B_{0}}{6M(r_{\star}-M)+3r_{\star}\,(r_{\star}-2M)\,\ln{\left(1-2Mr_{\star}^{-1}\right)}},
\end{equation}
where $B_{0}$ is the magnetic field at the neutron star equator,
$r_{\star}$ is the radius of the star surface.\footnote{The existence of
extremely compact stars with $r_{\star}\approx{}3M$ is unlikely, but not
excluded \citep{bah89,stu09}. Most of the realistic equations of state
imply a lower limit $r_{\star}\approx{}3.5M$ \citep{glend97}. On the
other hand, the models of Q-stars do allow a lower limit of
$r_{\star}\approx2.8M$ \citep{mil98}.} 

We assume eqs.\ (\ref{MetricSchw})--(\ref{rotdippot}) to hold
outside the star surface ($r>r_*$) and inside the light cylinder
($u^{\mu}u_{\mu}<0$). We
set $r_*=4M$ as the inner radial boundary of the particle motion.
As for the light cylinder, the mentioned condition results
in a relation $r^2\sin{\theta}^2\Omega^2=1-\frac{2M}{r}$, which
implicitly specifies the outer boundary.
The vector potential (\ref{rotdippot}) is valid inside the rigidly
corotating magnetospheric plasma, which we consider to be an excellent
conductor, so that the force-free condition $F^{\mu}_{\nu}u^{\nu}=0$
holds for the plasma for which $u^{\mu}=(u^t,0,0,u^{\phi})$ and
$\frac{u^{\phi}}{u^{t}}=\Omega$.

A general formula for the effective potential eq.\ (\ref{effpot})
simplifies to the form
\begin{eqnarray}
\label{EffectiveNS}
V_{\rm eff}&=&
-\frac{3\tilde{q}\mathcal{M}\mathcal{R}\Omega\sin^2{\theta}}{8M^3}\\
& &+\left(1-\frac{2M}{r}\right)^{\frac{1}{2}}\!\left[1+\left(\frac{\tilde{L}}{r\sin{\theta}}+\frac{3\tilde{q}\mathcal{M}\mathcal{R}\sin{\theta}}{8M^3r}\right)^2\right]^{\frac{1}{2}}.\nonumber
\end{eqnarray}

\begin{figure*}
\centering
\includegraphics[scale=0.49,clip=true]{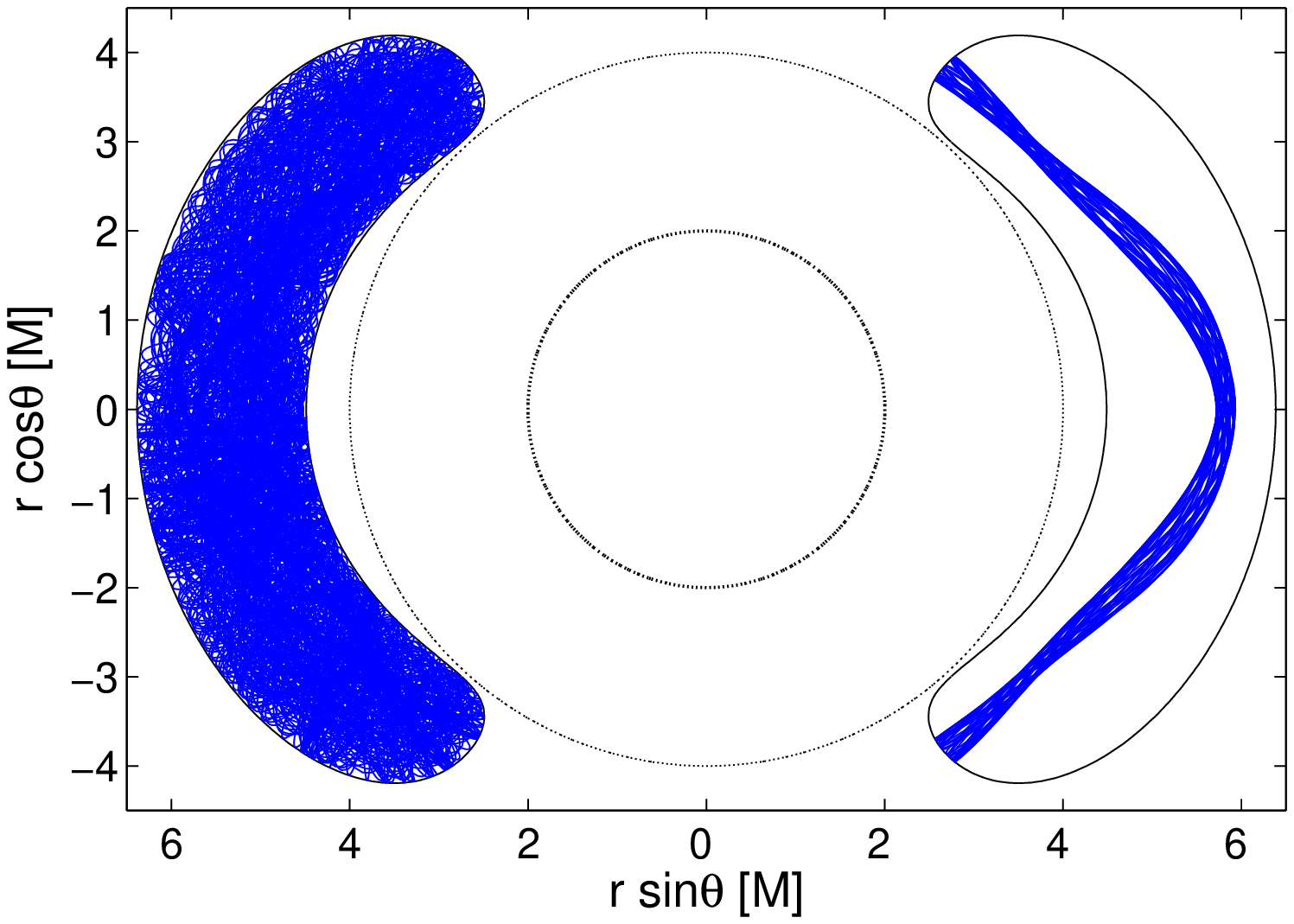}~~\includegraphics[scale=0.215, clip=true]{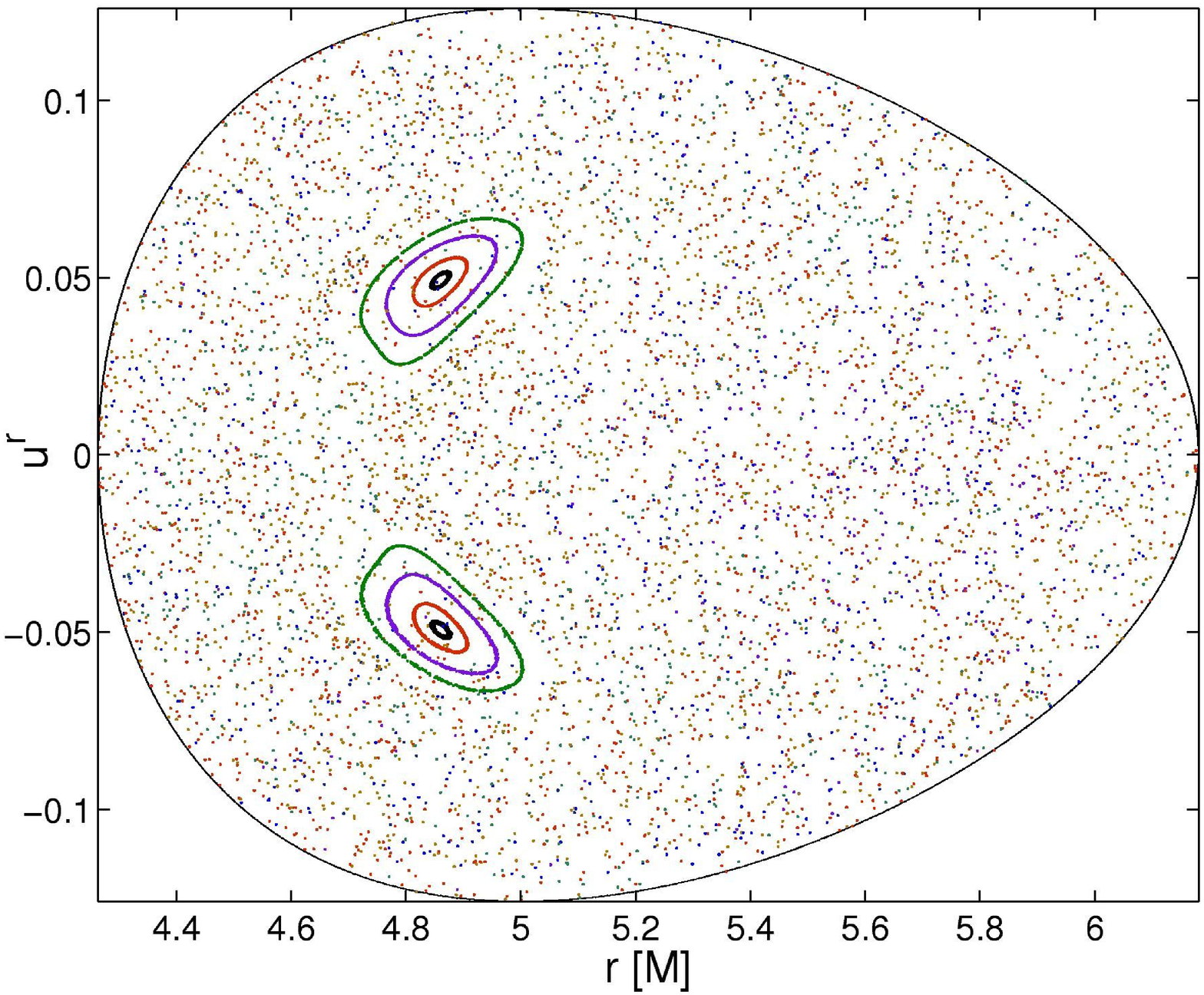}\\[10pt]
\includegraphics[scale=0.36, clip=true]{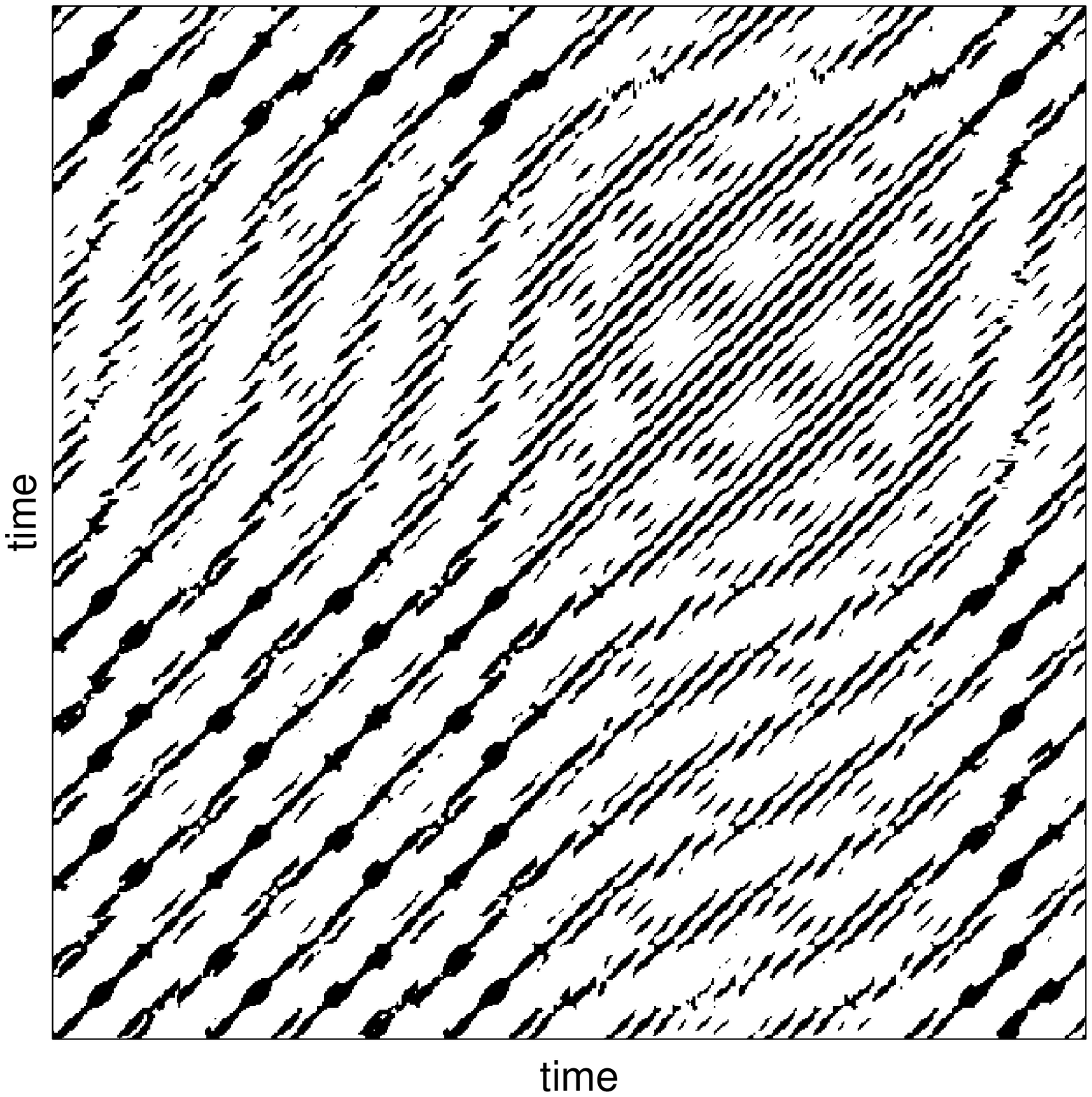}~~\includegraphics[scale=0.34, clip=true]{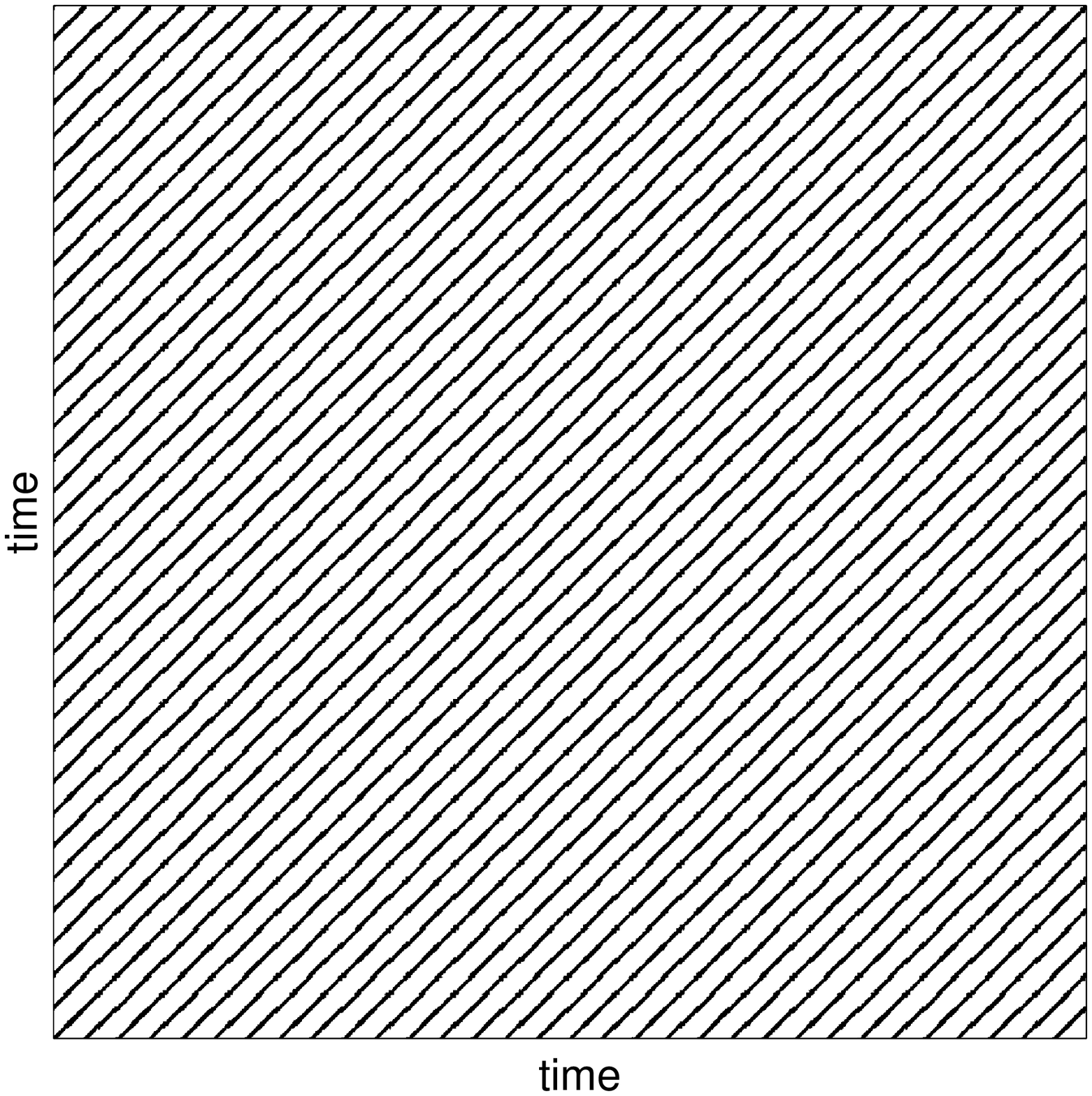}
\caption{For 
the energy level $\tilde{E}=0.857$ we obtain a broad lobe which almost
touches the surface of the star at $r=4\;M$. In the upper-left panel, 
we launch two particles with $\theta(0)=1.0492$, $r(0)=4.75\;M$. The
particle to the left of the star starts with $u^r(0)=0$ and moves
chaotically, while the other one with $u^r(0)=0.03$ follows a perfectly 
regular trajectory. The upper right-panel shows these two types of 
trajectory appear in the surface of section plot. The bottom panels 
demonstrate the difference between the two types in terms of 
Recurrence Plots.}
\label{rotdip1_3}
\end{figure*}

\subsection{Motion inside the potential lobes}
Our previous analysis \citep{halo2} revealed a number of distinct types
of possible topological structures of the effective potential. However
it appears that the system is not as rich in its dynamical properties.
The test particle trajectories share some similar features across
different classes of the effective potential. Therefore, we only present
surveys of particle dynamics in three exemplary types: Ia, IIa and IIIc 
(\rff{rotdip_abc}; see \citet{halo2} for the complete review).

Class Ia lobes grow with energy increasing. Once the level of the
equatorial saddle point is reached, the lobes merge with each other
across the equatorial plane. The single merged lobe eventually
intersects the surface of the star if the energy level is increased
sufficiently, letting the particle fall onto the surface. Lobes of IIa
type also merge via the equatorial plane but in contrast to the first
type the merged lobe opens toward the light cylinder (beyond which the
model becomes invalid). Lobes of the class IIIc first open via the
off-equatorial saddle points, allowing the particles fall onto the
star, before the lobes merge through the equatorial plane.

Now we study the three selected types of the effective potential
topology in more detail. We are primarily interested whether and how the
dynamic regime changes across the given range of specific energy
$\tilde{E}$. Especially, we shall address what happens with the dynamics
when the particle acquires enough energy to cross the saddle point.

The first survey (type Ia) begins at energy level $\tilde{E}=0.8482$,
corresponding to the closed lobe. In \rff{rotdip1_1p} we observe that
the motion inside the lobe is stable. No chaotic properties are detected
-- neither in Poincar\'e surfaces of section nor in the Recurrence
Plots.

As the energy increases to $\tilde{E}=0.8485$, the symmetrical lobes
merge via the equatorial plane. By inspecting a number of trajectories
in this case we find that chaos starts appearing at this point -- those
particles which notice the gate through the equatorial plane always fill
the entire allowed region and they move chaotically. Nevertheless, there
are still such particles which move regularly in one of the two parts of
merged lobe and never cross the equatorial plane. We can also find
transient trajectories corresponding to regular motion lasting for some
period of time in one part of the lobe, followed by chaotic motion over
the entire lobe once the particle finds and encounters the passage
across the equatorial plane. All of the mentioned cases are illustrated
in \rff{rotdip1_2}.

Increasing the energy further to $\tilde{E}=0.857$, we obtain a broad
potential lobe which almost touches the star surface. The situation
changes from the previous case where the gate connecting the
off-equatorial lobes was narrow. Now we do not find trajectories which
occupy only one part of the lobe, above (or below) the equatorial plane.
Chaotic trajectories densely filling the entire lobe are typical for
this setup. We also encounter perfectly regular trajectories forming
ribbon--like structures spanned between northern and southern borders of
the lobe. In Poincar\'e sections these appear as regular islands
surrounded by a chaotic ocean (upper panels of \rff{rotdip1_3}). We
notice that the RP of this regular trajectory is extraordinarily simple
and consists of almost perfect diagonal lines (bottom panels of
\rff{rotdip1_3}). Thus its dynamic properties are close to those of a
periodic system, which is in contrast with the neighboring fully chaotic
orbits.

\begin{figure*}
\centering
\includegraphics[scale=0.253, clip=true]{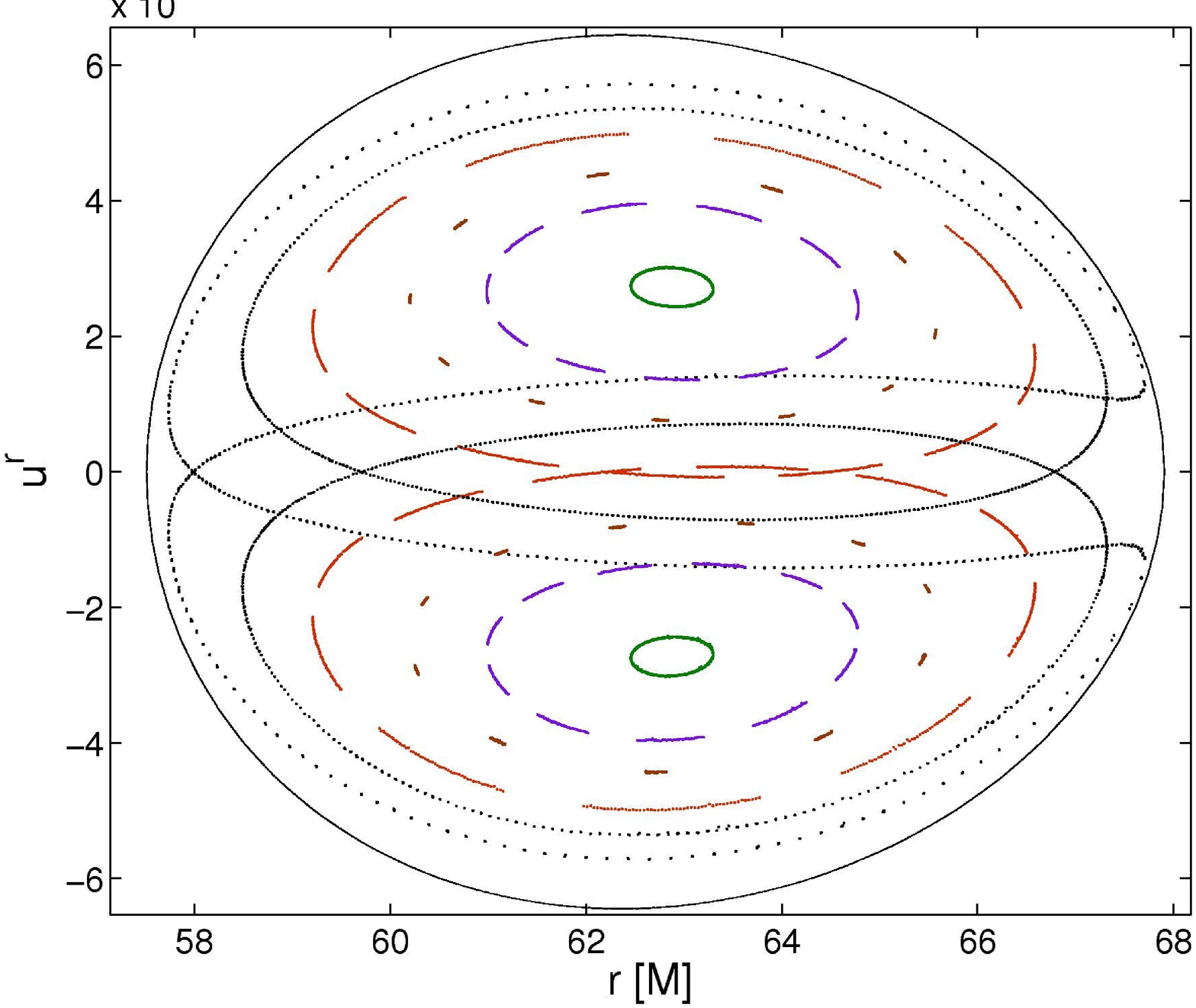}~~ 
\includegraphics[scale=0.58, clip=true]{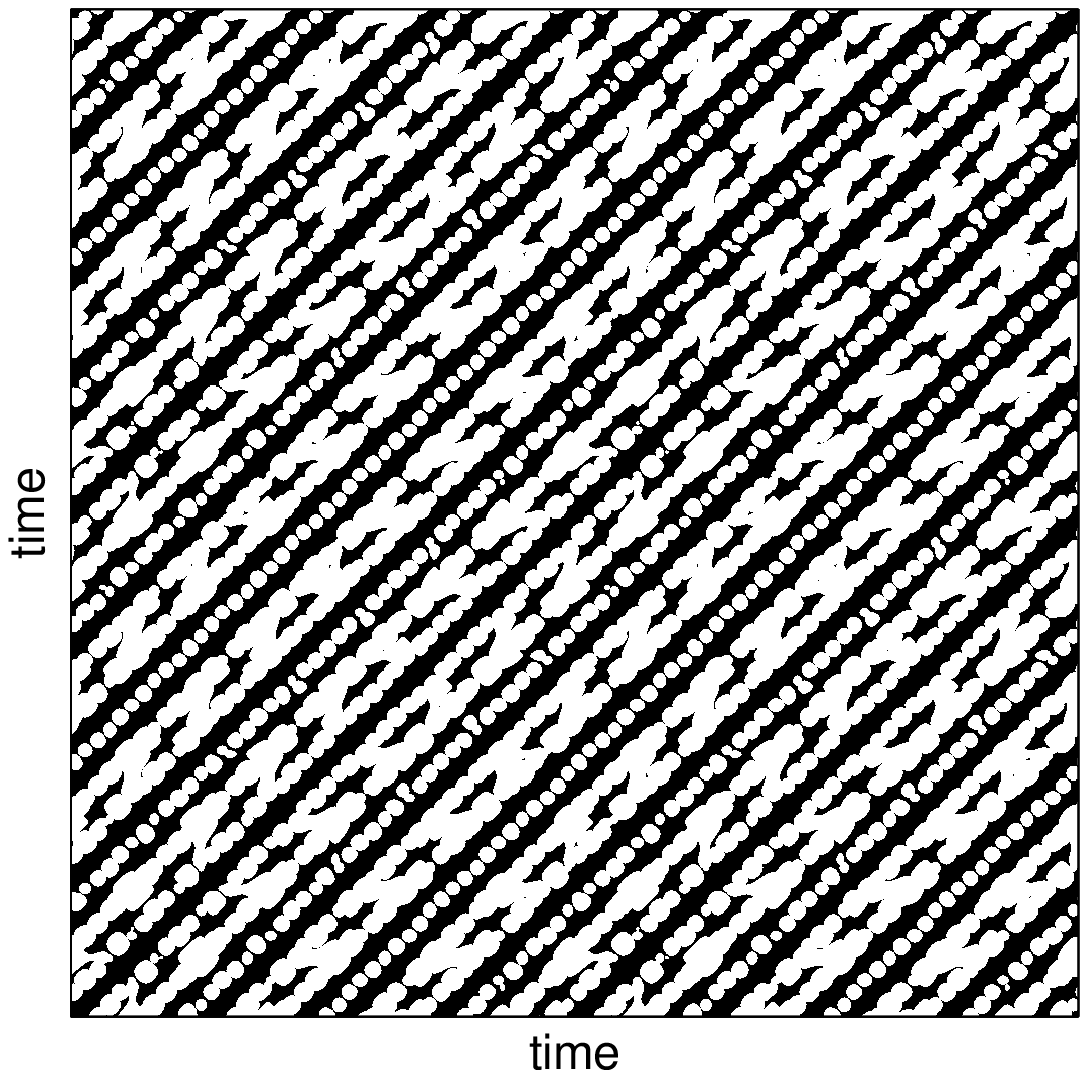} 
\caption{Regular motion in an
off-equatorial lobe of the third type. Parameters used:
$\tilde{E}=0.99579$, $\tilde{L}=6.25382\;M$,
$\tilde{q}\mathcal{M}=45.87368\; M^2$, $\tilde{L}=6.25382\; M$,
$\Omega=0.011485\; M^{-1}$. Particles are launched from latitude
$\theta(0)=\theta_{\rm{section}}=1.0492$.}
\label{rotdip3_1}
\end{figure*}

The second type (class IIa) of the effective potential topology of
off-equatorial lobes differs from the first one significantly as the
lobes do not open towards the star when the energy is raised
sufficiently. On the contrary, in this case we observe that the lobe's
boundary touches the light cylinder first.

The motion in the off-equatorial lobes proves to be regular while the
merging lobes bring chaos into play. Chaos becomes dominant for broader
lobes, however, stable regular orbits also persist. The results are
similar to those of the first type of potential topology discussed
above.

The last analyzed topology of the lobes (class IIIc) differs profoundly
from the preceding two cases, as can be seen in \rff{rotdip_abc}. We
find that stable motion dominates in this setup. This can be verified by
comparison with \rff{rotdip3_1}.

As we further increase the energy level, we obtain more complicated
shapes of the equipotentials that allow the particle to fall on the star
surface. On the other hand, opening the outflow gate energetically
precedes the merging point of both off-equatorial lobes. In \rff{rotdip3_3}, we discuss the motion governed by the largest possible
lobe which almost touches the light cylinder. We observe that stable
regular orbits are still possible for those particles that do not hit
the passage.

From the above-given discussion we conclude that the motion of charged
test particles in the off-equatorial lobes allowed by the test field of
the rotating magnetic dipole on the Schwarzschild background is largely
regular. Once the off-equatorial lobes merge with each other, chaos
may appear. Increasing the energy, the chaotic motion becomes typical
but, quite surprisingly, very stable orbits also exist under these
circumstances.

\section{Discussion and conclusions}
\label{concl}
We studied the regular and chaotic motion of electrically charged
particles near a magnetized rotating black hole or a compact star. We
employed the method of recurrence analysis in the phase space, which
allowed us to characterize the chaoticness of the system in a
quantitative manner. Unlike the method of Poincar\'e surfaces, the
Recurrence Plots have not yet been widely used to study the chaotic
systems in the regime of strong gravity.

The main motivation for these investigations is the question of whether
the matter around magnetized compact objects can exhibit chaotic motion,
or if instead the system is typically regular. One of the main
applications of our considerations concerns the putative envelopes of
charged particles enshrouding the central body in a form of a fall-back
corona, or plasma coronae extending above the accretion disk. While we
concentrated on the specifications of the RP method in circumstances
of a relativistic system, the assumed model cannot be considered as any
kind of a realistic scheme for a genuine corona. We simply imposed a
large-scale ordered magnetic field acting on particles in a combination
with strong gravity.

Various aspects of charged particle motion were addressed throughout
this paper. First of all, we investigated the motion in off-equatorial
lobes above the horizon of a rotating black hole (modeled
by Kerr metric equipped with the Wald test field), as well as above the
surface of a magnetic star (modeled by the Schwarzschild metric with the
rotating dipolar magnetic field). In both cases we conclude that the
motion of test particles is regular, which was confirmed for a
representative number of orbits across the wide range of parameters over
all topological types of off-equatorial potential wells. This result is
somewhat unexpected because the off-equatorial orbits require a
perturbation to be strong enough (in terms of strength of the
electromagnetic field), so that it can balance the vertical component of
the gravitational force. 

\begin{figure*}
\centering
\includegraphics[scale=0.44, clip=true]{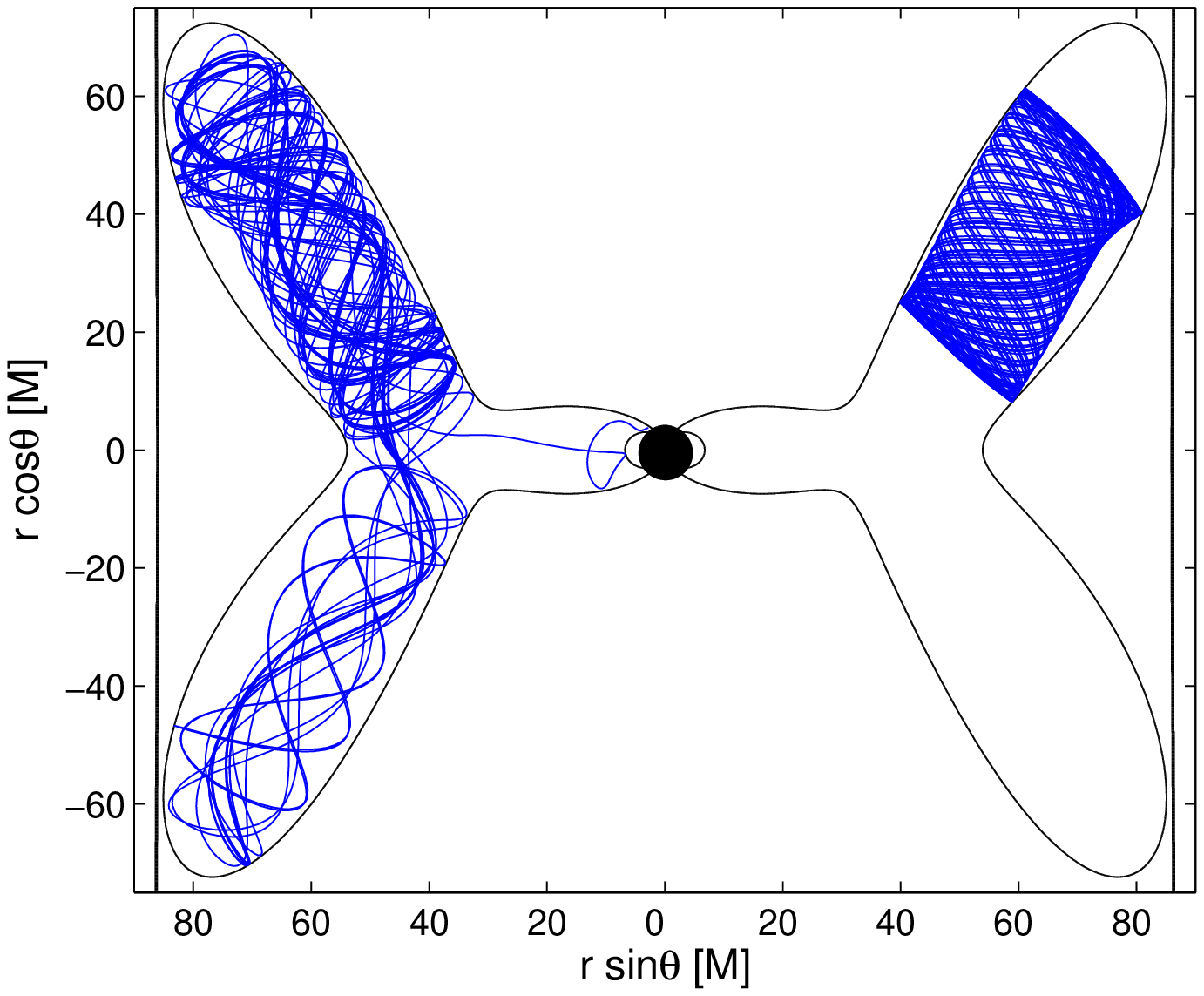}~~
\includegraphics[scale=0.197, clip=true]{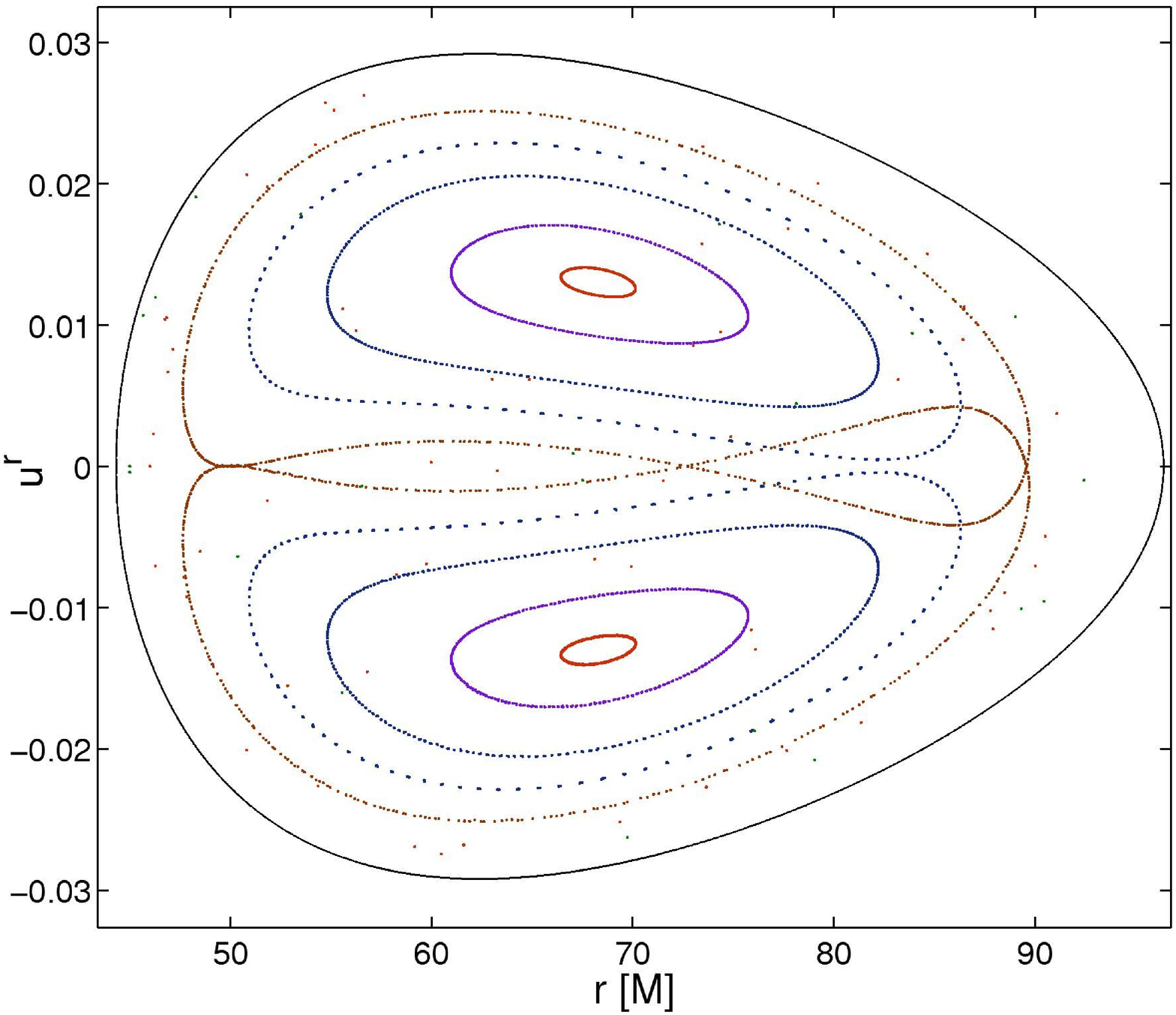}\\[10pt]
\includegraphics[scale=0.49, clip=true]{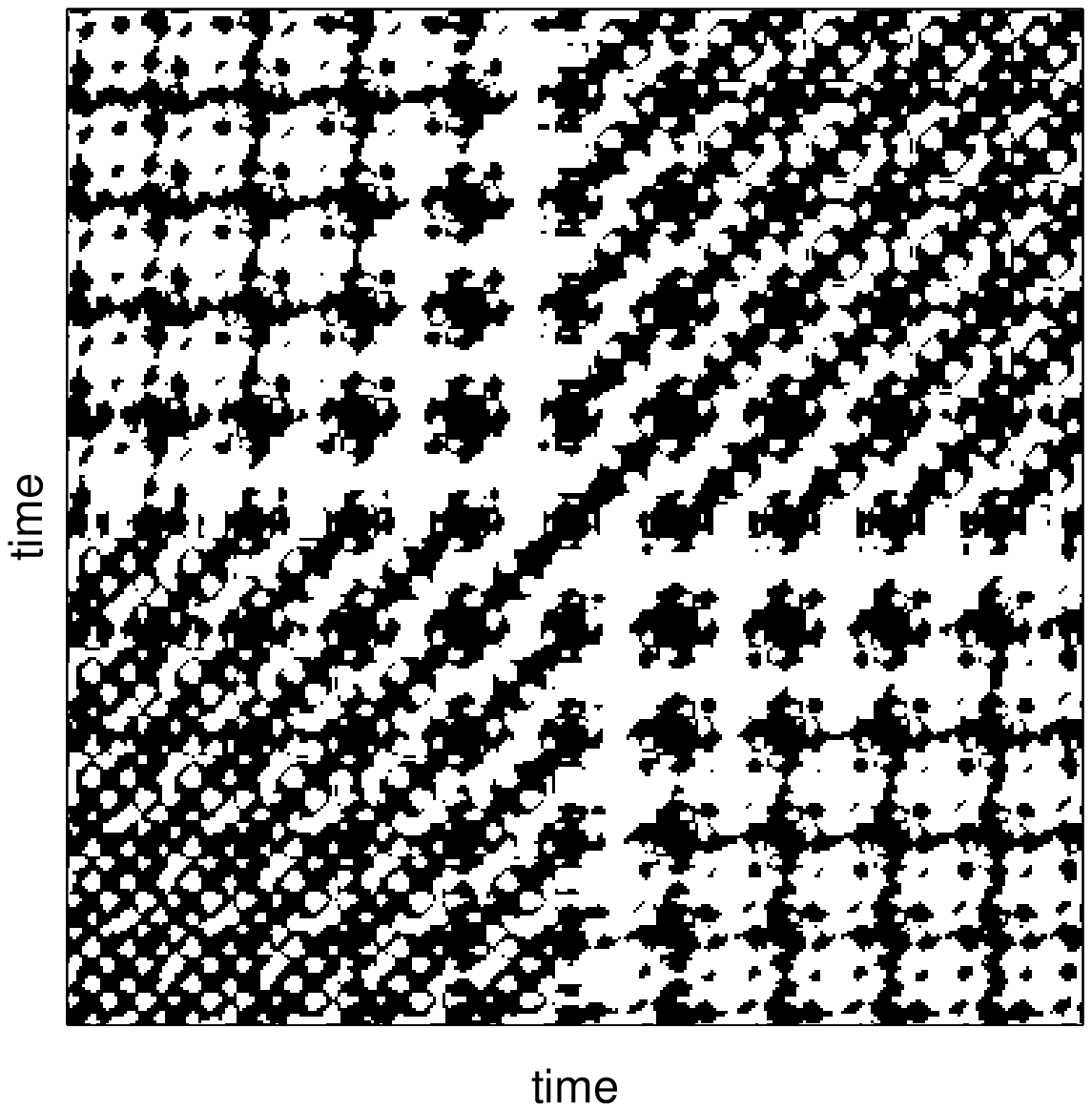}~~~~
\includegraphics[scale=.52, clip=true]{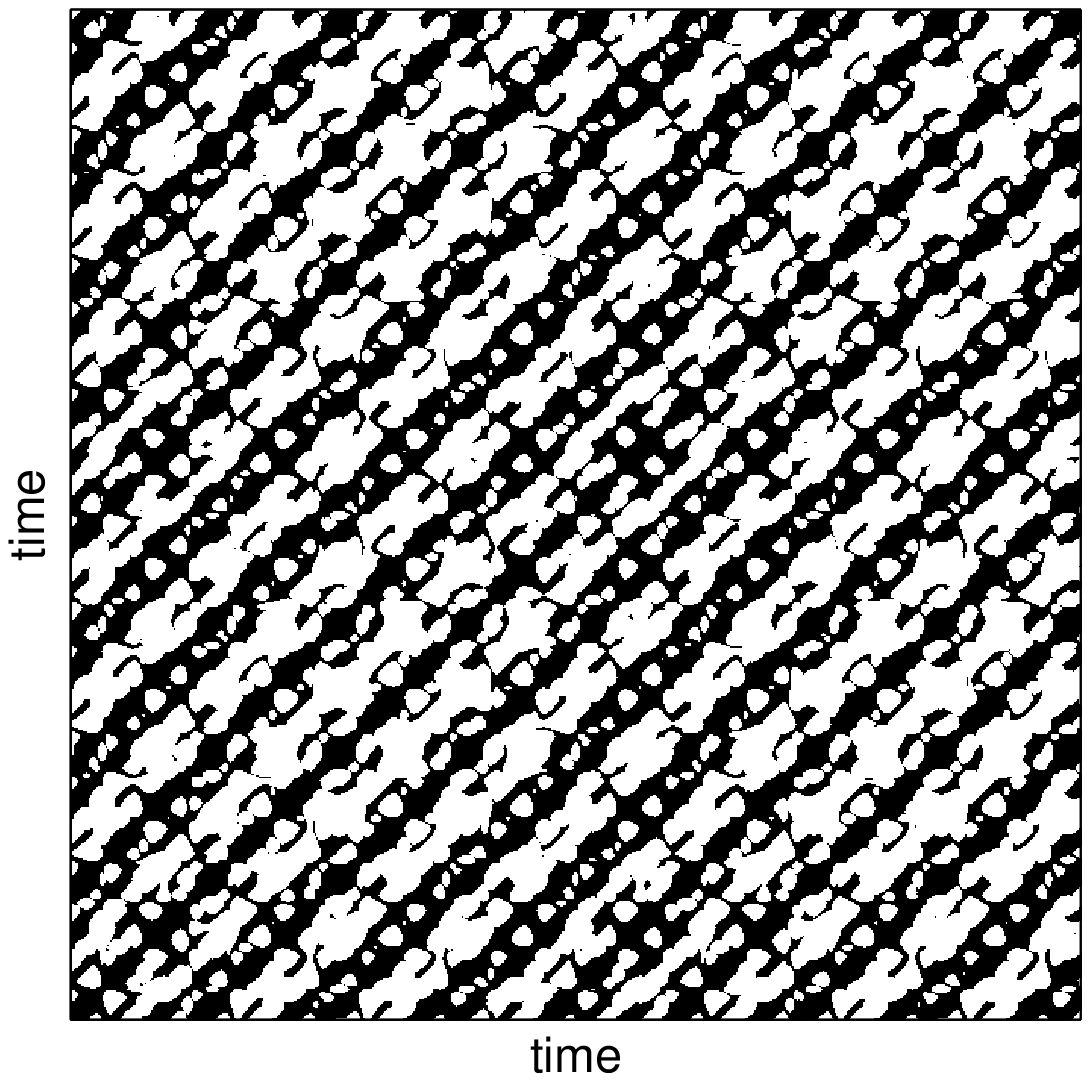} 
\caption{For the energy level $\tilde{E}=0.9962$ (other parameters as in 
\rff{rotdip3_1}) we obtain a large lobe that almost touches the
light cylinder and allows the particle to fall onto the surface of the
star via a narrow passage above and below the equatorial plane. The
upper left panel shows two trajectories launched from
$\theta(0)=1.0497$, $u^r(0)=0$. One of the particles (starting from
$r(0)=61.5\;M$) follows an unstable path and eventually falls on the
star surface. On the contrary, the other particle (starting from
$r(0)=72.5\;M$) moves regularly and never escapes any given part of the
lobe. The upper-right panel shows these two kinds of trajectory depicted
in the Poincar\'e surface of section. Chaotically dispersed points
belong to the escaping trajectory. The bottom-left panel shows the
Recurrence Plot of the escaping particle; this plot does not exhibit
typical chaotic behavior, although the large-scale structures are
present. The bottom-right panel presents the Recurrence Plot of stable
motion.}
\label{rotdip3_3}
\end{figure*}

Further, we investigated the response of the particle dynamics when the
energy level $\tilde{E}$ was raised gradually from the potential minimum
to values allowing cross-equatorial motion. We examined various
topological classes of the effective potential and came to the
conclusion that the cross-equatorial orbits are typically chaotic,
although very stable regular orbits may also persist for a certain
intermediate energy range. The classical work of \citet{henhei64} should be recalled in this context since it also identifies the energy as a trigger for chaotic motion in the analysed simple system. More recently the H\'{e}non--Heiles system was revisited in the relativistic context by \citet{vieira96}.

We also addressed the question of spin dependence of the stability of
motion for Kerr black hole in the Wald field. We noticed that this is a
rather subtle problem. The effective potential is by itself sensitive to
the spin value $a$ -- hence, we had to link the potential value roughly
linearly with the energy $\tilde{E}$ to maintain the potential lobe at a
given position. In other words, we did not find any clear and unique
indication of the spin dependence of the motion chaoticness. Most
trajectories exhibited regular behavior, which is also in agreement with
the previous results indicating that motion in off-equatorial lobes is
generally regular. On the other hand, in the case of the
cross-equatorial motion we observed that, for higher spins, more chaotic
features come into play when compared with the case of slow rotation.
This trend might be also attributed to simultaneous adjustments of
$\tilde{E}$. In other words, it appears impossible to give an
unambiguous conclusion about the spin dependence of the particles
dynamics. Instead, one has to deal with a complex, interrelated
dependence.

In the case of a Kerr black hole immersed in a large-scale magnetic
field, we observed the effect of confinement of particles regularly
oscillating around the equatorial plane. Escape of particles from the
plane is allowed for a given range of initial conditions since the
equipotentials do not close; they form an endless axial ``valley''
instead. The escaping trajectories create a narrow, collimated structure
parallel to the axis.

Asking the question about the dynamical properties of motion of matter
in the black hole coronae is of theoretical interest by its own, but it
is relevant also in view of precise spectroscopical and timing studies
with the forthcoming X-ray satellites. However, one more word of caution
is appropriate especially concerning our assumption of the ordered
magnetic field. Although the large-scale magnetic fields are adequate to
describe the initial background field of the magnetic star or a toroidal
current in the accretion disk
\citep[e.g.]{koide06,beckwith08,rothstein08}, a turbulent component is
known to develop quickly in the accreted plasma \citep{mckinney07}. These
will also perturb the motion of matter, and so the regular orbits may
quickly disappear. This should translate to the transient emergence and
subsequent disappearance of the periodic component in the observed
signal.

\acknowledgements
We thank Dr.\ Tom\'{a}\v{s} Pech\'{a}\v{c}ek for helpful advice
concerning the recurrence analysis, and Dr.\ Kendrah Murphy for her
critical comments and useful suggestions. OK acknowledges the doctoral
student program of the Czech Science Foundation (project No.\
205/09/H033). JK, VK and ZS thank the Czech Science Foundation (No.\
P209/10/P190, 205/07/0052 and 202/09/0772). Astronomical Institute and
the Institute of Physics have been operated under the projects MSM
AV0Z10030501 and 4781305903, and further supported by the Ministry of
Education project of Research Centres LC06014 in the Czech Republic.
\newpage

\appendix
\section{Choice of preferred observers}
\label{appa}
When analyzing the dynamics in the general relativistic context the fundamental question arises whether the distinction between chaotic and regular motion is coordinate dependent or not. To this end \citet{motter10} infers the transformation law for Lyapunov exponents. He concludes that although the Lyapunov exponents themselves are not invariant they transform in such a way that positive Lyapunov exponents remain positive and vice versa. In other words the distinction between regular and chaotic dynamics may be drawn invariantly.

In order to give the notion of recurrence a rigorous and, at the same
time, an intuitive sense, we can employ the $3+1$ formalism
\citep{thorne} that is based on an appropriate selection of a family of
spacetime-filling three-dimensional spacelike hypersurfaces (foliation)
of constant time $t$. The timelike curves orthogonal (in a spacetime
sense) to the hypersurfaces may be regarded as the world-lines of a
family of fiducial observers (FIDO) who naturally parameterize their
world line by proper time $\tau$ (whose rate of change generally differs
from that of $t$). FIDO identify each spatial hypersurface along his
world line as a slice of simultaneity. The geometry of this spacetime
slice is given by 3-metric $\gamma_{ij}$:
\begin{equation}
\label{3Dmetric}
\gamma_{ij}=g_{ij}+u_i{}u_j,
\end{equation}
where $u_i$ stands for the spatial part of FIDO's four-velocity and
$g_{ij}$ for the spatial part of the spacetime metric. Considering also
the time coordinate, $\gamma_{\mu\nu}$ can be regarded as a projector
onto the three-dimensional spatial hypersurface.

We select zero-angular momentum observers (ZAMO) \citep{bardeen}
as an appropriate representation of the fiducial observers. Their
orthonormal tetrad is given by
\begin{eqnarray}
\label{baze}
e_{(t)}&=& u\;=\;\frac{\sqrt{A}}{\sqrt{\Delta}\rho}\left(\frac{\partial}{\partial{}t}+\Omega\frac{\partial}{\partial\phi}\right)\\
e_{(r)}&=&\frac{\sqrt\Delta}{\rho}\frac{\partial}{\partial{}r}\\
e_{(\theta)}&=&\frac{1}{\rho}\frac{\partial}{\partial\theta}\\
e_{(\phi)}&=&\frac{\rho}{\sqrt{A}\sin\theta}\frac{\partial}{\partial\phi},
\end{eqnarray}
where $A\equiv(r^2+a^2)^2-a^2\Delta\sin^2\theta$ and
$\Omega=2aA^{-1}Mr$.

Projecting an arbitrary four-vector $C^{\mu}$ onto ZAMO's hypersurface of
simultaneity directly results in the following 3-dimensional quantities,
\begin{eqnarray}
\label{3Dslozky}
^{\rm{3D}}C^i&=&\gamma^{ij}C_j=\left({}C^r{},C^\theta{},\gamma^{\phi\phi}C_{\phi}\right),\\
^{\rm{3D}}C_i&=&\gamma_{ij}^{\rm{3D}}C^j=\left(C_r,C_\theta,g_{\phi\phi}C^{\phi}\right);
\end{eqnarray}
here, $\gamma_{\phi\phi}=g_{\phi\phi}$ ($u_\phi=0$ for ZAMO) and,
consequently, $\gamma^{\phi\phi}g_{\phi\phi}=1$.

ZAMO tetrad components $^{\rm{3D}}C^{(i)}$ are given as
\begin{eqnarray}
\label{projekce}
^{\rm{3D}}C^{(i)}=^{\rm{3D}}C_{(i)}=e_{(i)j}^{\rm{3D}}C_{j}\equiv
\left(\sqrt{g_{rr}}C^{r},\sqrt{g_{\theta\theta}}C^{\theta},\sqrt{g_{\phi\phi}}C^{\phi}\right).
\end{eqnarray}
The hypersurface components of the phase space constituents $x^{i}$ and $\pi_{i}$,
as measured by ZAMO, are then:
\begin{eqnarray}
\label{FIDOcomp}
^{\rm{3D}}x^{(i)}&=&(\sqrt{g_{rr}}r,\sqrt{g_{\theta\theta}}\theta,\sqrt{g_{\phi\phi}}\phi),\\
^{\rm{3D}}\pi_{(i)}&=&\left(\sqrt{g^{rr}}\pi_{r},\sqrt{g^{\theta\theta}}\pi_{\theta},\frac{1}{\sqrt{g_{\phi\phi}}}L\right).
\end{eqnarray}
The spatial 3-metric is
\begin{equation}
\label{ZAMOmetric}
ds^2=\delta_{(i)(j)}dx^{(i)}dx^{(j)}+O(|x^{(k)}|^2)dx^{(i)}dx^{(j)},
\end{equation}
where $x^{(k)}$ represents the spatial distance from the origin of the
tetrad, i.e.\ ZAMO's current location. ZAMO is not an inertial observer,
which generally causes the first order corrections $O(|x^{(k)}|)$ to the
Minkowskian metric $g_{(i)(j)}=\eta_{(i)(j)}$. But these  do not enter
the spatial part of the metric. Thus the 3-metric within the spatial
hypersurface is a Euclidean one, with the deviations of second order in
the distance from the spatial origin on ZAMO's world-line.

We will use ZAMO's metric at distances up to the value of the threshold
parameter $\varepsilon$. The Euclidean metric according to eq.\
(\ref{ZAMOmetric}) will be therefore justified if
$\frac{\varepsilon^2}{P^2}\ll{}1$, where $P$ stands for a constant (for
a given ZAMO). $P$ has the dimension of length and characterizes the
curvature of the hypersurface. We suggest setting $P\equiv{}K^{-1/4}$,
where $K=R^{\mu\nu\xi\pi}R_{\mu\nu\xi\pi}$ represents the Kretschmann
scalar evaluated from the Riemann curvature tensor. In the case of the
Kerr black hole the Kretschmann scalar may be expressed in a
surprisingly simple form \citep{henry}. While constructing the Recurrence
Plots, we check whether the condition $\frac{\varepsilon^2}{P^2}\ll{}1$
remains satisfied.

The above-mentioned adoption of preferred observers is needed in
order to maintain an operational criterion of chaos and be able to
formulate an explicit form of the equations for RQA measures in a
curved spacetime. Here the notion of the phase space distance plays
a role. In Kerr metric (or another axially symmetric stationary
spacetime), Fiducial Observers (a.k.a. FIDOs) represent a natural
selection of preferred observers. Obviously, this option is not
unique, and so a detailed appearance of the recurrence plots is also
ambiguous to certain extent. But not so the main conclusions that
we infer regarding the chaoticness versus regularity of the system
behavior, because this distinction can be eventually traced down to
the exponential versus polynomial growth of the separation with the
particle proper time along neighboring trajectories.

We can deduce the kind of transformation between different
families of observers that could affect our conclusions: these are
transformations involving exponential dependencies on
observer's phase-space position. For example transformation to
accelerated frames and spacetime points in the vicinity of
singularities may need a special consideration, as well as the
investigation of highly dynamical spacetimes that are lacking
symmetries. On the other hand, selecting LNRF to define ZAMOs in
(weakly perturbed) Kerr metric outside the black hole horizon
appears to be a well-substantiated choice.

Similar arguments for the adoption of preferred observers on the
basis of spacetime symmetries have been elaborated in greater detail
by \citet{karas92} in the context of Ernst's
magnetized black hole, which is another particularly simple (static)
exact solution of Einstein-Maxwell equations exhibiting the onset of
chaos as the magnetic field strength is increased.

\end{document}